\documentclass[11pt]{article}
\usepackage{cite}
\usepackage[numbers,sort&compress]{natbib}

\usepackage{amsmath}
\usepackage{graphicx}
\usepackage{amsfonts}
\usepackage{bm}
\usepackage{amssymb}
\usepackage{epsfig}
\usepackage{color}
\usepackage{psfrag}
\usepackage{mathrsfs}
\usepackage{esint}
\usepackage{pifont}
\usepackage{amsthm}
\usepackage{url}
\usepackage{nicefrac} % For comparison
\numberwithin{equation}{section}
\usepackage{stackengine}
\usepackage{comment}
\usepackage[mathscr]{euscript}
\usepackage[cal=esstix,scr=boondox,bb= pazo]{mathalfa}
\usepackage{stmaryrd,scalerel} % questa package genera la doppia parentesi quadra: \llbracket \rrbracket % \llparenthesis 
\usepackage{subcaption}

\newtheorem{theorem}{Theorem}[section]
\newtheorem{corollary}{Corollary}[theorem]
\newtheorem{lemma}[theorem]{Lemma}
\newtheorem{definition}[theorem]{Definition}
\usepackage{tikz}
\usetikzlibrary{calc}
\usepackage{relsize}
\usepackage{bm}
\usepackage{moresize}
\usepgflibrary{shapes.geometric}
\usetikzlibrary{chains, positioning, quotes}
\usetikzlibrary{shapes,snakes}
\usetikzlibrary{snakes}
\usetikzlibrary{automata}
\usetikzlibrary{positioning}
\usetikzlibrary{arrows,decorations.pathmorphing,patterns,backgrounds,positioning,fit,petri}
\usetikzlibrary{automata}
\usetikzlibrary{graphs}
\usetikzlibrary{shadows}
\definecolor{colordiagram1}{RGB}{155, 255, 217}
\definecolor{colordiagram2}{RGB}{255, 148, 37} %colordiagram2
\definecolor{myyellow}{RGB}{255, 207, 0}
\newcommand{\Os}{ {\color{black}{ {\mathcal O} }} }

\newcommand{\im}{{\rm i}}

\def\dashint{\,\ThisStyle{\ensurestackMath{\stackinset{c}{.2\LMpt}{c}{.5\LMpt}{\SavedStyle-}{\SavedStyle\phantom{\int}}}\setbox0=\hbox{$\SavedStyle\int\,$}\kern-\wd0}\int}
\begin{document} \setcounter{page}{0}
\topmargin 0pt
\oddsidemargin 5mm
\renewcommand{\thefootnote}{\arabic{footnote}}
\newpage
\setcounter{page}{1}
\topmargin 0pt
\oddsidemargin 5mm
\renewcommand{\thefootnote}{\arabic{footnote}}
\newpage

\begin{titlepage}
\begin{flushright}
%SISSA 40/2012/EP \\
%DFTT 9/2007
\end{flushright}
\vspace{0.5cm}
\begin{center}
{\large {\bf Multipoint correlation functions at phase separation. \\Exact results from field theory}}\\
\vspace{1.8cm}
{\large Alessio Squarcini$^{1,2,\natural}$ }\\
\vspace{0.5cm}
{\em $^1$Max-Planck-Institut f\"ur Intelligente Systeme,\\
Heisenbergstr. 3, D-70569, Stuttgart, Germany}\\
{\em $^2$IV. Institut f\"ur Theoretische Physik, Universit\"at Stuttgart,\\
Pfaffenwaldring 57, D-70569 
Stuttgart, Germany}\\
\end{center}
\vspace{1.2cm}

\renewcommand{\thefootnote}{\arabic{footnote}}
\setcounter{footnote}{0}

\begin{center}
\today
\end{center}

\begin{abstract}
We consider near-critical two-dimensional statistical systems with boundary conditions inducing phase separation on the strip. By exploiting low-energy properties of two-dimensional field theories, we compute arbitrary $n$-point correlation of the order parameter field. Finite-size corrections and mixed correlations involving the stress tensor trace are also discussed. As an explicit illustration of the technique, we provide a closed-form expression for a three-point correlation function and illustrate the explicit form of the long-ranged interfacial fluctuations as well as their confinement within the interfacial region.
\end{abstract}

\vfill
$^\natural$squarcio@is.mpg.de%, $^\flat$antonio.tinti@uniroma1.it
\end{titlepage}

\newpage
\tableofcontents

%==================================================================================
\section{Introduction}
\label{sec_1}

The characterization of the fluctuating interface separating coexisting phases is a longstanding problem in classical statistical mechanics \cite{Gallavotti}. A conspicuous amount of investigations in the field of interfacial phenomena has been stimulated by the need of a theoretical understanding and also because of numerous technological applications triggered by capillary forces and wetting effects at the nanoscale \cite{BEIMR}. The current understanding of interfacial behavior benefitted from several theoretical approaches based on microscopic descriptions formulated within lattice models, effective models, renormalization group, and numerical simulations; we refer to \cite{gennes_wetting_1985, dietrich_wetting_1988, FLN, BEIMR} for general reviews on the subject.

The two-dimensional case holds a central role because of the availability of non-perturbative techniques which lead to exact solvability of the strongly fluctuating regime. Among exactly solvable planar systems, the Ising model occupies a predominant position in the above framework due the existence of exact solutions based on the diagonalization of the transfer matrix for a wide class of boundary conditions, which, in turn, lead to the formation of interfaces in certain planar geometries \cite{AR_1974_phasesep, AR_1974_diagonal, Abraham1980, Abraham_review}. Results based on the scaling limit of exact solutions for the planar Ising model \cite{Abraham_review} motivated the introduction of phenomenological approaches relying on the analogy between interfaces and random walks \cite{fisher_walks_1984}. Such an analogy has been exploited in order to construct effective coarse-grained descriptions based on the so-called Solid-On-Solid (SOS) models \cite{Kroll,Burkhardt1981}. In these models, spin interfaces are identified with fluctuating Onsager-Temperley strings \cite{temperley_1952}. The equilibrium statistical mechanical problem in two dimensions is thus equivalent to a quantum mechanical problem in one space dimension. Within such an analogy, the construction of the partition function by summing over interfacial configurations is thus mapped onto the evaluation of path summations in the quantum-mechanical picture \cite{VL,Kroll,Burkhardt}.

In recent years, an additional analytical framework allowed for the full exploration of interfacial phenomena in near-critical systems belonging to a wider range of universality classes in two dimensions \cite{DV}, ranging from the Ising model, the $q$-state Potts model, as well as the Ashkin-Teller model, and other models which exhibit interfacial wetting and phase separation through intermediate phases \cite{DS_bubbles}. The versatility of the field-theoretical formalism then allowed also for the investigation of the interplay between interfacial fluctuations and entropic repulsion due to a flat wall \cite{DS_wetting}, a defect line \cite{Delfino_localization}, a wedge-shaped boundary \cite{DS_wedge,DS_wedgebubble}, and its corresponding wetting/filling transition. Some of the above mentioned exact findings have been successfully tested  by means of high-precision Monte Carlo simulations \cite{DSS1}.

In most of the cases, as the ones mentioned above, the knowledge of one-point correlation functions in certain geometries is informative enough and therefore it suffices for an adequate description. A more refined characterization of the interfacial behavior, however, demands the knowledge of higher-order correlation functions, with the pair correlation as the simplest nontrivial representative. The occurrence of long-range correlations within the interfacial region has been established within the context of theories of inhomogeneous fluids \cite{Wertheim_1976, Weeks_1977, Evans_79, BW_1985}, effective models such as the so-called capillary wave theory \cite{BLS_1965}, and full scale numerical simulations \cite{HD_2015,MD}. The fact that the above studies refer either to space dimension $d \geqslant 3$, or to effective interface models arising from specific assumptions, prevents the straight application of them to the strongly fluctuating regime which characterizes the scenario in $d=2$.

In order to avoid the introduction of ad-hoc assumptions -- which intrinsically characterize any effective model -- the exact investigation of the two-dimensional case has to be inevitably built on a formalism which is based on the truly fundamental degrees of freedom of the system. This first-principles-based viewpoint is at the basis of the results of \cite{DS_twopoint}; there, it has been showed how field theory can be used in order to extract the exact form of interfacial correlations in real space. As a result, by examining the structure of the pair correlation function of the order parameter, it has been possible to exhibit the specific form of long-range correlations generated by phase separation.% Moreover, since the field-theoretical approach developed in \cite{DS_twopoint} applies to a wide range of models, the corresponding results for long range correlations are universal. This means that such results are shared by those systems which exhibit separation of phases through a single interface.

In this paper, we show how the field-theoretic formalism \cite{DS_twopoint} can be extended to the calculation of $n$-point correlation functions for arbitrary $n$. We will illustrate firstly the case of correlations of the order parameter field, for which we will single out those contributions which are originated by interfacial fluctuations from those which are genuinely due to bulk fluctuations. Then, we will identify the exact analytic form of order parameter correlations and interpret them according to a probabilistic picture. Closed form expressions for certain three-point correlation functions are written as an explicit illustration of the technique, which then is further specialized to the cases of the Ising and $q$-state Potts models. The theoretical analysis of correlation functions is then pushed in order to capture subleading finite-size corrections. It is then showed how the treatment of such corrections can be systematized in a power series of the small parameter $(\xi_{\rm b}/R)^{1/2}$, with $\xi_{\rm b}$ the bulk correlation length and $R$ the separation between the interface endpoints. Specific results about corrections at order $R^{-1/2}$ and $R^{-1}$ are discussed in great detail and are related to the probabilistic picture, the latter amounts to interpret those terms as effects due to interface structure. The stress tensor trace $\Theta$ and its $n$-point correlations are also considered, as well as mixed correlators involving both $\Theta$ and spin fields.

This paper is organized as follows. In Sec. \ref{sec_2}, we laid the basis for the calculation of $n$-point correlation functions of the order parameter profile. The calculation is thus broken into successive steps which are structured into subsections. The connected part of the correlation function, which is examined in Sec.~\ref{sec_2_1}, is expressed through $n$-body cluster functions, the latter are constructed explicitly in Sec.~\ref{sec_2_2}. The disconnected parts of the correlation function are computed in Sec.~\ref{sec_2_3}. The full result for the correlation function is thus supplied in Sec.~\ref{sec_2_4} in general terms together with explicit applications to Ising and $q$-state Potts models. Secs.~\ref{sec_3} and \ref{sec_4} deal with subleading corrections at orders $R^{-1/2}$ and $R^{-1}$, respectively, and their emergence within the probabilistic picture. Conclusive remarks and a summary with a description of future perspectives is outlined in Sec.~\ref{sec_5}. Two appendices contain additional mathematical details related to the buildout of the material covered in Sec.~\ref{sec_2}.

%==================================================================================
\section{Spin correlations: field theoretical derivation}
\label{sec_2}
\begin{figure}[htbp]
\centering
\begin{tikzpicture}[thick, line cap=round, >=latex, scale=0.5]
\tikzset{fontscale/.style = {font=\relsize{#1}}}
\draw[thin, fill=green!30] (-1.5, 3.3) circle (3pt) node[right] {${\sigma_{1}(x_{1},y_{1})}$};;
\draw[thin, fill=green!30] (1.8, 1.0) circle (3pt) node[right] {${\sigma_{2}(x_{2},y_{2})}$};;
\path (0,0.0) node () {} (1.8, -0.7) node (s4) { $\vdots$ };
\draw[thin, fill=green!30] (1.8, -1.8) circle (3pt) node[right] {${\sigma_{n-1}(x_{n-1},y_{n-1})}$};;
\draw[thin, fill=green!30] (0.6, -3.0-0.5) circle (3pt) node[right] {${\sigma_{n}(x_{n},y_{n})}$};;
\draw[thin, dashed, ->] (-10, 0) -- (11, 0) node[below] {$x$};
\draw[thin, dashed, -] (0, -5) -- (0, 2.5) node[left] {};
\draw[thin, dashed, -] (0, 4.1) -- (0, 5) node[left] {};
\draw[thin, dashed, ->] (0, 6) -- (0, 7) node[left] {$y$};
\draw[very thick, red, -] (0, 5) -- (10, 5) node[] {};
\draw[very thick, red, -] (0, -5) -- (10, -5) node[] {};
\draw[very thick, blue, -] (-10, 5) -- (0, 5) node[] {};
\draw[very thick, blue, -] (-10, -5) -- (0, -5) node[] {};
\draw[thin, fill=white] (0, -5) circle (3pt) node[below] {$(0,-R/2)$};;
\draw[thin, fill=white] (0, 5) circle (3pt) node[above] {$(0,R/2)$};;
\draw[thin, fill=white] (-5, 5) circle (0pt) node[above] {${\color{blue}{a}}$};;
\draw[thin, fill=white] (-5, -5) circle (0pt) node[below] {${\color{blue}{a}}$};;
\draw[thin, fill=white] (5, 5) circle (0pt) node[above] {${\color{red}{b}}$};;
\draw[thin, fill=white] (5, -5) circle (0pt) node[below] {${\color{red}{b}}$};;
\end{tikzpicture}
\caption{The strip geometry with $ab$ boundary conditions. The spin fields which define the correlation function $G_{n}$ are illustrated with green circles.}
\label{fig_geometry}
\end{figure}
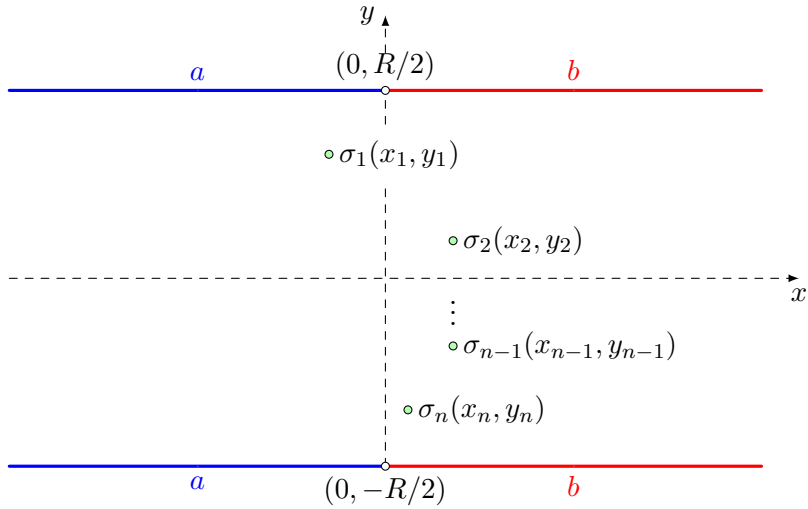

We illustrate the calculation of the $n$-point correlation function of the spin field on the finite strip showed in Fig.~\ref{fig_geometry}. The quantities of interest are the correlation functions
\begin{equation}
\label{01}
G_{n}(\bm{x}_{1}, \dots, \bm{x}_{n}) = \langle \sigma_{1}(\bm{x}_{1}) \cdots \sigma_{n}(\bm{x}_{n}) \rangle_{ab} \, ,
\end{equation}
in which the notation $\langle \cdots \rangle_{ab}$ stands for the statistical average on the strip with $ab$ boundary conditions (see Fig.~\ref{fig_geometry}) and $\sigma_{j}(x_{j},y_{j}) \equiv \sigma_{j}$ is the spin field in the point $\bm{x}_{j} = (x_{j},y_{j}) \in \mathbb{R} \times (-R/2,R/2)$ on the Euclidean plane. The following ordering is considered in (\ref{01}): $y_{1} > y_{2} > \dots > y_{n}$. The subscript $j$ in $\sigma_{j}$ labels the $j$-th spin field; in the most generic case the spin field carries a color index such as in the field theory associated to the scaling $q$-state Potts model (see \cite{DC98} and Sec.~\ref{sec_triplet}).

The switching of boundary condition from $a$ to $b$ at $x=0$ along the edges $y=\pm R/2$ is implemented within the field-theoretical language through the boundary state formalism. The correlation function (\ref{01}) is thus written as follows
\begin{equation}
\label{02}
G_{n}(\bm{x}_{1}, \dots, \bm{x}_{n}) = \frac{1}{\mathcal{Z}_{ab}(R)} \langle B_{ab}(0,\im R/2) \vert \sigma_{1}(\bm{x}_{1}) \cdots \sigma_{n}(\bm{x}_{n}) \vert B_{ab}(0,-\im R/2) \rangle \, ,
\end{equation}
where
\begin{equation}
\label{03}
\mathcal{Z}_{ab}(R) = \langle B_{ab}(0,\im R/2) \vert B_{ab}(0,-\im R/2) \rangle
\end{equation}
is the partition function and $\vert B_{ab}\rangle$ is the boundary state with the inhomogeneous boundary condition shown in Fig.~\ref{fig_geometry}. We refer to \cite{GZ} for translationally invariant boundaries in the framework of massive integrable field theories. The boundary state is expanded in the basis of bulk excitations compatible with the topological charge of the boundary, namely
\begin{equation}
\label{04}
\vert B_{ab}(x,t) \rangle = \textrm{e}^{\im xP- \im t H} \biggl[ \int\frac{\textrm{d}\theta}{2\pi}f_{ab}(\theta) \vert K_{ab}(\theta) \rangle + \sum_{c \neq a,b} \int\frac{\textrm{d}\theta\textrm{d}\theta^{\prime}}{(2\pi)^{2}} f_{acb} \vert K_{ac}(\theta) K_{cb}(\theta^{\prime}) \rangle + \dots \biggr] \, ,
\end{equation}
where $\vert K_{ab}(\theta) \rangle$ is the kink state corresponding to a topological particle with mass $m$ which interpolates between vacua $\vert \Omega_{a} \rangle$ and $\vert \Omega_{b} \rangle$, and $\theta$ is the rapidity variable. Within the dictionary of phase separation, the vacuum $\vert\Omega_{a}\rangle$ is identified with the homogeneous system filled by phase $a$ (pure phase). The second and subsequent terms on the right hand side of (\ref{04}) stem from the propagation of multi-kink states. The simplest of them -- the double-kink state $\vert K_{ac} K_{cb} \rangle$ -- plays the dominant role only when the single-kink state $\vert K_{ab} \rangle$ is absent in the expansion of the boundary state. For those models and boundary conditions in which such an instance happens, the double-kink state is responsible for the formation of a double interface with an intermediate layer of phase $c$ adsorbed on the $ab$ interface \cite{DS_bubbles}. We refer to \cite{Delfino_localization} for a further characterization of models which exhibits such a phenomenology.

The amplitudes $f_{ab}(\theta)$, $f_{abc}(\theta,\theta^{\prime})$ depend on both the bulk and boundary universality classes and are known for certain integrable field theories \cite{LS, BH_2016, BPT_BFF}. For the purposes of this paper, however, it is sufficient to know only the infrared properties of the amplitude $f_{ab}(\theta)$, which is responsible for the emission of a single-kink state from the boundary. To this end, we only need the Taylor expansion around $\theta=0$, which reads $f_{ab}(\theta) = f_{ab}(0) + \Os(\theta^{2})$ for small rapidities. The absence of the linear term in $\theta$ follows by reflection symmetry around the vertical axis, $f_{ab}(\theta)=f_{ba}(-\theta)$, in conjunction with the symmetric role played by $a$ and $b$, i.e., $f_{ab}(\theta)=f_{ba}(\theta)$.

In the limit $mR \gg1$, which we consider from now on, the partition function (\ref{03}) can be computed straightforwardly thanks to a saddle-point calculation. The result is
\begin{equation}
\label{05}
\mathcal{Z}_{ab}(R) = \frac{|f_{ab}(0)|^{2}}{\sqrt{2\pi m R}} \textrm{e}^{-mR} + \Os\left(\textrm{e}^{-2mR}\right) \, ,
\end{equation}
up to higher order terms due to multi-kink states. From the boundary state formalism it is possible to identify the surface tension $\Sigma_{ab}$ associated to the creation of an interface separating the coexisting phases $a$ and $b$. The surface tension is computed as the excess free energy\footnote{In units of $k_{\textrm{B}}T$.} per unit length and is defined through the limit
\begin{equation}
\label{2007}
\Sigma_{ab} = - \lim_{R\rightarrow\infty} \frac{1}{R} \ln\frac{\mathcal{Z}_{ab}(R)}{\mathcal{Z}_{a}(R)} \, ,
\end{equation}
where $\mathcal{Z}_{a}(R)$ stands for the partition function of the strip with uniform boundary condition $a$. The latter can be determined from the boundary state $\vert B_{a} \rangle$ for a uniform boundary. These states are known exactly for several models including, among the most relevant examples for the purposes of this paper, the Ising model with a surface field \cite{GZ} and the $q$-state Potts model \cite{Chim_1995}. The state with lowest mass entering the expansion of the uniform boundary is actually the vacuum $\vert\Omega_{a}\rangle$; hence,\footnote{The symbol $\simeq$ stands for the omission of exponentially suppressed terms as in (\ref{05}).} $\mathcal{Z}_{a}(R) = \langle \Omega_{a} \vert \Omega_{a} \rangle \simeq 1$ and the interfacial tension equals the kink mass, i.e.,
\begin{equation}
\label{2008}
\Sigma_{ab} = m \, .
\end{equation}
In the symmetry broken phase the bulk correlation length $\xi_{\rm b}$ is related to the kink mass $m$ via $\xi_{\rm b}=1/(2m)$. Therefore, (\ref{2008}) implies $\Sigma_{ab} \xi_{\rm b} = 1/2$. It is worth emphasizing that such a relationship is compatible with Widom scaling \cite{Widom_1972} in two dimensions. More interestingly, thanks to calculations based on the exact solution of the planar Ising model, the relation $\Sigma_{ab} \xi_{\rm b} = 1/2$ has been proved to be valid for all subcritical temperatures. This result follows as an application of duality \cite{FF67,AGM73,Abraham_review}; see also \cite{AMSV}.

Let us move on the correlation function (\ref{02}). The time ordering of spin fields we discussed above is actually implemented in a more strict sense since consecutive spin fields in (\ref{01}) are separated by a distance much larger than the bulk correlation length $\xi_{\rm b}$ and, in a similar fashion, spin fields are also taken to be far from the boundaries, hence, $\sqrt{(x_{i}-x_{j})^{2} + (y_{i}-y_{j})^{2}} \gg 1/m$, $R/2-y_{1} \gg 1/m$ and $y_{n}+R/2 \gg 1/m$. 

We apply a spectral decomposition which amounts to insert a resolution of the identity between each pair of spin operators. The identity operator is itself expanded in terms of kink states, something that we write as follows
\begin{equation}
\label{06}
\mathbb{I} = \sum_{c_{1},\dots,c_{n} \neq a,b} \sum_{n} \int \frac{\textrm{d}\theta_{1}}{2\pi} \cdots \frac{\textrm{d}\theta_{n}}{2\pi} \frac{1}{n!} \vert K_{ac_{1}}(\theta_{1}) \cdots K_{c_{n}b}(\theta_{n}) \rangle \langle K_{bc_{n}}(\theta_{n}) \cdots K_{c_{1}a}(\theta_{1}) \vert \, ,
\end{equation}
with kink states normalized according to
\begin{equation}
\label{07}
\langle K_{ba}(\theta) \vert K_{a^{\prime}b^{\prime}}(\theta) \rangle = 2\pi \delta_{aa^{\prime}}\delta_{bb^{\prime}} \delta(\theta-\theta^{\prime}) \, .
\end{equation}

By inserting the resolution of the identity between each spin field appearing in (\ref{02}), and recalling translational invariance for bulk fields
\begin{equation}
\label{08}
\Phi(x,y) = \textrm{e}^{\im xP+yH} \Phi(0,0) \textrm{e}^{- \im xP-yH} \, ,
\end{equation}
where $H$ and $P$ are the Hamiltonian and momentum operators in relativistic quantum field theory, i.e.
\begin{equation}
\begin{aligned}
\label{09}
H \vert K_{ab}(\theta) \rangle & = m \cosh\theta \vert K_{ab}(\theta) \rangle \, , \\
P \vert K_{ab}(\theta) \rangle & = m \sinh\theta \vert K_{ab}(\theta) \rangle \, ,
\end{aligned}
\end{equation}
we can extract the space-time dependence of matrix elements and write
\begin{equation}
\label{10}
G_{n}(\bm{x}_{1}, \dots, \bm{x}_{n}) \simeq \frac{1}{\mathcal{Z}_{ab}(R)} \int_{\mathbb{R}^{n+1}} \prod_{j=1}^{n+1} \frac{\textrm{d}\theta_{j}}{2\pi} f_{ab}^{*}(\theta_{1}) f_{ab}(\theta_{n+1}) \left( \prod_{j=1}^{n} \mathcal{M}_{ab}^{\sigma_{j}}(\theta_{j} \vert \theta_{j+1}) \right) \mathcal{U}_{n}(\theta_{1},\dots,\theta_{n+1}) \, ,
\end{equation}
where $\mathcal{M}_{ab}^{\sigma_{j}}(\theta_{j}\vert\theta_{j+1})$ is the matrix element of the spin field $\sigma_{j}$
\begin{equation}
\begin{aligned}
\label{}
\mathcal{M}_{ab}^{\sigma_{j}}(\theta_{j}\vert\theta_{j+1}) & \equiv \langle K_{ba}(\theta_{j}) \vert \sigma_{j}(0,0) \vert K_{ab}(\theta_{j+1}) \rangle \, .
\end{aligned}
\end{equation}
The dependence through the coordinates is encoded in the function
\begin{equation}
\begin{aligned}
\label{12}
\mathcal{U}_{n}(\theta_{1},\dots,\theta_{n+1}) & = \textrm{e}^{-m\left(R/2-y_{1}\right)\cosh\theta_{1}} \left( \prod_{j=1}^{n-1} \textrm{e}^{-m(y_{j}-y_{j+1})\cosh\theta_{j+1} } \right) \textrm{e}^{-m\left(y_{n}+R/2\right)\cosh\theta_{n+1}} \times \\ & \times \prod_{j=1}^{n} \textrm{e}^{\im m(\sinh\theta_{j}-\sinh\theta_{j+1})x_{j}} \, .
\end{aligned}
\end{equation}
Matrix elements of the spin field $\mathcal{M}_{ab}^{\sigma_{j}}$ are decomposed into a connected part and a disconnected one. The connected part is expressed in terms of the spin field two-particle form factor $F_{aba}^{\sigma_{j}}(\theta_{j}-\theta_{j+1}+\im\pi)$, thus
\begin{equation}
\begin{aligned}
\label{14012021_1627}
\mathcal{M}_{ab}^{\sigma_{j}}(\theta_{j}\vert\theta_{j+1}) & =  F_{aba}^{\sigma_{j}}(\theta_{j}-\theta_{j+1}+\im\pi) + \left\{
\begin{array}{ll}
2\pi \langle\sigma_{j}\rangle_{a} \, \delta(\theta_{j}-\theta_{j+1}) \,, \quad \text{right}\,, \\
\\
2\pi \langle\sigma_{j}\rangle_{b} \, \delta(\theta_{j}-\theta_{j+1}) \,, \quad \text{left}\,. \\
\end{array} 
\right.
\end{aligned}
\end{equation}
By adopting a pictorial representation for matrix elements, (\ref{14012021_1627}) reads
\begin{equation}
\label{14012021_1619}
\mathcal{M}_{ab}^{\sigma_{j}}(\theta_{j} \vert \theta_{j+1}) \,\,
=
\vcenter{\hbox{
\begin{tikzpicture}[baseline={([yshift=-.6ex]current bounding box.center)},vertex/.style={anchor=base, circle, minimum size=50mm, inner sep=0pt}]
\draw[-,black]   (0, -1.8) -- (0, -0.6);
\draw[-,black]   (0, 0.6) -- (0, 1.8);
\path (0,0.0) node[circle, draw, fill=green!30] (s1) {$\sigma_{j}$};
\path (0,0.0) node () {} (-0.75,0) node (s4) { ${\color{blue}{a}}$ };
\path (0,0.0) node () {} (0.75,0) node (s4) { ${\color{red}{b}}$ };
\path (0,0.0) node () {} (0.26,1.6) node (s4) { ${\color{black}{\theta_{j}}}$ };
\path (0,0.0) node () {} (0.47,-1.6) node (s4) { ${\color{black}{\theta_{j+1}}}$ };
\end{tikzpicture}
}}
=
\vcenter{\hbox{
\begin{tikzpicture}[baseline={([yshift=-.6ex]current bounding box.center)},vertex/.style={anchor=base, circle, minimum size=50mm, inner sep=0pt}]
\draw[-,black]   (0, -1.8) -- (0, 1.8);
\path (0,0.0) node[circle, draw, fill=green!30] (s1) {$\sigma_{j}$};
\path (0,0.0) node () {} (-0.75,0) node (s4) { ${\color{blue}{a}}$ };
\path (0,0.0) node () {} (0.75,0) node (s4) { ${\color{red}{b}}$ };
\path (0,0.0) node () {} (0.26,1.6) node (s4) { ${\color{black}{\theta_{j}}}$ };
\path (0,0.0) node () {} (0.47,-1.6) node (s4) { ${\color{black}{\theta_{j+1}}}$ };
\end{tikzpicture}
}}
\quad
+
\vcenter{\hbox{
\begin{tikzpicture}[baseline={([yshift=-.6ex]current bounding box.center)},vertex/.style={anchor=base, circle, minimum size=50mm, inner sep=0pt}]
\def\u{0.95}
\def\v{0.65}
\path (0,0.0) node[circle, draw, fill=green!30] (s1) {$\sigma_{j}$};
\path (0,0.0) node () {} (-0.75,0) node (s4) { ${\color{blue}{a}}$ };
\path (0,0.0) node () {} (1.05,0) node (s4) { ${\color{red}{b}}$ };
\path (0,0.0) node () {} (0.26,1.6) node (s4) { ${\color{black}{\theta_{j}}}$ };
\path (0,0.0) node () {} (0.47,-1.6) node (s4) { ${\color{black}{\theta_{j+1}}}$ };
\draw[]
(0, -1.8) ..controls +(0, \u) and ( $(0.8, 0) - (0, +\v)$ )..
(0.8, 0) ..controls +(0, \v) and ( $(0, 1.8) - (0, +\u)$ )..
(0, 1.8);
\end{tikzpicture}
}} \, .
\end{equation}

The disconnected part originates a Dirac delta corresponding to particle annihilation. The vacuum expectation value which multiplies the Dirac delta is $\langle\sigma_{j}\rangle_{a}$ if the two kinks are annihilated by passing right aside the spin field, as depicted in (\ref{14012021_1619}). Conversely, the overall vacuum expectation value is $\langle\sigma_{j}\rangle_{b}$ for the passage aside left. The right-left alternative is ultimately responsible for the presence of the annihilation pole\footnote{See \cite{Smirnov} for a comprehensive treatment of form factors in integrable quantum field theories.} \cite{DC98}, whose behavior at small rapidity differences reads
\begin{equation}
\label{27052020_01}
F_{aba}^{\sigma_{j}}(\theta_{j}-\theta_{j+1} + \im \pi) = \frac{ \im \Delta\langle\sigma_{j}\rangle }{ \theta_{j}-\theta_{j+1} } + c_{ab}^{(\sigma_{j})} + \Os(\theta_{j}-\theta_{j+1}) \, ,
\end{equation}
where
\begin{equation}
\label{29012021_0814}
\Delta\langle \sigma_{j} \rangle \equiv \langle \sigma_{j} \rangle_{a} - \langle \sigma_{j} \rangle_{b}
\end{equation}
is the jump of vacuum expectation values across the $ab$ interface. The coefficient $c_{ab}^{(\sigma_{j})}$ and the subsequent ones appearing in (\ref{27052020_01}) are known for integrable field theories \cite{YZ, DC98}. By plugging the expansion (\ref{14012021_1619}) into the product $\prod_{j=1}^{n} \mathcal{M}_{ab}^{\sigma_{j}}(\theta_{j}\vert\theta_{j+1})$, we obtain a decomposition of the $n$-point correlation function $G_{n}$ which comprises a connected part and a sequence of disconnected parts.

The rest of this section is organized as follows: in Sec.~\ref{sec_2_1}, we address the calculation of the connected part. Such a task will require the introduction of a certain class of special functions -- which will be termed \emph{$n$-body cluster functions} -- whose definition and main properties will be provided in Sec. \ref{sec_2_2}; there, we will also provide the result for the connected part of $G_{n}$. The disconnected parts of the correlation function will be investigated in Sec.~\ref{sec_2_3} and the \emph{full} result for $G_{n}$ will be supplied Sec.~\ref{sec_2_4}.

%==================================================================================
\subsection{Connected part and cluster functions}
\label{sec_2_1}
The connected part of the $n$-point correlation function $G_{n}(\bm{x}_{1}, \dots, \bm{x}_{n})$ is obtained from the product of the two-particle form factors $F_{aba}^{\sigma_{j}}(\theta_{j}-\theta_{j+1}+\im\pi)$ in (\ref{10}). Diagrammatically, this operation corresponds to stack the decomposition (\ref{14012021_1619}) for $j=1,\dots,n$ and retain the diagram in which all spin fields are connected, as illustrated in (\ref{14012021_1724}). The open necklace shown in the right hand side of (\ref{14012021_1724}) is the diagram which we will examine in this section. 
\begin{equation}
\label{14012021_1724}
\prod_{j=1}^{n} \mathcal{M}_{ab}^{\sigma_{j}}(\theta_{j} \vert \theta_{j+1})
\quad
=
\quad
\vcenter{\hbox{
%\begin{tikzpicture}[baseline={([yshift=-.6ex]current bounding box.center)},vertex/.style={anchor=base, circle, fill=green!30, minimum size=8.2pt, inner sep=0pt}]
\begin{tikzpicture}[baseline={([yshift=-.6ex]current bounding box.center)},vertex/.style={anchor=base, circle, minimum size=50mm, inner sep=0pt}]
\path (0, 4.0) node[circle, draw, fill=green!30] (s1) {$\sigma_{1}$} (0, 2.0) node[circle, draw, fill=green!30](s2) {$\sigma_{2}$};
\path (0, 0.0) node[circle, draw, fill=green!30] (s3) {$\sigma_{3}$} (0, -4.0) node[circle, draw, fill=green!30](sn) {$\sigma_{n}$};
\draw[-,black]   (0, 4+0.6) -- (0, 5+0.6);
\draw[-,black]   (0, 4-0.6) -- (0, 2+0.6);
\draw[-,black]   (0, 2-0.6) -- (0, 0+0.6);
\draw[-,black]   (0, 0-0.6) -- (0, -2+0.6);
\draw[-,black]   (0, -4+0.6) -- (0, -4+0.6+0.8);
\draw[-,black]   (0, -4-0.6) -- (0, -5-0.6);
%\draw[-,black]   (s1) -- (s2);
%\draw[-,black]   (s1) -- (0,2);
%\draw[-,black]   (s2) -- (0,-2);
\path (0,0.0) node () {} (-0.75,0) node (s4) { ${\color{blue}{a}}$ };
\path (0,0.0) node () {} (0.75,0) node (s4) { ${\color{red}{b}}$ };
\path (0,0.0) node () {} (0.25, 5.1) node { ${\color{black}{\theta_{1}}}$ };
\path (0,0.0) node () {} (0.25, 3.0) node { ${\color{black}{\theta_{2}}}$ };
\path (0,0.0) node () {} (0.25, 1.0) node { ${\color{black}{\theta_{3}}}$ };
\path (0,0.0) node () {} (0.25, -1.9) node { ${\color{black}{\vdots}}$ };
\path (0,0.0) node () {} (0.25, -3.0) node { ${\color{black}{\theta_{n}}}$ };
\path (0,0.0) node () {} (0.44, -5.015) node { ${\color{black}{\theta_{n+1}}}$ };
%\node[vertex,draw, line width=0.5pt, densely dotted, fill=white!30] (s1) at (-1,0)  { \ssmall{1} };
\end{tikzpicture}
}}
\quad
=
\quad
\vcenter{\hbox{
%\begin{tikzpicture}[baseline={([yshift=-.6ex]current bounding box.center)},vertex/.style={anchor=base, circle, fill=green!30, minimum size=8.2pt, inner sep=0pt}]
\begin{tikzpicture}[baseline={([yshift=-.6ex]current bounding box.center)},vertex/.style={anchor=base, circle, minimum size=50mm, inner sep=0pt}]
\draw[-,black]   (0, 5+0.6) -- (0, -2+0.6);
\draw[-,black]   (0, -5-0.6) -- (0, -4+0.6+0.8);
\path (0, 4.0) node[circle, draw, fill=green!30] (s1) {$\sigma_{1}$} (0, 2.0) node[circle, draw, fill=green!30](s2) {$\sigma_{2}$};
\path (0, 0.0) node[circle, draw, fill=green!30] (s3) {$\sigma_{3}$} (0, -4.0) node[circle, draw, fill=green!30](sn) {$\sigma_{n}$};
%\draw[-,black]   (s1) -- (s2);
%\draw[-,black]   (s1) -- (0,2);
%\draw[-,black]   (s2) -- (0,-2);
\path (0,0.0) node () {} (-0.75,0) node (s4) { ${\color{blue}{a}}$ };
\path (0,0.0) node () {} (0.75,0) node (s4) { ${\color{red}{b}}$ };
\path (0,0.0) node () {} (0.25, 5.1) node { ${\color{black}{\theta_{1}}}$ };
\path (0,0.0) node () {} (0.25, 3.0) node { ${\color{black}{\theta_{2}}}$ };
\path (0,0.0) node () {} (0.25, 1.0) node { ${\color{black}{\theta_{3}}}$ };
\path (0,0.0) node () {} (0.25, -1.9) node { ${\color{black}{\vdots}}$ };
\path (0,0.0) node () {} (0.25, -3.0) node { ${\color{black}{\theta_{n}}}$ };
\path (0,0.0) node () {} (0.44, -5.015) node { ${\color{black}{\theta_{n+1}}}$ };
%\node[vertex,draw, line width=0.5pt, densely dotted, fill=white!30] (s1) at (-1,0)  { \ssmall{1} };
\end{tikzpicture}
}}
\quad
+
\textrm{disconnected} \, .
\end{equation}

According to (\ref{27052020_01}), the leading low-energy behavior of the connected diagram on the right hand side of (\ref{14012021_1724}) is captured by the product of kinematical poles, therefore
\begin{equation}
\label{29012021_1058}
\prod_{j=1}^{n} \mathcal{M}_{ab}^{\sigma_{j}}(\theta_{j} \vert \theta_{j+1}) = \prod_{j=1}^{n} \frac{ \im \Delta\langle\sigma_{j}\rangle }{ \theta_{j}-\theta_{j+1} } + \Os(\{\theta\}^{-n+1}) \, ,
\end{equation}
where $\Os(\{\theta\}^{-n+1})$ stands for terms which are homogeneous functions of order $-n+1$ in the rapidity variables. Subsequent terms in the above expansion lead to subleading finite-size corrections of the correlation function whose systematic analysis will be carried out in Sec. \ref{sec_3}.

The computation of the integrals in (\ref{10}) proceeds by expanding the function $\mathcal{U}_{n}$ at small rapidities. To this end it is convenient to rescale rapidities through the change of variables $\theta_{j} \rightarrow \sqrt{2/(mR)} \theta_{j}$. The function $\mathcal{U}_{n}$ becomes $\mathcal{U}_{n}(\theta_{1},\dots,\theta_{n+1}) \rightarrow \textrm{e}^{-mR} \mathcal{Y}_{n}(\theta_{1},\dots,\theta_{n+1})$, where
\begin{equation}
\label{ }
\mathcal{Y}_{n}(\theta_{1},\dots,\theta_{n+1}) = \left( \prod_{j=0}^{n} \textrm{e}^{-(\tau_{j}-\tau_{j+1})\theta_{j+1}^{2}/2 } \right) \prod_{j=1}^{n} \textrm{e}^{\im (\theta_{j}-\theta_{j+1})\eta_{j}} \, ,
\end{equation}
with $\tau_{0} \equiv 1$, $\tau_{n+1} \equiv -1$ and
\begin{equation}
\label{06012021_1029}
\eta_{j} = x_{j} / \lambda \, , \qquad \lambda = \sqrt{R/(2m)} \, , \qquad \tau_{j}=2y_{j}/R \, .
\end{equation}
In order to ease the calculations, we introduce a compact notation for the evaluation of $(n+1)$-fold integrals with respect rapidities; we define
\begin{equation}
\label{14012021_1631}
\Lbag \Psi(\theta_{1},\dots,\theta_{n+1}) \Rbag_{\eta_{1},\tau_{1}; \dots ;\eta_{n},\tau_{n}} \equiv 2\sqrt{\pi} \dashint_{\mathbb{R}^{n+1}}\prod_{j=1}^{n+1}\frac{\textrm{d}\theta_{j}}{2\pi} \, \Psi(\theta_{1},\dots,\theta_{n+1}) \mathcal{Y}_{n}(\theta_{1},\dots,\theta_{n+1}) \, , %\OK
\end{equation}
where $\Psi(\theta_{1},\dots,\theta_{n+1})$ is a function of the rapidities. The symbol $\dashint$ stands for the principal value of the integral. The need for the principal value follows from the fact that spin field matrix elements exhibit a kinematical pole \cite{DV}. The result of the integrations in (\ref{14012021_1631}) is, in general, a function of the rescaled coordinates $\{ \eta_{j} , \tau_{j} \}$ which are indicated as a subscript. For the sake of simplicity, we will omit the subscripts when there is no ambiguity and we shall write $\Lbag \Psi \Rbag$ in place of (\ref{14012021_1631}).

The connected part of the correlation function is computed as follows
\begin{equation}
\label{27052020_02}
G_{n}^{\rm CP}(\bm{x}_{1}, \dots, \bm{x}_{n}) = \scaleleftright[1.2ex]{\Lbag}{ \prod_{j=1}^{n} \frac{ \im \Delta\langle\sigma_{j}\rangle }{\theta_{j}-\theta_{j+1}}    }{\Rbag} + \Os(R^{-1/2}) \, . %\OK
\end{equation}
As it has been illustrated in \cite{DV} and \cite{DS_twopoint}, a simple route for the calculation of spin field matrix elements is to take a first derivative with respect to the horizontal coordinate. Thanks to this procedure it is possible to get rid of the kinematical pole $1/(\theta_{j}-\theta_{j+1})$ by taking the first derivative with respect to $\eta_{j}$. Hence, by applying the differential operator $\partial_{x_{1}} \cdots \partial_{x_{n}}$, we get rid of kinematical poles and we are left with
\begin{equation}
\label{27052020_04}
\partial_{x_{1}} \cdots \partial_{x_{n}} G_{n}^{\rm CP}(\bm{x}_{1}, \dots, \bm{x}_{n}) = \lambda^{-n}  \, \biggl[ \prod_{j=1}^{n} (-\Delta\langle\sigma_{j}\rangle) \biggr] \Lbag 1 \Rbag \, . %\OK
\end{equation}
The quantity indicated with $\Lbag 1 \Rbag$ amounts to compute a $(n+1)$-fold gaussian integral. Anticipating some results, it is convenient to write the outcome of the integration in the following way
\begin{equation}
\label{27052020_05}
\Lbag 1 \Rbag_{\eta_{1},\tau_{1}; \dots ;\eta_{n},\tau_{n}} = \lambda^{n} P_{n}(x_{1},y_{1};\dots ;x_{n},y_{n}) \, . %\OK
\end{equation}
We will show in Sec.~\ref{sec_4} that $P_{n}$ is the joint probability density of a Brownian bridge. This means that $P_{n} \textrm{d}x_{1} \cdots \textrm{d}x_{n}$ is the probability for the interface -- regarded as the trajectory of a Brownian particle -- to pass through \emph{all} intervals $(x_{j},x_{j}+\textrm{d}x_{j})$ at $y=y_{j}$, for $j=1,\dots,n$. Leaving the details in Appendix \ref{Appendix_Bridges}, the passage probability reads
\begin{equation}
\label{06012021_1046}
P_{n}(x_{1}, y_{1}; \dots ; x_{n}, y_{n}) = \frac{2^{n/2}}{\lambda^{n}\prod_{j=1}^{n}\kappa_{j}} \Pi_{n}( \sqrt{2}\chi_{1},\dots,\sqrt{2}\chi_{n} \vert \textsf{R}_{1 \dots n} ) \, .
\end{equation}
Some comments are in order. The dependence through the coordinates is encoded in the rescaled coordinates $\chi_{j}$ and $\tau_{j}$, which are defined by
\begin{equation}
\label{ }
\chi_{j} = \frac{x_{j}}{\kappa_{j}\lambda} \, , \quad \kappa_{j}=\sqrt{1-\tau_{j}^{2}} \, ,
\end{equation}
for $j=1,\dots,n$. Then, $\Pi_{n}$ indicates the multivariate normal distribution with correlation matrix $\textsf{R}_{1\dots n}$. Further details on the multivariate normal distribution are collected in Appendix \ref{Appendix_Bridges}. The correlation matrix is a $n \times n$ symmetric matrix with $1$ along the main diagonal and with entries
\begin{equation}
\label{06012021_1031}
\rho_{ij} = \sqrt{ \frac{1-\tau_{i}}{1+\tau_{i}} \frac{1+\tau_{j}}{1-\tau_{j}} } \, , \qquad i \leqslant j
\end{equation}
in the upper triangle. The joint passage probability is normalized such that
\begin{equation}
\label{ }
\int_{\mathbb{R}^{n}}\prod_{j=1}^{n}\textrm{d}x_{j} \, P_{n}(x_{1}, y_{1}; \dots ; x_{n}, y_{n}) = 1 \, .
\end{equation}

Marginal passage probabilities are obtained upon integration with respect to a subset of the $n$ coordinates. For instance,
\begin{equation}
\label{29012021_0917}
P_{k}(x_{1}, y_{1}; \dots ; x_{k}, y_{k}) = \int_{\mathbb{R}^{n-k}}\prod_{j=k+1}^{n}\textrm{d}x_{j} \, P_{n}(x_{1}, y_{1}; \dots ; x_{n}, y_{n}) \, .
\end{equation}
Note that $P_{k}(x_{1}, y_{1} ; \dots ; x_{k}, y_{k})$ is characterized by a $k \times k$ correlation matrix obtained by removing the last $n-k$ rows and columns of $\textsf{R}_{1 \dots n}$. Analogously, by integrating $P_{n}$ with respect to $x_{k}$, we obtain a joint passage probability in the variables with labels $1,\dots,k-1,k+1,\dots,n$ and whose correlation matrix is obtained by removing the $k^{\rm th}$ row and the $k^{\rm th}$ column of $\textsf{R}_{1 \dots n}$.

Coming back to the correlation function, (\ref{27052020_04}) reads
\begin{equation}
\label{06012021_2105}
\partial_{x_{1}} \cdots \partial_{x_{n}} G_{n}^{\rm CP}(\bm{x}_{1}, \dots, \bm{x}_{n}) = \biggl[ \prod_{j=1}^{n} \left( -\Delta\langle\sigma_{j}\rangle \right) \biggr] P_{n}(x_{1},y_{1}; \dots ; x_{n},y_{n}) \, . %\OK
\end{equation}
The above equation is actually satisfied also by the ``full'' correlation function $G_{n}$ and not necessarily by its connected part. The reason is that, as we are going to show, the action of $\partial_{x_{1}} \cdots \partial_{x_{n}}$ on disconnected terms gives zero. Such a property follows from the fact that disconnected terms depend only on a subset of coordinates $x_{1},\dots,x_{n}$.

In order to make further progresses it is convenient to adopt a systematic notation for connected correlation functions. We introduce cluster functions of order $n$ by means of the following integral representation
\begin{equation}
\begin{aligned}
\label{05012021_1653}
\mathcal{G}_{n}(\bm{x}_{1}, \dots, \bm{x}_{n}) & = \frac{ 1 }{(\pi \im)^{n} \sqrt{\pi}} \dashint_{\mathbb{R}} \textrm{d}\theta_{1} \cdots \dashint_{\mathbb{R}} \textrm{d}\theta_{n+1} \, \frac{ \mathcal{Y}_{n}(\theta_{1},\dots,\theta_{n+1}) }{ \prod_{j=1}^{n}(\theta_{j}-\theta_{j+1}) } \, , \\
& = 2^{n}\, \, \scaleleftright[1.2ex]{\Lbag}{ \prod_{j=1}^{n} \frac{ -\im }{\theta_{j}-\theta_{j+1}}    }{\Rbag} {\vphantom{\Bigg\vert}}_{\eta_{1},\tau_{1}; \dots ;\eta_{n},\tau_{n}}  \, , %\OK
\end{aligned}
\end{equation}
thanks to which it is possible to write the connected correlation function as follows
\begin{equation}
\label{07062020_02}
G_{n}^{\rm CP}(\bm{x}_{1}, \dots, \bm{x}_{n}) = \biggl[ \prod_{j=1}^{n} (-1) \widehat{\langle\sigma_{j}\rangle} \biggr] \mathcal{G}_{n}(\bm{x}_{1}, \dots, \bm{x}_{n}) \, , %\OK
\end{equation}
with
\begin{equation}
\label{29012021_0848}
\widehat{\langle\sigma_{j}\rangle}  \equiv \frac{\langle \sigma_{j} \rangle_{a} - \langle \sigma_{j} \rangle_{b}}{2}
\end{equation}
the half jump of vacuum expectation values across the $ab$ interface. The product enclosed by square brackets in (\ref{07062020_02}) contains the jumps of vacuum expectation values, which are intrinsically model-dependent quantities. Conversely, the cluster function $\mathcal{G}_{n}$ is universal in the sense that it is shared by \emph{all} models in which a single interface separates coexisting phases $a$ and $b$. We further anticipate that, thanks to the normalization in (\ref{05012021_1653}), each cluster function tends to $+1$ when all the arguments are sent to $+\infty$.

By combining the above equations, (\ref{27052020_04}) becomes
\begin{equation}
\label{27052020_06}
\partial_{x_{1}} \cdots \partial_{x_{n}} \mathcal{G}_{n}(\bm{x}_{1}, \dots, \bm{x}_{n}) = 2^{n} P_{n}(x_{1},y_{1}; \dots ; x_{n},y_{n}) \, .
\end{equation}
In order to find the cluster function $\mathcal{G}_{n}$, we integrate back with respect to $x_{1},\dots,x_{n}$. This procedure however must be followed carefully because (\ref{27052020_06}) defines cluster functions up to arbitrary functions of a subset of coordinates. For instance, the cluster function $\mathcal{G}_{2}$ would be determined up to functions of $x_{1}$ and $x_{2}$. In order to fix the cluster function in a \emph{unique} fashion, we impose a set of constraints which ensure the clustering property of correlation functions.

The above discussion can be rephrased under a slightly different angle by using the identity
\begin{equation}
\label{05012021_1713}
\frac{ \textrm{e}^{\im \eta\theta} }{ \theta } = \im \int\textrm{d}\eta \, \textrm{e}^{ \im \eta\theta} \, ,
\end{equation}
which allows us to replace each simple pole in (\ref{05012021_1653}) with an integration with respect to an auxiliary variable conjugated to a rapidity difference. Such a variable can be identified by noting that $x_{j}$, or its rescaled counterpart, $\eta_{j}$, is coupled to the rapidity difference $\theta_{j}-\theta_{j+1}$ in the function $\mathcal{Y}_{n}$. By inserting (\ref{05012021_1713}) into (\ref{05012021_1653}) and carrying out the integrations with respect to the rapidities, we find 
\begin{equation}
\begin{aligned}
\label{06012021_1010}
\mathcal{G}_{n}(\bm{x}_{1}, \dots, \bm{x}_{n}) & = \frac{1}{\pi^{n+1/2}} \int_{-\infty}^{x_{1}/\lambda}\textrm{d}\eta_{1} \cdots \int_{-\infty}^{x_{n}/\lambda}\textrm{d}\eta_{n} \dashint_{\mathbb{R}}\textrm{d}\theta_{1} \cdots \dashint_{\mathbb{R}}\textrm{d}\theta_{n+1} \, \mathcal{Y}_{n}(\theta_{1},\dots,\theta_{n+1}) + \mathcal{R}_{n} \, , \\
& = 2^{n} \int_{-\infty}^{x_{1}/\lambda}\textrm{d}\eta_{1} \cdots \int_{-\infty}^{x_{n}/\lambda}\textrm{d}\eta_{n} \, \Lbag 1 \Rbag_{\eta_{1},\tau_{1}; \dots ;\eta_{n},\tau_{n}} + \mathcal{R}_{n}(\bm{x}_{1}, \dots, \bm{x}_{n}) \, , \\
& = 2^{n} \int_{-\infty}^{x_{1}}\textrm{d}x_{1} \cdots \int_{-\infty}^{x_{n}}\textrm{d}x_{n} P_{n}(x_{1},y_{1};\dots ; x_{n},y_{n}) + \mathcal{R}_{n}(\bm{x}_{1}, \dots, \bm{x}_{n}) \, .
\end{aligned}
\end{equation}
The lower integration extremes in the auxiliary variables $\eta_{j}$ are conventionally set to $-\infty$. The residual term $\mathcal{R}_{n}(\bm{x}_{1}, \dots, \bm{x}_{n})$, which satisfies $\partial_{x_{1}} \cdots \partial_{x_{n}} \mathcal{R}_{n}(\bm{x}_{1}, \dots, \bm{x}_{n}) = 0$ because of the property (\ref{27052020_06}), will be identified in the following section.

%==================================================================================
\subsection{Construction of $n$-body cluster functions}
\label{sec_2_2}
By expressing the passage probability $P_{n}$ in terms of the multivariate normal distribution $\Pi_{n}$, (\ref{06012021_1010}) becomes
\begin{equation}
\label{17012021_1113}
\mathcal{G}_{n}(\bm{x}_{1}, \dots, \bm{x}_{n}) = 2^{n} \int_{-\infty}^{\sqrt{2}\chi_{1}}\textrm{d}v_{1} \cdots \int_{-\infty}^{\sqrt{2}\chi_{n}}\textrm{d}v_{n} \, \Pi_{n}(v_{1},\dots,v_{n} \vert \textsf{R}_{1\dots n}) + \mathcal{R}_{n}(\bm{x}_{1}, \dots, \bm{x}_{n}) \, .
\end{equation}
We introduce the cumulative distribution function (CDF) of the multivariate normal distribution and denote it as follows
\begin{equation}
\label{17012021_1114}
\Phi_{n}(x_{1},\dots,x_{n} \vert \textsf{R}_{1\dots n} ) = \int_{-\infty}^{x_{1}}\textrm{d}u_{1} \dots \int_{-\infty}^{x_{n}}\textrm{d}u_{n} \, \Pi_{n}(u_{1},\dots,u_{n} \vert \textsf{R}_{1\dots n}) \, .
\end{equation}
Thanks to (\ref{17012021_1114}), (\ref{17012021_1113}) becomes
\begin{equation}
\label{18012021_1028}
\mathcal{G}_{n}(\bm{x}_{1},\dots,\bm{x}_{n}) = 2^{n} \Phi_{n}(\sqrt{2}\chi_{1},\dots,\sqrt{2}\chi_{n} \vert \textsf{R}_{1\dots n}) + \mathcal{R}_{n}(\bm{x}_{1}, \dots, \bm{x}_{n}) \, .
\end{equation}

The functions $\mathcal{R}_{n}$ can be fixed by requiring that $G_{n}$ satisfies the clustering property of correlation functions when at least one of its arguments is sent to infinity. The correct clustering is achieved provided that $\mathcal{G}_{n}( \bm{x}_{1}, \dots, \bm{x}_{n} ) \rightarrow \mathcal{G}_{n-1}( \bm{x}_{1}, \dots, \widehat{\bm{x}_{j}} \dots, \bm{x}_{n} )$ when $x_{j} \rightarrow + \infty$, where $\widehat{\bm{x}_{j}}$ denotes the removal of $\bm{x}_{j}$. It is important to stress that such an information emerges from the analysis of the \emph{full} correlation function $G_{n}$, which includes both the connected part and the disconnected ones. Thus, in order to facilitate the exposition, we shall use the above input in order to construct the $n$-body cluster functions and we will check \emph{a posteriori} that such a prescription is indeed the correct one. This consistency check is actually the content of Theorem \ref{clusteringGn}, which will be enunciated at the end of Sec.~\ref{sec_2_4}.

In order to illustrate the approach in a constructive fashion, we consider the simplest case, $n=1$, which will guide our further considerations towards the case of arbitrary $n$. The function $\mathcal{R}_{1}$ is inevitably a constant since it has to satisfy $\partial_{x_{1}}\mathcal{R}_{1}=0$. The value of such constant is determined by imposing $\mathcal{G}_{1}(\bm{x}_{1}) \rightarrow \pm 1$ for $x \rightarrow \pm \infty$. Thus, the one-body cluster function is $\mathcal{G}_{1}(\bm{x}_{1}) = \mathscr{G}_{1}(\sqrt{2}\chi_{1})$, with
\begin{equation}
\label{17012021_1200}
\mathscr{G}_{1}(x_{1}) = 2 \Phi_{1}(x_{1}) -1 \, .
\end{equation}
Equivalently, we can write $\mathscr{G}_{1}(x_{1})=\textrm{erf}(x_{1}/\sqrt{2})$, where $\textrm{erf}(z)=(2/\sqrt{\pi})\int_{0}^{z}\textrm{d}t \, \textrm{e}^{-t^{2}}$ is the error function \cite{Temme}, and therefore $\mathcal{G}_{1}(\bm{x}_{1}) = \textrm{erf}(\chi_{1})$. The analysis of the case $n=2$ reveals that $\mathcal{G}_{2}(\bm{x}_{1},\bm{x}_{2}) = \mathscr{G}_{2}(\sqrt{2}\chi_{1},\sqrt{2}\chi_{2} \vert \textsf{R}_{12})$, where
\begin{equation}
\label{17012021_1201}
\mathscr{G}_{2}(x_{1}, x_{2} \vert \textsf{R}_{12} ) = 4 \Phi_{2}(x_{1},x_{2} \vert \textsf{R}_{12}) - 2\Phi_{1}(x_{1}) -2\Phi_{1}(x_{2}) +1\, .
\end{equation}
The function $\mathcal{R}_{2} = - 2\Phi_{1}(x_{1}) -2\Phi_{1}(x_{2}) +1$ clearly satisfies $\partial_{x_{1}}\partial_{x_{2}}\mathcal{R}_{2}=0$. The clustering follows by observing that $\lim_{x_{2} \rightarrow - \infty}\Phi_{2}(x_{1},x_{2} \vert \textsf{R}_{12})=0$, and $\lim_{x_{2} \rightarrow + \infty} \Phi_{2}(x_{1},x_{2} \vert \textsf{R}_{12}) = \Phi_{1}(x_{1})$; therefore
\begin{equation}
\label{ }
\lim_{x_{2} \rightarrow \pm \infty} \mathscr{G}_{2}(x_{1}, x_{2} \vert \textsf{R}_{12} ) = \pm \mathscr{G}_{1}(x_{1}) \, .
\end{equation}
Analogously to the case $n=1$, the $2$-body cluster function $\mathscr{G}_{2}$ can be expressed in closed form by introducing a suitable set of special functions -- Owen's $T$ function \cite{Owen1956,Owen1980}; we refer to \cite{DS_twopoint} for a detailed account on this aspect.

The results (\ref{17012021_1200}) and (\ref{17012021_1201}) already suggest what the mathematical structure of the $n$-body cluster function for arbitrary $n$ should be. In order to construct a formal expression for $\mathscr{G}_{n}$, we need some preparatory definitions; we begin with the following one
\begin{definition}[block functions]
\label{def_block_functions}
Let $p$ be an integer such that $0 \leqslant p \leqslant n$. For $p \geqslant 1$ the $(p,n)$ block function $\mathcal{B}_{p,n}$ is defined by
\[ \mathcal{B}_{p,n}(x_{1},\dots,x_{n} \vert \textsf{R}_{1\dots n}) = \sum_{ 1 \leqslant k_{1} < \dots < k_{p} \leqslant n} \Phi_{p}( x_{k_{1}}, \dots, x_{k_{p}} \vert \textsf{R}_{k_{1} \dots k_{p}} ) \, ; \]
the sum in the above runs over the set of ordered $p$-tuples with respect to the natural ordering ($<$) of integers. For $p=0$, $\mathcal{B}_{0,n} = 1$, then, $\mathcal{B}_{n,n} = \Phi_{n}$.
\end{definition}

It is useful to write some explicit examples. For $n=2$: $\mathcal{B}_{1,2}(x_{1},x_{2} \vert \textsf{R}_{12}) = \Phi_{1}(x_{1}) + \Phi_{1}(x_{2})$, for $n=3$: $\mathcal{B}_{1,3}(x_{1}, x_{2}, x_{3} \vert \textsf{R}_{123}) = \Phi_{1}(x_{1}) + \Phi_{1}(x_{2}) + \Phi_{1}(x_{3})$, and $\mathcal{B}_{2,3}(x_{1}, x_{2}, x_{3} \vert \textsf{R}_{123}) = \Phi_{2}(x_{1},x_{2} \vert \textsf{R}_{12}) + \Phi_{2}(x_{1},x_{3} \vert \textsf{R}_{13}) + \Phi_{2}(x_{2},x_{3} \vert \textsf{R}_{23})$. Note that the dependence on the correlation matrix, and so on the correlation coefficients, occurs for $p\geqslant2$.

The structure of the $n$-body cluster functions is formalized by the following theorem
\begin{theorem}[$n$-body cluster functions]
\label{thm_cluster_functions}
The $n$-body cluster function is expressed in terms of block functions by means of
\begin{equation}
\label{18012021_1026}
\mathcal{G}_{n}(\bm{x}_{1},\dots,\bm{x}_{n}) = \mathscr{G}_{n}(\sqrt{2}\chi_{1},\dots,\sqrt{2}\chi_{n} \vert \textsf{R}_{1\dots n}) \, ,
\end{equation}
with
\begin{equation}
\label{18012021_1027}
\mathscr{G}_{n}(x_{1},\dots,x_{n} \vert \textsf{R}_{1\dots n} ) = \sum_{p=0}^{n} (-1)^{p} 2^{n-p} \mathcal{B}_{n-p,n}(x_{1},\dots,x_{n} \vert \textsf{R}_{1\dots n} ) \, .
\end{equation}
\end{theorem}

The functions $\mathcal{R}_{n}$ are identified by means of the next corollary
\begin{corollary}
Since the term with $p=0$ in (\ref{18012021_1027}) gives the first term in the right hand side of (\ref{18012021_1028}), it follows that $\mathcal{R}_{n}$ is identified as the sum of the terms with $p=1,\dots,n$ in (\ref{18012021_1027}).
\end{corollary}

We stress that $\mathcal{G}_{n}$ depends on both the horizontal and vertical coordinates, $x_{j}$ and $y_{j}$. Such a dependence is codified by the rescaled coordinates $\chi_{j}$ and by the rescaled vertical coordinates $\tau_{j}$, both introduced in (\ref{06012021_1029}). In particular, the dependence on $\tau_{j}$ occurs also through the correlation coefficients $\rho_{ij} = ( \textsf{R}_{1\dots n})_{ij}$; see (\ref{06012021_1031}).

It is useful to introduce a graphical notation. We represent the CDF of the $n$-variate normal distribution by means of the following \emph{block diagram}
\begin{equation}
\label{06012021_1747}
\Phi_{n}(x_{1},\dots,x_{n} \vert \textsf{R}_{1\dots n} ) \equiv 
\begin{tikzpicture}[baseline={([yshift=-.5ex]current bounding box.center)}, vertex/.style={anchor=center, circle, rounded corners=3, fill=yellow!30, minimum size=4mm, inner sep=0pt}]
\draw[rounded corners=8pt,  fill=black!10]
(0,0) rectangle ++(2+1.4, 0.8);
\node[vertex,draw, line width=0.5pt, fill=colordiagram1!100] (G1) at (0.4, 0.4)  { {\footnotesize{$1$}} };
\node[vertex,draw, line width=0.5pt, fill=colordiagram1!100] (G1) at (1.0, 0.4)  { {\footnotesize{$2$}} };
\node[] (G1) at (2.00, 0.4)  { {\footnotesize{$\cdots$}} };
\node[vertex,draw, line width=0.5pt, fill=colordiagram1!100] (G2) at (2+1.0,0.4)   { {\footnotesize{$n$}} };
\end{tikzpicture}
\, ,
\end{equation}
which consists of $n$ circles corresponding to the arguments shown in the left hand side of (\ref{06012021_1747}). The block functions introduced in Def.~\ref{def_block_functions} are thus depicted as follows
\begin{equation}
\label{06012021_1748}
\mathcal{B}_{p, n}(x_{1},\dots,x_{n} \vert \textsf{R}_{1\dots n} ) \equiv \sum_{ 1 \leqslant i_{1} < \cdots < i_{p} \leqslant n}
\begin{tikzpicture}[baseline={([yshift=-.5ex]current bounding box.center)}, vertex/.style={anchor=center, circle, rounded corners=3, fill=yellow!30, minimum size=4mm, inner sep=0pt}]
\draw[rounded corners=8pt,  fill=black!10]
(0,0) rectangle ++(2+1.4, 0.8);
\node[vertex,draw, line width=0.5pt, fill=colordiagram1!100] (G1) at (0.4, 0.4)  { {\footnotesize{$i_{1}$}} };
\node[vertex,draw, line width=0.5pt, fill=colordiagram1!100] (G1) at (1.0, 0.4)  { {\footnotesize{$i_{2}$}} };
\node[] (G1) at (2.00, 0.4)  { {\footnotesize{$\cdots$}} };
\node[vertex,draw, line width=0.5pt, fill=colordiagram1!100] (G2) at (2+1.0,0.4)   { {\footnotesize{$i_{p}$}} };
\end{tikzpicture}
\, .
\end{equation}

Thanks to the diagrammatic representation provided by (\ref{06012021_1747}) and (\ref{06012021_1748}), cluster functions admit the following graphical rewriting
\begin{equation}
\begin{aligned}
\label{}
\mathscr{G}_{1}(x_{1}) & = 
2 \,
\begin{tikzpicture}[baseline={([yshift=-.6ex]current bounding box.center)}, vertex/.style={anchor=center, circle, rounded corners=3, fill=yellow!30, minimum size=4mm, inner sep=0pt}]
\draw[rounded corners=8pt,  fill=black!10]
(0,0) rectangle ++(0.8, 0.8);
\node[vertex,draw, line width=0.5pt, fill=colordiagram1!100] (G1) at (0.4, 0.4)  { {\footnotesize{1}} };
\end{tikzpicture}
\,\,
-1 \, , \\
\mathscr{G}_{2}(x_{1}, x_{2} \vert \textsf{R}_{12} ) & = 
4 \, 
\begin{tikzpicture}[baseline={([yshift=-.6ex]current bounding box.center)}, vertex/.style={anchor=center, circle, rounded corners=3, fill=yellow!30, minimum size=4mm, inner sep=0pt}]
\draw[rounded corners=8pt,  fill=black!10]
(0,0) rectangle ++(1.4, 0.8);
\node[vertex,draw, line width=0.5pt, fill=colordiagram1!100] (G1) at (0.4, 0.4)  { {\footnotesize{1}} };
\node[vertex,draw, line width=0.5pt, fill=colordiagram1!100] (G2) at (1.0,0.4)   { {\footnotesize{2}} };
\end{tikzpicture}
\,\,
- 2 \,
\begin{tikzpicture}[baseline={([yshift=-.6ex]current bounding box.center)}, vertex/.style={anchor=center, circle, rounded corners=3, fill=yellow!30, minimum size=4mm, inner sep=0pt}]
\draw[rounded corners=8pt,  fill=black!10]
(0,0) rectangle ++(0.8, 0.8);
\node[vertex,draw, line width=0.5pt, fill=colordiagram1!100] (G1) at (0.4, 0.4)  { {\footnotesize{1}} };
\end{tikzpicture}
\,\,
- 2 \,
\begin{tikzpicture}[baseline={([yshift=-.6ex]current bounding box.center)}, vertex/.style={anchor=center, circle, rounded corners=3, fill=yellow!30, minimum size=4mm, inner sep=0pt}]
\draw[rounded corners=8pt,  fill=black!10]
(0,0) rectangle ++(0.8, 0.8);
\node[vertex,draw, line width=0.5pt, fill=colordiagram1!100] (G1) at (0.4, 0.4)  { {\footnotesize{2}} };
\end{tikzpicture}
\,\,
+1 \, , \\
\label{18012021_1031}
\mathscr{G}_{3}(x_{1}, x_{2}, x_{3} \vert \textsf{R}_{123} ) & = 
8 \, 
\begin{tikzpicture}[baseline={([yshift=-.6ex]current bounding box.center)}, vertex/.style={anchor=center, circle, rounded corners=3, fill=yellow!30, minimum size=4mm, inner sep=0pt}]
\draw[rounded corners=8pt,  fill=black!10]
(0,0) rectangle ++(2.0, 0.8);
\node[vertex,draw, line width=0.5pt, fill=colordiagram1!100] (G1) at (0.4, 0.4)  { {\footnotesize{1}} };
\node[vertex,draw, line width=0.5pt, fill=colordiagram1!100] (G2) at (1.0,0.4)   { {\footnotesize{2}} };
\node[vertex,draw, line width=0.5pt, fill=colordiagram1!100] (G2) at (1.6,0.4)   { {\footnotesize{3}} };
\end{tikzpicture}
\,\,
- 4 \, 
\begin{tikzpicture}[baseline={([yshift=-.6ex]current bounding box.center)}, vertex/.style={anchor=center, circle, rounded corners=3, fill=yellow!30, minimum size=4mm, inner sep=0pt}]
\draw[rounded corners=8pt,  fill=black!10]
(0,0) rectangle ++(1.4, 0.8);
\node[vertex,draw, line width=0.5pt, fill=colordiagram1!100] (G1) at (0.4, 0.4)  { {\footnotesize{1}} };
\node[vertex,draw, line width=0.5pt, fill=colordiagram1!100] (G2) at (1.0,0.4)   { {\footnotesize{2}} };
\end{tikzpicture}
\,\,
- 4 \, 
\begin{tikzpicture}[baseline={([yshift=-.6ex]current bounding box.center)}, vertex/.style={anchor=center, circle, rounded corners=3, fill=yellow!30, minimum size=4mm, inner sep=0pt}]
\draw[rounded corners=8pt,  fill=black!10]
(0,0) rectangle ++(1.4, 0.8);
\node[vertex,draw, line width=0.5pt, fill=colordiagram1!100] (G1) at (0.4, 0.4)  { {\footnotesize{1}} };
\node[vertex,draw, line width=0.5pt, fill=colordiagram1!100] (G2) at (1.0,0.4)   { {\footnotesize{3}} };
\end{tikzpicture}
\,\,
- 4 \, 
\begin{tikzpicture}[baseline={([yshift=-.6ex]current bounding box.center)}, vertex/.style={anchor=center, circle, rounded corners=3, fill=yellow!30, minimum size=4mm, inner sep=0pt}]
\draw[rounded corners=8pt,  fill=black!10]
(0,0) rectangle ++(1.4, 0.8);
\node[vertex,draw, line width=0.5pt, fill=colordiagram1!100] (G1) at (0.4, 0.4)  { {\footnotesize{2}} };
\node[vertex,draw, line width=0.5pt, fill=colordiagram1!100] (G2) at (1.0,0.4)   { {\footnotesize{3}} };
\end{tikzpicture}
\,\, \\
& + 2 \, 
\begin{tikzpicture}[baseline={([yshift=-.6ex]current bounding box.center)}, vertex/.style={anchor=center, circle, rounded corners=3, fill=yellow!30, minimum size=4mm, inner sep=0pt}]
\draw[rounded corners=8pt,  fill=black!10]
(0,0) rectangle ++(0.8, 0.8);
\node[vertex,draw, line width=0.5pt, fill=colordiagram1!100] (G1) at (0.4, 0.4)  { {\footnotesize{1}} };
\end{tikzpicture}
\,\,
+ 2 \, 
\begin{tikzpicture}[baseline={([yshift=-.6ex]current bounding box.center)}, vertex/.style={anchor=center, circle, rounded corners=3, fill=yellow!30, minimum size=4mm, inner sep=0pt}]
\draw[rounded corners=8pt,  fill=black!10]
(0,0) rectangle ++(0.8, 0.8);
\node[vertex,draw, line width=0.5pt, fill=colordiagram1!100] (G1) at (0.4, 0.4)  { {\footnotesize{2}} };
\end{tikzpicture}
\,\,
+ 2 \, 
\begin{tikzpicture}[baseline={([yshift=-.6ex]current bounding box.center)}, vertex/.style={anchor=center, circle, rounded corners=3, fill=yellow!30, minimum size=4mm, inner sep=0pt}]
\draw[rounded corners=8pt,  fill=black!10]
(0,0) rectangle ++(0.8, 0.8);
\node[vertex,draw, line width=0.5pt, fill=colordiagram1!100] (G1) at (0.4, 0.4)  { {\footnotesize{3}} };
\end{tikzpicture}
\,\,
-1 \, .
\end{aligned}
\end{equation}
For mathematical convenience, we can take $\mathcal{G}_{0} = \mathscr{G}_{0} \equiv 1$ as the seed for the recursive hierarchy of cluster functions. The explicit form of $\mathscr{G}_{3}$ reads
\begin{equation}
\begin{aligned}
\label{}
\mathscr{G}_{3}(x_{1}, x_{2}, x_{3} \vert \textsf{R}_{123}) & = 8 \Phi_{3}(x_{1},x_{2},x_{3} \vert \textsf{R}_{123}) - 4 \Phi_{2}(x_{1},x_{2} \vert \textsf{R}_{12}) - 4 \Phi_{2}(x_{1},x_{3} \vert \textsf{R}_{13}) \\
& - 4 \Phi_{2}(x_{2},x_{3} \vert \textsf{R}_{23}) + 2 \Phi_{1}(x_{1}) + 2 \Phi_{1}(x_{2}) + 2 \Phi_{1}(x_{3}) -1 \, .
\end{aligned}
\end{equation}
The correlation matrix $\textsf{R}_{123}$ is characterized by three independent correlation coefficients: $\rho_{12}$, $\rho_{13}$ and $\rho_{23}$; the remaining one is obtained by virtue of the Markov property (see e.g. \cite{Doob,McFadden}): $\rho_{13}=\rho_{12}\rho_{23}$. The function $\Phi_{3}$ can be expressed in closed-form in terms of Steck's $S$ and Owen's $T$ functions\footnote{We refer to \cite{Owen1980} for a detailed exposition on the functions $T$ and $S$, as well as for a thorough examination of integrals arising from Gaussian distributions.}. Consequently, even for $n=3$ it is possible to write the cluster function in an analytic form which involves single integrals instead of three-fold ones \cite{ST_threepoint}.

Carrying on the above procedure, we can write the cluster function for the four-point correlation function, namely
\begin{equation}
\begin{aligned}
\label{}
\mathscr{G}_{4}(x_{1}, x_{2}, x_{3}, x_{4} \vert \textsf{R}_{1234}) & = 16 \Phi_{4}(x_{1},x_{2},x_{3},x_{4} \vert \textsf{R}_{1234}) - 8 \Phi_{3}(x_{1},x_{2},x_{3} \vert \textsf{R}_{123}) \\
& - 8 \Phi_{3}(x_{1},x_{2},x_{4} \vert \textsf{R}_{124}) - 8 \Phi_{3}(x_{1},x_{3},x_{4} \vert \textsf{R}_{134}) - 8 \Phi_{3}(x_{2},x_{3},x_{4} \vert \textsf{R}_{234}) \\
& + 4 \Phi_{2}(x_{1},x_{2} \vert \textsf{R}_{12}) + 4 \Phi_{2}(x_{1},x_{3} \vert \textsf{R}_{13}) + 4 \Phi_{2}(x_{1},x_{4} \vert \textsf{R}_{14}) \\
& + 4 \Phi_{2}(x_{2},x_{3} \vert \textsf{R}_{23}) + 4 \Phi_{2}(x_{2},x_{4} \vert \textsf{R}_{24}) + 4 \Phi_{2}(x_{3},x_{4} \vert \textsf{R}_{34}) - 2 \Phi_{1}(x_{1}) \\
& - 2 \Phi_{1}(x_{2}) - 2 \Phi_{1}(x_{3}) - 2 \Phi_{1}(x_{4}) +1 \, ,
\end{aligned}
\end{equation}
or equivalently, within the pictorial form, we have
\begin{equation}
\begin{aligned}
\label{}
\mathscr{G}_{4}(x_{1}, x_{2}, x_{3}, x_{4} \vert \textsf{R}_{1234}) & = 16
\,\,
\begin{tikzpicture}[baseline={([yshift=-.6ex]current bounding box.center)}, vertex/.style={anchor=center, circle, rounded corners=3, fill=yellow!30, minimum size=4mm, inner sep=0pt}]
\draw[rounded corners=8pt,  fill=black!10]
(0,0) rectangle ++(2.6, 0.8);
\node[vertex,draw, line width=0.5pt, fill=colordiagram1!100] (G1) at (0.4, 0.4)  { {\footnotesize{1}} };
\node[vertex,draw, line width=0.5pt, fill=colordiagram1!100] (G2) at (1.0,0.4)   { {\footnotesize{2}} };
\node[vertex,draw, line width=0.5pt, fill=colordiagram1!100] (G2) at (1.6,0.4)   { {\footnotesize{3}} };
\node[vertex,draw, line width=0.5pt, fill=colordiagram1!100] (G2) at (2.2,0.4)   { {\footnotesize{4}} };
\end{tikzpicture}
\,\,
-
8
\,\,
\begin{tikzpicture}[baseline={([yshift=-.6ex]current bounding box.center)}, vertex/.style={anchor=center, circle, rounded corners=3, fill=yellow!30, minimum size=4mm, inner sep=0pt}]
\draw[rounded corners=8pt,  fill=black!10]
(0,0) rectangle ++(2.0, 0.8);
\node[vertex,draw, line width=0.5pt, fill=colordiagram1!100] (G1) at (0.4, 0.4)  { {\footnotesize{1}} };
\node[vertex,draw, line width=0.5pt, fill=colordiagram1!100] (G2) at (1.0,0.4)   { {\footnotesize{2}} };
\node[vertex,draw, line width=0.5pt, fill=colordiagram1!100] (G2) at (1.6,0.4)   { {\footnotesize{3}} };
\end{tikzpicture}
\,\,
-
8
\,\,
\begin{tikzpicture}[baseline={([yshift=-.6ex]current bounding box.center)}, vertex/.style={anchor=center, circle, rounded corners=3, fill=yellow!30, minimum size=4mm, inner sep=0pt}]
\draw[rounded corners=8pt,  fill=black!10]
(0,0) rectangle ++(2.0, 0.8);
\node[vertex,draw, line width=0.5pt, fill=colordiagram1!100] (G1) at (0.4, 0.4)  { {\footnotesize{1}} };
\node[vertex,draw, line width=0.5pt, fill=colordiagram1!100] (G2) at (1.0,0.4)   { {\footnotesize{2}} };
\node[vertex,draw, line width=0.5pt, fill=colordiagram1!100] (G2) at (1.6,0.4)   { {\footnotesize{4}} };
\end{tikzpicture}
\\
&
-
8
\,\,
\begin{tikzpicture}[baseline={([yshift=-.6ex]current bounding box.center)}, vertex/.style={anchor=center, circle, rounded corners=3, fill=yellow!30, minimum size=4mm, inner sep=0pt}]
\draw[rounded corners=8pt,  fill=black!10]
(0,0) rectangle ++(2.0, 0.8);
\node[vertex,draw, line width=0.5pt, fill=colordiagram1!100] (G1) at (0.4, 0.4)  { {\footnotesize{1}} };
\node[vertex,draw, line width=0.5pt, fill=colordiagram1!100] (G2) at (1.0,0.4)   { {\footnotesize{3}} };
\node[vertex,draw, line width=0.5pt, fill=colordiagram1!100] (G2) at (1.6,0.4)   { {\footnotesize{4}} };
\end{tikzpicture}
\,\,
-
8
\,\,
\begin{tikzpicture}[baseline={([yshift=-.6ex]current bounding box.center)}, vertex/.style={anchor=center, circle, rounded corners=3, fill=yellow!30, minimum size=4mm, inner sep=0pt}]
\draw[rounded corners=8pt,  fill=black!10]
(0,0) rectangle ++(2.0, 0.8);
\node[vertex,draw, line width=0.5pt, fill=colordiagram1!100] (G1) at (0.4, 0.4)  { {\footnotesize{2}} };
\node[vertex,draw, line width=0.5pt, fill=colordiagram1!100] (G2) at (1.0,0.4)   { {\footnotesize{3}} };
\node[vertex,draw, line width=0.5pt, fill=colordiagram1!100] (G2) at (1.6,0.4)   { {\footnotesize{4}} };
\end{tikzpicture}
\,\
+
4
\,\,
\begin{tikzpicture}[baseline={([yshift=-.6ex]current bounding box.center)}, vertex/.style={anchor=center, circle, rounded corners=3, fill=yellow!30, minimum size=4mm, inner sep=0pt}]
\draw[rounded corners=8pt,  fill=black!10]
(0,0) rectangle ++(1.4, 0.8);
\node[vertex,draw, line width=0.5pt, fill=colordiagram1!100] (G1) at (0.4, 0.4)  { {\footnotesize{1}} };
\node[vertex,draw, line width=0.5pt, fill=colordiagram1!100] (G2) at (1.0,0.4)   { {\footnotesize{2}} };
\end{tikzpicture}
\,\,
+ 4
\,\,
\begin{tikzpicture}[baseline={([yshift=-.6ex]current bounding box.center)}, vertex/.style={anchor=center, circle, rounded corners=3, fill=yellow!30, minimum size=4mm, inner sep=0pt}]
\draw[rounded corners=8pt,  fill=black!10]
(0,0) rectangle ++(1.4, 0.8);
\node[vertex,draw, line width=0.5pt, fill=colordiagram1!100] (G1) at (0.4, 0.4)  { {\footnotesize{1}} };
\node[vertex,draw, line width=0.5pt, fill=colordiagram1!100] (G2) at (1.0,0.4)   { {\footnotesize{3}} };
\end{tikzpicture}
\\
&
+ 4
\,\,
\begin{tikzpicture}[baseline={([yshift=-.6ex]current bounding box.center)}, vertex/.style={anchor=center, circle, rounded corners=3, fill=yellow!30, minimum size=4mm, inner sep=0pt}]
\draw[rounded corners=8pt,  fill=black!10]
(0,0) rectangle ++(1.4, 0.8);
\node[vertex,draw, line width=0.5pt, fill=colordiagram1!100] (G1) at (0.4, 0.4)  { {\footnotesize{1}} };
\node[vertex,draw, line width=0.5pt, fill=colordiagram1!100] (G2) at (1.0,0.4)   { {\footnotesize{4}} };
\end{tikzpicture}
\,\,
+ 4
\,\,
\begin{tikzpicture}[baseline={([yshift=-.6ex]current bounding box.center)}, vertex/.style={anchor=center, circle, rounded corners=3, fill=yellow!30, minimum size=4mm, inner sep=0pt}]
\draw[rounded corners=8pt,  fill=black!10]
(0,0) rectangle ++(1.4, 0.8);
\node[vertex,draw, line width=0.5pt, fill=colordiagram1!100] (G1) at (0.4, 0.4)  { {\footnotesize{2}} };
\node[vertex,draw, line width=0.5pt, fill=colordiagram1!100] (G2) at (1.0,0.4)   { {\footnotesize{3}} };
\end{tikzpicture}
\,\,
+ 4
\,\,
\begin{tikzpicture}[baseline={([yshift=-.6ex]current bounding box.center)}, vertex/.style={anchor=center, circle, rounded corners=3, fill=yellow!30, minimum size=4mm, inner sep=0pt}]
\draw[rounded corners=8pt,  fill=black!10]
(0,0) rectangle ++(1.4, 0.8);
\node[vertex,draw, line width=0.5pt, fill=colordiagram1!100] (G1) at (0.4, 0.4)  { {\footnotesize{2}} };
\node[vertex,draw, line width=0.5pt, fill=colordiagram1!100] (G2) at (1.0,0.4)   { {\footnotesize{4}} };
\end{tikzpicture}
\,\,
+ 4
\,\,
\begin{tikzpicture}[baseline={([yshift=-.6ex]current bounding box.center)}, vertex/.style={anchor=center, circle, rounded corners=3, fill=yellow!30, minimum size=4mm, inner sep=0pt}]
\draw[rounded corners=8pt,  fill=black!10]
(0,0) rectangle ++(1.4, 0.8);
\node[vertex,draw, line width=0.5pt, fill=colordiagram1!100] (G1) at (0.4, 0.4)  { {\footnotesize{3}} };
\node[vertex,draw, line width=0.5pt, fill=colordiagram1!100] (G2) at (1.0,0.4)   { {\footnotesize{4}} };
\end{tikzpicture}\\
&
- 2
\,\,
\begin{tikzpicture}[baseline={([yshift=-.6ex]current bounding box.center)}, vertex/.style={anchor=center, circle, rounded corners=3, fill=yellow!30, minimum size=4mm, inner sep=0pt}]
\draw[rounded corners=8pt,  fill=black!10]
(0,0) rectangle ++(0.8, 0.8);
\node[vertex,draw, line width=0.5pt, fill=colordiagram1!100] (G1) at (0.4, 0.4)  { {\footnotesize{1}} };
\end{tikzpicture}
\,\,
- 2
\,\,
\begin{tikzpicture}[baseline={([yshift=-.6ex]current bounding box.center)}, vertex/.style={anchor=center, circle, rounded corners=3, fill=yellow!30, minimum size=4mm, inner sep=0pt}]
\draw[rounded corners=8pt,  fill=black!10]
(0,0) rectangle ++(0.8, 0.8);
\node[vertex,draw, line width=0.5pt, fill=colordiagram1!100] (G1) at (0.4, 0.4)  { {\footnotesize{2}} };
\end{tikzpicture}
\,\,
- 2
\,\,
\begin{tikzpicture}[baseline={([yshift=-.6ex]current bounding box.center)}, vertex/.style={anchor=center, circle, rounded corners=3, fill=yellow!30, minimum size=4mm, inner sep=0pt}]
\draw[rounded corners=8pt,  fill=black!10]
(0,0) rectangle ++(0.8, 0.8);
\node[vertex,draw, line width=0.5pt, fill=colordiagram1!100] (G1) at (0.4, 0.4)  { {\footnotesize{3}} };
\end{tikzpicture}
\,\,
- 2
\,\,
\begin{tikzpicture}[baseline={([yshift=-.6ex]current bounding box.center)}, vertex/.style={anchor=center, circle, rounded corners=3, fill=yellow!30, minimum size=4mm, inner sep=0pt}]
\draw[rounded corners=8pt,  fill=black!10]
(0,0) rectangle ++(0.8, 0.8);
\node[vertex,draw, line width=0.5pt, fill=colordiagram1!100] (G1) at (0.4, 0.4)  { {\footnotesize{4}} };
\end{tikzpicture}
\,\,
+
1 \, .
\end{aligned}
\end{equation}
We observe that for $n=4$ the independent correlation coefficients are $\rho_{12}$, $\rho_{23}$ and $\rho_{34}$. The other correlation coefficients follow from the Markov property: $\rho_{13}=\rho_{12}\rho_{23}$, $\rho_{14}=\rho_{12}\rho_{23}\rho_{34}$, and $\rho_{24}=\rho_{23}\rho_{34}$.

The next task we need to carry out is to prove Theorem \ref{thm_cluster_functions}. To this end, we need to recall the asymptotic properties satisfied by block functions.
\begin{lemma}
\label{lemma}
The block function $\mathcal{B}_{p,n}$ vanishes when at least one of its argument is sent to $-\infty$, e.g., 
\[ \lim_{x_{n} \rightarrow - \infty} \mathcal{B}_{p,n}(x_{1},\dots,x_{n} \vert \textsf{R}_{1\dots n}) = 0 \, . \]
For $x_{n} \rightarrow + \infty$, the block function $\mathcal{B}_{1,n}$ satisfies the following property
\[ \lim_{x_{n} \rightarrow + \infty} \mathcal{B}_{1,n}(x_{1},\dots,x_{n} \vert \textsf{R}_{1\dots n}) = 1 + \mathcal{B}_{1,n-1}(x_{1},\dots,x_{n-1} \vert \textsf{R}_{1\dots n-1}) \, , \]
while for $2 \leqslant p \leqslant n-1$
\[ \lim_{x_{n} \rightarrow + \infty} \mathcal{B}_{p,n}(x_{1},\dots,x_{n} \vert \textsf{R}_{1\dots n}) = \mathcal{B}_{p,n-1}(x_{1},\dots,x_{n-1} \vert \textsf{R}_{1\dots n-1}) + \mathcal{B}_{p-1,n-1}(x_{1},\dots,x_{n-1} \vert \textsf{R}_{1\dots n-1}) \, , \]
and for $p=n$
\[ \lim_{x_{n} \rightarrow + \infty} \mathcal{B}_{n,n}(x_{1},\dots,x_{n} \vert \textsf{R}_{1\dots n}) = \mathcal{B}_{n-1,n-1}(x_{1},\dots,x_{n-1} \vert \textsf{R}_{1\dots n-1}) \, . \]
The limits in which $x_{j} \rightarrow \pm \infty$ with $j \neq n$ are treated along the same lines by replacing $\textsf{R}_{1\dots n-1}$ with the correlation matrix $\textsf{R}_{1\dots \hat{j} \dots n}$, where $\textsf{R}_{1\dots \hat{j} \dots n}$ is obtained by removing the $j$-th row and the $j$-th column of $\textsf{R}_{1\dots n}$, and $\hat{j}$ stands for the removed label.
\end{lemma}
\emph{Proof}. The derivation of the asymptotic relations listed in Lemma \ref{lemma} follows by using elementary properties of cumulative distribution functions. Let us consider the first of the properties listed in Lemma \ref{lemma}. The limit in which $x_{n} \rightarrow - \infty$ is established thanks to
\begin{equation}
\label{05012021_2029e}
\lim_{x_{n} \rightarrow - \infty} \Phi_{n}(x_{1}, \dots, x_{n} \vert \textsf{R}_{1\dots n} ) = 0 \, ,
\end{equation}
because the lower integration extrema in the CDFs are $-\infty$. Let us consider the case $p=n$. The limit $x_{n} \rightarrow + \infty$ can be analyzed by using the identity
\begin{equation}
\label{ }
\lim_{x_{n} \rightarrow + \infty} \Phi_{n}(x_{1},\dots,x_{n} \vert \textsf{R}_{1\dots n}) = \Phi_{n-1}(x_{1},\dots,x_{n-1} \vert \textsf{R}_{1\dots n-1}) \, ,
\end{equation}
which is a natural consequence of the marginalization property of the probability distribution $P_{n}$; see (\ref{29012021_0917}). The properties with $1 \leqslant p \leqslant n-1$ follow straightforwardly. \hfill $\square$

\begin{theorem}[clustering]
\label{thm_clustering}
The $n$-body cluster function satisfies the following clustering properties
\begin{equation}
\begin{aligned}
\label{05012021_2029g}
\lim_{x_{n} \rightarrow -\infty} \mathscr{G}_{n}(x_{1}, \dots , x_{n} \vert \textsf{R}_{1\dots n} ) & = - \mathscr{G}_{n-1}(x_{1}, \dots , x_{n-1} \vert \textsf{R}_{1\dots n-1} ) \\
\lim_{x_{n} \rightarrow +\infty} \mathscr{G}_{n}(x_{1}, \dots , x_{n} \vert \textsf{R}_{1\dots n} ) & = \mathscr{G}_{n-1}(x_{1}, \dots , x_{n-1} \vert \textsf{R}_{1\dots n-1} ) \, .
\end{aligned}
\end{equation}
\end{theorem}
The proof of Theorem \ref{thm_clustering} follows as a straightforward application of Lemma \ref{lemma}.

As an explicit illustration of the asymptotic properties, we consider the asymptotic properties of the block functions $\mathcal{B}_{p,3}$, which constitute the building blocks of the three-point correlation function. For $p=3$, we have
\begin{equation}
\begin{aligned}
\label{}
\lim_{x_{3} \rightarrow +\infty} 
\begin{tikzpicture}[baseline={([yshift=-.6ex]current bounding box.center)}, vertex/.style={anchor=center, circle, rounded corners=3, fill=yellow!30, minimum size=4mm, inner sep=0pt}]
\draw[rounded corners=8pt,  fill=black!10]
(0,0) rectangle ++(2.0, 0.8);
\node[vertex,draw, line width=0.5pt, fill=colordiagram1!100] (G1) at (0.4, 0.4)  { {\footnotesize{1}} };
\node[vertex,draw, line width=0.5pt, fill=colordiagram1!100] (G2) at (1.0,0.4)   { {\footnotesize{2}} };
\node[vertex,draw, line width=0.5pt, fill=colordiagram1!100] (G2) at (1.6,0.4)   { {\footnotesize{3}} };
\end{tikzpicture}
\,\,
& =  
\begin{tikzpicture}[baseline={([yshift=-.6ex]current bounding box.center)}, vertex/.style={anchor=center, circle, rounded corners=3, fill=yellow!30, minimum size=4mm, inner sep=0pt}]
\draw[rounded corners=8pt,  fill=black!10]
(0,0) rectangle ++(1.4, 0.8);
\node[vertex,draw, line width=0.5pt, fill=colordiagram1!100] (G1) at (0.4, 0.4)  { {\footnotesize{1}} };
\node[vertex,draw, line width=0.5pt, fill=colordiagram1!100] (G2) at (1.0,0.4)   { {\footnotesize{2}} };
\end{tikzpicture}
\, \, ,
\end{aligned}
\end{equation}
which is actually the asymptotic property of the cumulative distribution function $\Phi_{3}$. Then, for $p=2$ one finds
\begin{equation}
\begin{aligned}
\label{}
\lim_{x_{3} \rightarrow +\infty} \mathcal{B}_{2,3} & = \lim_{x_{3} \rightarrow +\infty} \left(
\begin{tikzpicture}[baseline={([yshift=-.6ex]current bounding box.center)}, vertex/.style={anchor=center, circle, rounded corners=3, fill=yellow!30, minimum size=4mm, inner sep=0pt}]
\draw[rounded corners=8pt,  fill=black!10]
(0,0) rectangle ++(1.4, 0.8);
\node[vertex,draw, line width=0.5pt, fill=colordiagram1!100] (G1) at (0.4, 0.4)  { {\footnotesize{1}} };
\node[vertex,draw, line width=0.5pt, fill=colordiagram1!100] (G2) at (1.0,0.4)   { {\footnotesize{2}} };
\end{tikzpicture}
\,\,
+
\begin{tikzpicture}[baseline={([yshift=-.6ex]current bounding box.center)}, vertex/.style={anchor=center, circle, rounded corners=3, fill=yellow!30, minimum size=4mm, inner sep=0pt}]
\draw[rounded corners=8pt,  fill=black!10]
(0,0) rectangle ++(1.4, 0.8);
\node[vertex,draw, line width=0.5pt, fill=colordiagram1!100] (G1) at (0.4, 0.4)  { {\footnotesize{1}} };
\node[vertex,draw, line width=0.5pt, fill=colordiagram1!100] (G2) at (1.0,0.4)   { {\footnotesize{3}} };
\end{tikzpicture}
\,\,
+
\begin{tikzpicture}[baseline={([yshift=-.6ex]current bounding box.center)}, vertex/.style={anchor=center, circle, rounded corners=3, fill=yellow!30, minimum size=4mm, inner sep=0pt}]
\draw[rounded corners=8pt,  fill=black!10]
(0,0) rectangle ++(1.4, 0.8);
\node[vertex,draw, line width=0.5pt, fill=colordiagram1!100] (G1) at (0.4, 0.4)  { {\footnotesize{2}} };
\node[vertex,draw, line width=0.5pt, fill=colordiagram1!100] (G2) at (1.0,0.4)   { {\footnotesize{3}} };
\end{tikzpicture}
\right) \\
& =
\begin{tikzpicture}[baseline={([yshift=-.6ex]current bounding box.center)}, vertex/.style={anchor=center, circle, rounded corners=3, fill=yellow!30, minimum size=4mm, inner sep=0pt}]
\draw[rounded corners=8pt,  fill=black!10]
(0,0) rectangle ++(1.4, 0.8);
\node[vertex,draw, line width=0.5pt, fill=colordiagram1!100] (G1) at (0.4, 0.4)  { {\footnotesize{1}} };
\node[vertex,draw, line width=0.5pt, fill=colordiagram1!100] (G2) at (1.0,0.4)   { {\footnotesize{2}} };
\end{tikzpicture}
\,\,
+
\begin{tikzpicture}[baseline={([yshift=-.6ex]current bounding box.center)}, vertex/.style={anchor=center, circle, rounded corners=3, fill=yellow!30, minimum size=4mm, inner sep=0pt}]
\draw[rounded corners=8pt,  fill=black!10]
(0,0) rectangle ++(0.8, 0.8);
\node[vertex,draw, line width=0.5pt, fill=colordiagram1!100] (G1) at (0.4, 0.4)  { {\footnotesize{1}} };
\end{tikzpicture}
\,\,
+
\begin{tikzpicture}[baseline={([yshift=-.6ex]current bounding box.center)}, vertex/.style={anchor=center, circle, rounded corners=3, fill=yellow!30, minimum size=4mm, inner sep=0pt}]
\draw[rounded corners=8pt,  fill=black!10]
(0,0) rectangle ++(0.8, 0.8);
\node[vertex,draw, line width=0.5pt, fill=colordiagram1!100] (G1) at (0.4, 0.4)  { {\footnotesize{2}} };
\end{tikzpicture} \\
& = \mathcal{B}_{2,2} + \mathcal{B}_{1,2}
\end{aligned}
\end{equation}
while for $p=1$
\begin{equation}
\begin{aligned}
\label{}
\lim_{x_{3} \rightarrow +\infty} \mathcal{B}_{1,3} & = \lim_{x_{3} \rightarrow +\infty} \left(
\begin{tikzpicture}[baseline={([yshift=-.6ex]current bounding box.center)}, vertex/.style={anchor=center, circle, rounded corners=3, fill=yellow!30, minimum size=4mm, inner sep=0pt}]
\draw[rounded corners=8pt,  fill=black!10]
(0,0) rectangle ++(0.8, 0.8);
\node[vertex,draw, line width=0.5pt, fill=colordiagram1!100] (G1) at (0.4, 0.4)  { {\footnotesize{1}} };
\end{tikzpicture}
\,\,
+
\begin{tikzpicture}[baseline={([yshift=-.6ex]current bounding box.center)}, vertex/.style={anchor=center, circle, rounded corners=3, fill=yellow!30, minimum size=4mm, inner sep=0pt}]
\draw[rounded corners=8pt,  fill=black!10]
(0,0) rectangle ++(0.8, 0.8);
\node[vertex,draw, line width=0.5pt, fill=colordiagram1!100] (G1) at (0.4, 0.4)  { {\footnotesize{2}} };
\end{tikzpicture}
\,\,
+
\begin{tikzpicture}[baseline={([yshift=-.6ex]current bounding box.center)}, vertex/.style={anchor=center, circle, rounded corners=3, fill=yellow!30, minimum size=4mm, inner sep=0pt}]
\draw[rounded corners=8pt,  fill=black!10]
(0,0) rectangle ++(0.8, 0.8);
\node[vertex,draw, line width=0.5pt, fill=colordiagram1!100] (G1) at (0.4, 0.4)  { {\footnotesize{3}} };
\end{tikzpicture}
\right) \\
& = 1 + \,\,
\begin{tikzpicture}[baseline={([yshift=-.6ex]current bounding box.center)}, vertex/.style={anchor=center, circle, rounded corners=3, fill=yellow!30, minimum size=4mm, inner sep=0pt}]
\draw[rounded corners=8pt,  fill=black!10]
(0,0) rectangle ++(0.8, 0.8);
\node[vertex,draw, line width=0.5pt, fill=colordiagram1!100] (G1) at (0.4, 0.4)  { {\footnotesize{1}} };
\end{tikzpicture}
\,\,
+
\,\,
\begin{tikzpicture}[baseline={([yshift=-.6ex]current bounding box.center)}, vertex/.style={anchor=center, circle, rounded corners=3, fill=yellow!30, minimum size=4mm, inner sep=0pt}]
\draw[rounded corners=8pt,  fill=black!10]
(0,0) rectangle ++(0.8, 0.8);
\node[vertex,draw, line width=0.5pt, fill=colordiagram1!100] (G1) at (0.4, 0.4)  { {\footnotesize{2}} };
\end{tikzpicture}\\
& = 1 + \mathcal{B}_{1,2} \, .
\end{aligned}
\end{equation}

\begin{comment}
\footnote{\begin{eqnarray} \nonumber
\mathcal{G}_{n} & = & 2^{n}\varphi_{n,n} + \sum_{k=1}^{n-2}2^{n-k}(-1)^{k}\varphi_{n-k,n} + 2(-1)^{n-1}\varphi_{1,n} + (-1)^{n} \\ \nonumber
& \rightarrow & 2^{n} \varphi_{n-1,n-1} + \sum_{k=1}^{n-2}2^{n-k}(-1)^{k} \left( \varphi_{n-k,n-1} + \varphi_{n-k-1,n-1} \right) + 2 (-1)^{n-1} \left( 1 + \varphi_{1,n-1} \right) + (-1)^{n} \\ \nonumber
& = & 2^{n} \varphi_{n-1,n-1} + \sum_{k=1}^{n}2^{n-k}(-1)^{k}\varphi_{n-k,n-1} + \sum_{k=1}^{n-1}2^{n-k}(-1)^{k}\varphi_{n-k-1,n-1} \\ \nonumber
& = & 2^{n} \varphi_{n-1,n-1} - 2^{n-1} \varphi_{n-1,n-1} + \sum_{m=1}^{n-1}2^{n-(m+1)}(-1)^{m+1}\varphi_{n-m-1,n-1} + \sum_{m=1}^{n-1}2^{n-m}(-1)^{m}\varphi_{n-m-1,n-1} \\ \nonumber
& = & \sum_{m=0}^{n-1}2^{n-1-m}(-1)^{m} \varphi_{n-1-m,n-1} \\
& = & \mathcal{G}_{n-1}
\end{eqnarray}}.
\end{comment}

We conclude this section by adding some considerations about the construction of block functions. According to Definition \ref{def_block_functions} the $(p,n)$ block function is constructed by summing cumulative functions $\Phi_{p}$ in which the $p$ arguments are ordered $p$-tuples drawn from of the set $A_{n} = \{1,\dots,n\}$. This observation allows us to rationalize the construction of block functions by putting them in touch with Hasse diagrams \cite{SW_1986,PS_2003} in discrete mathematics.

In order to proceed along this direction, we recall the definition of power set. The power set of $A_{n}$, denoted $\mathcal{P}(A_{n})$, is the set which contains all subsets of $A_{n}$. Let us write a few examples. The power set of $A_{2}$ is $\mathcal{P}(A_{2}) = \{ \O, \{1\}, \{2\}, \{1,2\} \}$ while the power set of $A_{3}$ is $\mathcal{P}(A_{3}) = \{ \O, \{1\}, \{2\}, \{3\}, \{1,2\}, \{1,3\}, \{2,3\}, \{1,2,3\} \}$, with $\O$ the empty set.

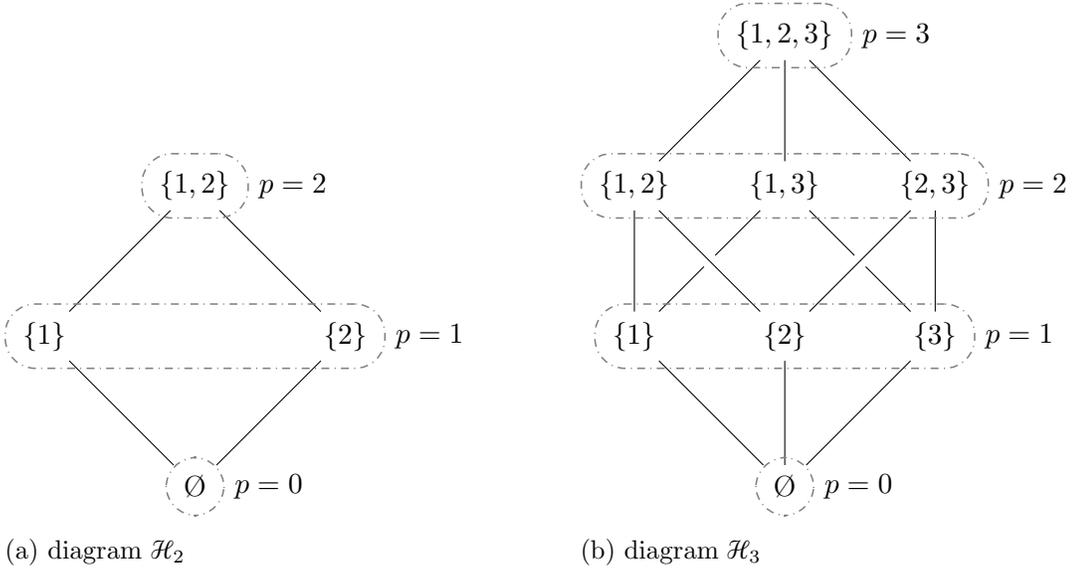
\begin{figure*}[htbp]
\centering
\hspace{-40mm}
\begin{subfigure}[b]{0.15\textwidth}
\centering
\begin{tikzpicture}[y=1.0cm, every fit/.style={inner sep=1mm, gray, dashdotted, rounded corners=4mm, draw, line width=0.2mm}]
\node (a) at (0,2) {$\{1,2\}$};
\node (b) at (-2,0) {$\{1\}$};
\node (c) at (2,0) {$\{2\}$};
\node (d) at (0,-2) {$\O$};
\draw (a) -- (b) -- (d) -- (c) -- (a);
\node[fit=(a)](f1){};
\node[right] at (f1.east) {$p=2$};
\node[fit=(b)(c)](f2){};
\node[right] at (f2.east) {$p=1$};
\node[fit=(d)](f3){};
\node[right] at (f3.east) {$p=0$};
\end{tikzpicture}
\caption[]%
{{\small diagram $\mathcal{H}_{2}$}}
\label{fig_Hasse_2}
\end{subfigure}
%\hfill
\hspace{50mm}
\begin{subfigure}[b]{0.15\textwidth}
\centering
\begin{tikzpicture}[y=1.0cm, every fit/.style={inner sep=1mm, gray, dashdotted, rounded corners=4mm, draw, line width=0.2mm}]
\node (max) at (0,4) {$\{1,2,3\}$};
\node (a) at (-2,2) {$\{1,2\}$};
\node (b) at (0,2) {$\{1,3\}$};
\node (c) at (2,2) {$\{2,3\}$};
\node (d) at (-2,0) {$\{1\}$};
\node (e) at (0,0) {$\{2\}$};
\node (f) at (2,0) {$\{3\}$};
\node (min) at (0,-2) {$\O$};
\draw (min) -- (d) -- (a) -- (max) -- (b) -- (f)
(e) -- (min) -- (f) -- (c) -- (max)
(d) -- (b);
\draw[preaction={draw=white, -,line width=6pt}] (a) -- (e) -- (c);
\node[fit=(max)](f1){};
\node[right] at (f1.east) {$p=3$};
\node[fit=(a)(c)](f2){};
\node[right] at (f2.east) {$p=2$};
\node[fit=(d)(f)](f3){};
\node[right] at (f3.east) {$p=1$};
\node[fit=(min)](f4){};
\node[right] at (f4.east) {$p=0$};
\end{tikzpicture}
\caption[]%
{{\small diagram $\mathcal{H}_{3}$}}
\label{fig_Hasse_3}
\end{subfigure}
%\hspace{20mm}
\vskip\baselineskip
\caption[]
{\small Hasse diagram $\mathcal{H}_{n}$ for the poset $(\mathcal{P}(A_{n}), \subseteq)$ with $n=2$ (a) and $n=3$ (b).}
\label{fig_Hasse}
\end{figure*}

Subsets of $A_{n}$ are naturally ordered by set inclusion ($\subseteq$). The partially ordered set (\emph{poset}) $(\mathcal{P}(A_{n}), \subseteq)$ can be visualized by means of a graph in which the largest element is placed at the top, the smallest at the bottom, and other elements are allocated in between according; the notion of large/small has to be interpreted in terms of the cardinality. Two vertices are connected by an edge if the elements are ordered by set inclusion ($\subseteq$) \cite{SW_1986}. Coming back to the example quoted one moment ago, the Hasse diagram for the power set of $A_{2}$ and $A_{3}$ are shown in Fig.~\ref{fig_Hasse_2} and Fig.~\ref{fig_Hasse_3}, respectively.

Hasse diagrams $\mathcal{H}_{n}$ comprise $n+1$ levels in which elements share the same cardinality. Levels with cardinality $p$ are indicated with a dash-dotted notation in Fig.~\ref{fig_Hasse}. It then follows a level with cardinality $p$ contains $\binom{n}{p}$ elements. Then, the level $p$ contains the elements which define the block function $\mathcal{B}_{p,n}$. The asymptotic clustering properties listed in Lemma \ref{lemma} can be viewed in terms of Hasse diagrams. For instance, the limit $x_{3} \rightarrow \infty$ amounts to remove the node with label $3$ and those bonds attached to it in the diagram $\mathcal{H}_{3}$. Hasse diagrams can be used also to classify the disconnected diagrams arising from disconnected parts of matrix elements, as will be clear in the next section.

\newpage
%==================================================================================
\subsection{Disconnected parts}
\label{sec_2_3}
We can now illustrate how to compute the contribution of disconnected matrix elements to the $n$-point correlation function of the spin field.

The disconnected parts of matrix elements which appear in the right hand side of (\ref{14012021_1724}) are constructed by contracting legs according to the procedure outlined for $n=1$ in (\ref{14012021_1619}). The decomposition of matrix elements given in (\ref{14012021_1724}) is then written as follows
\begin{equation}
\label{06012021_1015}
\prod_{j=1}^{n} \mathcal{M}_{ab}^{\sigma_{j}}(\theta_{j}\vert\theta_{j+1}) = \sum_{m=0}^{n} \mathcal{D}_{ab}^{(m)}(\theta_{1},\dots,\theta_{n+1}) \, ,
\end{equation}
where the sum runs over the number $m$ of disconnected spin fields. Thus, $\mathcal{D}_{ab}^{(0)}$ stands for the fully connected matrix element corresponding to the diagram in the right hand side of (\ref{14012021_1724}) and whose contribution yields the connected part of the $n$-point correlation function. Conversely, $\mathcal{D}_{ab}^{(n)}$ is the fully disconnected matrix element. The decomposition (\ref{06012021_1015}) of matrix elements induces an analogous expansion of the $n$-point correlation function, which reads
\begin{equation}
\label{18012021_1317}
G_{n}(\bm{x}_{1},\dots,\bm{x}_{n}) = \sum_{m=0}^{n} \left( G_{n}(\bm{x}_{1},\dots,\bm{x}_{n}) \right)^{[ \mathcal{D}_{ab}^{(m)} ]} \, ,
\end{equation}
in which the term with superscript $[ \mathcal{D}_{ab}^{(m)} ]$ indicates the contribution of $\mathcal{D}_{ab}^{(m)}$ to $G_{n}$.

The term $\mathcal{D}_{ab}^{(m)}$ with $m=1,\dots,n$ in (\ref{06012021_1015}) indicates the sum of \emph{all} disconnected matrix elements in which $m$ spin fields have been disconnected according to (\ref{14012021_1619}). In graphical terms, the diagrams which contribute to $\mathcal{D}_{ab}^{(m)}$ are those in which $m$ spin fields in the necklace diagram of (\ref{14012021_1724}) are replaced by disconnected lines as those shown in the second term on the right hand side of (\ref{14012021_1619}). Correspondingly, matrix elements which contribute to $\mathcal{D}_{ab}^{(m)}$ contain the product of $m$ Dirac deltas of type $\delta(\theta_{j}-\theta_{j+1})$ and the product of $n-m$ two-particle form factors $F_{aba}^{\sigma_{j}}(\theta_{j}-\theta_{j+1}+\im \pi)$. 

Let us consider some illustrative examples. The diagrams which appear in the disconnected matrix elements $\mathcal{D}_{ab}^{(1)}$ admit the graphical depiction shown in (\ref{05012021_2131}). The term ``perm'' in (\ref{05012021_2131}) indicates the sum over permutations of diagrams formed by detaching -- one at the time -- the spin field with label $j=1,\dots,n-1$.
\begin{equation}
\label{05012021_2131}
\mathcal{D}_{ab}^{(1)}
\,\,\,
=
\vcenter{\hbox{
\begin{tikzpicture}[baseline={([yshift=-.6ex]current bounding box.center)},vertex/.style={anchor=base, circle, minimum size=50mm, inner sep=0pt}]
\draw[-,black]   (0, -0.6) -- (0, -0.6+0.8);
\draw[-,black]   (0, 1.6) -- (0, 1.6-0.8);
\draw[-,black]   (0, 2.4) -- (0, 3.2);
\def\u{0.4}
\def\v{0.6}
\draw[black] %ultra thick
(0, -4.4 ) ..controls +(0, \v) and ( $(0.9, -2.9) - (0, +\v)$ )..
(0.9, -2.9) ..controls +(0, \v) and ( $(0, -1.2) - (0, +\v)$ )..
(0, -1.4);
\path (0, -3.0) node[circle, draw, fill=green!30] (s1) {$\sigma_{n}$} (0, 2.0) node[circle, draw, fill=green!30](s2) {$\sigma_{1}$};
\path (0, -1.0) node[circle, draw, fill=green!30](sn) {${\color{green!30}{\sigma_{n}}}$} (0, -1.0) node[fill=none] (s3) {$\sigma_{n-1}$};
%\draw[-,black]   (0, -2+0.6) -- (0, -4+0.6+0.8);
%\draw[-,red]   (0, -2-0.6) -- (0, -3-0.6);
%\draw[-,black]   (s1) -- (s2);
%\draw[-,black]   (s1) -- (0,2);
%\draw[-,black]   (s2) -- (0,-2);
\path (0,0.0) node () {} (-0.75, -3.0) node (s4) { ${\color{blue}{a}}$ };
\path (0,0.0) node () {} (1.195, -3.0) node (s4) { ${\color{red}{b}}$ };
%\path (0,0.0) node () {} (0.25+0.0, 5.1) node { ${\color{black}{\theta_{1}}}$ };
%\path (0,0.0) node () {} (0.25+0.0, 3.0) node { ${\color{black}{\theta_{2}}}$ };
\path (0,0.0) node () {} (0.25, 1.25+1.5) node { ${\color{black}{\theta_{1}}}$ };
\path (0,0.0) node () {} (0.25, 1.25) node { ${\color{black}{\theta_{2}}}$ };
\path (0,0.0) node () {} (0.25, 0.6) node { ${\color{black}{\vdots}}$ };
\path (0,0.0) node () {} (0.44, -0.3) node { ${\color{black}{\theta_{n-1}}}$ };
\path (0,0.0) node () {} (0.75, -1.915) node { ${\color{black}{\theta_{n}}}$ };
\path (0,0.0) node () {} (0.75, -4.0) node { ${\color{black}{\theta_{n+1}}}$ };
%\node[vertex,draw, line width=0.5pt, densely dotted, fill=white!30] (s1) at (-1,0)  { \ssmall{1} };
\end{tikzpicture}
}}
\,\,\,
+
\,\,\,
\vcenter{\hbox{
\begin{tikzpicture}[baseline={([yshift=-.6ex]current bounding box.center)},vertex/.style={anchor=base, circle, minimum size=50mm, inner sep=0pt}]
\draw[-,black]   (0, -0.6) -- (0, -0.6+0.8);
\draw[-,black]   (0, 1.6) -- (0, 1.6-0.8);
\draw[-,black]   (0, 2.4) -- (0, 3.2);
\def\u{0.4}
\def\v{0.6}
\draw[black] %ultra thick
(0, -4.4 ) ..controls +(0, \v) and ( $(-0.9, -2.9) - (0, +\v)$ )..
(-0.9, -2.9) ..controls +(0, \v) and ( $(0, -1.2) - (0, +\v)$ )..
(0, -1.4);
\path (0, -3.0) node[circle, draw, fill=green!30] (s1) {$\sigma_{n}$} (0, 2.0) node[circle, draw, fill=green!30](s2) {$\sigma_{1}$};
\path (0, -1.0) node[circle, draw, fill=green!30](sn) {${\color{green!30}{\sigma_{n}}}$} (0, -1.0) node[fill=none] (s3) {$\sigma_{n-1}$};
%\draw[-,black]   (0, -2+0.6) -- (0, -4+0.6+0.8);
%\draw[-,red]   (0, -2-0.6) -- (0, -3-0.6);
%\draw[-,black]   (s1) -- (s2);
%\draw[-,black]   (s1) -- (0,2);
%\draw[-,black]   (s2) -- (0,-2);
\path (0,0.0) node () {} (-1.25, -3.0) node (s4) { ${\color{blue}{a}}$ };
\path (0,0.0) node () {} (1.1, -3.0) node (s4) { ${\color{red}{b}}$ };
%\path (0,0.0) node () {} (0.25+0.0, 5.1) node { ${\color{black}{\theta_{1}}}$ };
%\path (0,0.0) node () {} (0.25+0.0, 3.0) node { ${\color{black}{\theta_{2}}}$ };
\path (0,0.0) node () {} (0.25, 1.25+1.5) node { ${\color{black}{\theta_{1}}}$ };
\path (0,0.0) node () {} (0.25, 1.25) node { ${\color{black}{\theta_{2}}}$ };
\path (0,0.0) node () {} (0.25, 0.6) node { ${\color{black}{\vdots}}$ };
\path (0,0.0) node () {} (0.44, -0.3) node { ${\color{black}{\theta_{n-1}}}$ };
\path (0,0.0) node () {} (0.25-0.1, -1.915) node { ${\color{black}{\theta_{n}}}$ };
\path (0,0.0) node () {} (0.44-0.1, -3.9) node { ${\color{black}{\theta_{n+1}}}$ };
%\node[vertex,draw, line width=0.5pt, densely dotted, fill=white!30] (s1) at (-1,0)  { \ssmall{1} };
\end{tikzpicture}
}}
\,\,\,
+
\,\,\,
\textrm{perm.}
\end{equation}

The graphical construction which gives the disconnected matrix elements formed by detaching two spin fields proceeds in an analogous manner. Thus, $\mathcal{D}_{ab}^{(2)}$ admits the graphical decomposition shown in (\ref{05012021_2151}).
\begin{equation}
\label{}
\label{05012021_2151}
\mathcal{D}_{ab}^{(2)}
\,\,\,
=
\vcenter{\hbox{
\begin{tikzpicture}[baseline={([yshift=-.6ex]current bounding box.center)},vertex/.style={anchor=base, circle, minimum size=50mm, inner sep=0pt}]
\draw[-,black]   (0, 2.8) -- (0, 5.2);
\draw[-,black]   (0, 1.4) -- (0, 2.0);
\def\u{1.4}
\def\v{0.6}
\draw[black] %ultra thick
(0, -4.4 ) ..controls +(0, \v) and ( $(1.0, -1.9) - (0, +\u)$ )..
(1.0, -1.9) ..controls +(0, \u) and ( $(0, 0.6) - (0, +\v)$ )..
(0, 0.6);
\path (0, -3.0) node[circle, draw, fill=green!30] (s1) {$\sigma_{n}$} (0, 3.8) node[circle, draw, fill=green!30](s2) {$\sigma_{1}$};
\path (0, -1.0) node[circle, draw, fill=green!30](sn) {${\color{green!30}{\sigma_{n}}}$} (0, -1.0) node[fill=none] (s3) {$\sigma_{n-1}$};
\path (0, 1.0) node[circle, draw, fill=green!30](sn) {${\color{green!30}{\sigma_{n}}}$} (0, 1.0) node[fill=none] (s3) {$\sigma_{n-2}$};
\path (0,0.0) node () {} (-0.75, -2.0) node (s4) { ${\color{blue}{a}}$ };
\path (0,0.0) node () {} (1.25, -2.0) node (s4) { ${\color{red}{b}}$ };
\path (0,0.0) node () {} (0.25, 4.7) node { ${\color{black}{\theta_{1}}}$ };
\path (0,0.0) node () {} (0.25, 3.1) node { ${\color{black}{\theta_{2}}}$ };
\path (0,0.0) node () {} (0.25, 2.5) node { ${\color{black}{\vdots}}$ };
\path (0,0.0) node () {} (0.44, 1.65) node { ${\color{black}{\theta_{n-2}}}$ };
\path (0,0.0) node () {} (0.984, -0.3) node { ${\color{black}{\theta_{n-1}}}$ };
\path (0,0.0) node () {} (0.6, -4.1) node { ${\color{black}{\theta_{n+1}}}$ };
\end{tikzpicture}
}}
\,\,\,
+
\,\,\,
\vcenter{\hbox{
\begin{tikzpicture}[baseline={([yshift=-.6ex]current bounding box.center)},vertex/.style={anchor=base, circle, minimum size=50mm, inner sep=0pt}]
\draw[-,black]   (0, 2.8) -- (0, 5.2);
\draw[-,black]   (0, 1.4) -- (0, 2.0);
\def\u{1.4}
\def\v{0.6}
\draw[black] %ultra thick
(0, -4.4 ) ..controls +(0, \v) and ( $(-1.0, -1.9) - (0, +\u)$ )..
(-1.0, -1.9) ..controls +(0, \u) and ( $(0, 0.6) - (0, +\v)$ )..
(0, 0.6);
\path (0, -3.0) node[circle, draw, fill=green!30] (s1) {$\sigma_{n}$} (0, 3.8) node[circle, draw, fill=green!30](s2) {$\sigma_{1}$};
\path (0, -1.0) node[circle, draw, fill=green!30](sn) {${\color{green!30}{\sigma_{n}}}$} (0, -1.0) node[fill=none] (s3) {$\sigma_{n-1}$};
\path (0, 1.0) node[circle, draw, fill=green!30](sn) {${\color{green!30}{\sigma_{n}}}$} (0, 1.0) node[fill=none] (s3) {$\sigma_{n-2}$};
\path (0,0.0) node () {} (-1.35, -2.0) node (s4) { ${\color{blue}{a}}$ };
\path (0,0.0) node () {} (1.05, -2.0) node (s4) { ${\color{red}{b}}$ };
\path (0,0.0) node () {} (0.25, 4.7) node { ${\color{black}{\theta_{1}}}$ };
\path (0,0.0) node () {} (0.25, 3.1) node { ${\color{black}{\theta_{2}}}$ };
\path (0,0.0) node () {} (0.25, 2.5) node { ${\color{black}{\vdots}}$ };
\path (0,0.0) node () {} (0.44, 1.65) node { ${\color{black}{\theta_{n-2}}}$ };
\path (0,0.0) node () {} (0.284, -0.1) node { ${\color{black}{\theta_{n-1}}}$ };
\path (0,0.0) node () {} (0.6, -4.1) node { ${\color{black}{\theta_{n+1}}}$ };
\end{tikzpicture}
}}
\,\,\,
+
\,\,\,
\vcenter{\hbox{
\begin{tikzpicture}[baseline={([yshift=-.6ex]current bounding box.center)},vertex/.style={anchor=base, circle, minimum size=50mm, inner sep=0pt}]
\draw[-,black]   (0, 2.8) -- (0, 5.2);
\draw[-,black]   (0, 1.4) -- (0, 2.0);
\def\u{0.6}
\def\v{0.6}
\draw[black] %ultra thick
(0, -4.4 ) ..controls +(0, \v) and ( $(-1.0, -4.4+1.67) - (0, +\u)$ )..
(-1.0, -4.4+1.67) ..controls +(0, \u) and ( $(1.0, -4.4+3.3) - (0, +\u)$ )..
(1.0, -4.4+3.3) ..controls +(0, \u) and ( $(0, 0.6) - (0, +\v)$ )..
(0, 0.6);
\path (0, -3.0) node[circle, draw, fill=green!30] (s1) {$\sigma_{n}$} (0, 3.8) node[circle, draw, fill=green!30](s2) {$\sigma_{1}$};
\path (0, -1.0) node[circle, draw, fill=green!30](sn) {${\color{green!30}{\sigma_{n}}}$} (0, -1.0) node[fill=none] (s3) {$\sigma_{n-1}$};
\path (0, 1.0) node[circle, draw, fill=green!30](sn) {${\color{green!30}{\sigma_{n}}}$} (0, 1.0) node[fill=none] (s3) {$\sigma_{n-2}$};
\path (0,0.0) node () {} (-0.75, -2.0) node (s4) { ${\color{blue}{a}}$ };
\path (0,0.0) node () {} (0.75, -2.0) node (s4) { ${\color{red}{b}}$ };
\path (0,0.0) node () {} (0.25, 4.7) node { ${\color{black}{\theta_{1}}}$ };
\path (0,0.0) node () {} (0.25, 3.1) node { ${\color{black}{\theta_{2}}}$ };
\path (0,0.0) node () {} (0.25, 2.5) node { ${\color{black}{\vdots}}$ };
\path (0,0.0) node () {} (0.44, 1.65) node { ${\color{black}{\theta_{n-2}}}$ };
\path (0,0.0) node () {} (0.984, -0.10) node { ${\color{black}{\theta_{n-1}}}$ };
\path (0,0.0) node () {} (0.6, -4.1) node { ${\color{black}{\theta_{n+1}}}$ };
\end{tikzpicture}
}}
\,\,\,
+
\,\,\,
\vcenter{\hbox{
\begin{tikzpicture}[baseline={([yshift=-.6ex]current bounding box.center)},vertex/.style={anchor=base, circle, minimum size=50mm, inner sep=0pt}]
\draw[-,black]   (0, 2.8) -- (0, 5.2);
\draw[-,black]   (0, 1.4) -- (0, 2.0);
\def\u{0.6}
\def\v{0.6}
\draw[black] %ultra thick
(0, -4.4 ) ..controls +(0, \v) and ( $(1.0, -4.4+1.67) - (0, +\u)$ )..
(1.0, -4.4+1.67) ..controls +(0, \u) and ( $(-1.0, -4.4+3.3) - (0, +\u)$ )..
(-1.0, -4.4+3.3) ..controls +(0, \u) and ( $(0, 0.6) - (0, +\v)$ )..
(0, 0.6);
\path (0, -3.0) node[circle, draw, fill=green!30] (s1) {$\sigma_{n}$} (0, 3.8) node[circle, draw, fill=green!30](s2) {$\sigma_{1}$};
\path (0, -1.0) node[circle, draw, fill=green!30](sn) {${\color{green!30}{\sigma_{n}}}$} (0, -1.0) node[fill=none] (s3) {$\sigma_{n-1}$};
\path (0, 1.0) node[circle, draw, fill=green!30](sn) {${\color{green!30}{\sigma_{n}}}$} (0, 1.0) node[fill=none] (s3) {$\sigma_{n-2}$};
\path (0,0.0) node () {} (-0.75, -2.0) node (s4) { ${\color{blue}{a}}$ };
\path (0,0.0) node () {} (0.75, -2.0) node (s4) { ${\color{red}{b}}$ };
\path (0,0.0) node () {} (0.25, 4.7) node { ${\color{black}{\theta_{1}}}$ };
\path (0,0.0) node () {} (0.25, 3.1) node { ${\color{black}{\theta_{2}}}$ };
\path (0,0.0) node () {} (0.25, 2.5) node { ${\color{black}{\vdots}}$ };
\path (0,0.0) node () {} (0.44, 1.65) node { ${\color{black}{\theta_{n-2}}}$ };
\path (0,0.0) node () {} (0.284, -0.10) node { ${\color{black}{\theta_{n-1}}}$ };
\path (0,0.0) node () {} (0.6, -4.1) node { ${\color{black}{\theta_{n+1}}}$ };
\end{tikzpicture}
}}
\,\,
+ \textrm{perm.}
\end{equation}
By applying the above rules it is straightforward to construct diagrams corresponding to disconnected matrix elements with an arbitrary number of disconnected spin fields.

We can now address the calculation of $\left( G_{n}(\bm{x}_{1},\dots,\bm{x}_{n}) \right)^{[ \mathcal{D}_{ab}^{(m)} ]}$. In order to simplify the exposition, we show the calculation for a particular type of disconnected diagrams, denoted $\breve{\mathscr{D}}_{ab}^{(m)}$, which are obtained by disconnecting the last $m$ spin fields with labels $j=n-m+1,\dots,n$. The diagrams which contribute to $\breve{\mathscr{D}}_{ab}^{(1)}$ are precisely those depicted in the right hand side of (\ref{05012021_2131}). Analogously, the diagrams which contribute to $\breve{\mathscr{D}}_{ab}^{(2)}$ are those depicted in the right hand side of (\ref{05012021_2151}). There is actually no loss of generality in this choice since an arbitrary disconnected diagram can be obtained by a permutation of the spin and coordinate labels. Therefore, we shall focus on the following matrix element
\begin{equation}
\label{18082021_1103}
\breve{\mathscr{D}}_{ab}^{(m)}(\theta_{1},\dots,\theta_{n+1}) = \biggl[ \prod_{j=1}^{n-m} F_{aba}^{\sigma_{j}}(\theta_{j}-\theta_{j+1}+\im \pi) \biggr] \biggl[ \prod_{j=n-m+1}^{n} 2\pi \widetilde{\langle \sigma_{j} \rangle} \delta(\theta_{j}-\theta_{j+1}) \biggr] \, ,
\end{equation}
where
\begin{equation}
\label{ }
\widetilde{\langle \sigma_{j} \rangle} = \frac{ \langle \sigma_{j} \rangle_{a}  + \langle \sigma_{j} \rangle_{b} }{2} \, .
\end{equation}
The anatomy of (\ref{18082021_1103}) follows by noting that the first product is originated by those spin fields which form the connected part of the diagram, while the second product stems by tying together those legs which are detached from the disconnected spin fields. The occurrence of $\widetilde{\langle \sigma_{j} \rangle}$ follows since the arithmetic average between diagrams obtained within the left and right annihilations has to be performed \cite{DS_twopoint}.

By keeping the leading-order term in the small rapidity expansion, we can write the matrix element in the factorized form $\breve{\mathscr{D}}_{ab}^{(m)} = D_{ab}^{(m)} \mathfrak{D}^{(m)}$, with a rapidity-dependent part
\begin{equation}
\label{07062020_01}
\mathfrak{D}^{(m)}(\theta_{1},\dots,\theta_{n+1}) = \biggl[ \prod_{j=1}^{n-m}  \frac{ - 2\im }{ \theta_{j}-\theta_{j+1} } \biggr]  \biggl[ \prod_{j=n-m+1}^{n} 2\pi \delta(\theta_{j}-\theta_{j+1}) \biggr] \, ,
\end{equation}
and $D_{ab}^{(m)}$ an overall factor which depends solely on the vacuum expectation values
\begin{equation}
\label{ }
D_{ab}^{(m)} = \biggl[ \prod_{j=1}^{n-m}  (- \Delta\langle \sigma_{j} \rangle/2 ) \biggr] \biggl[ \prod_{j=n-m+1}^{n} \widetilde{\langle \sigma_{j} \rangle} \biggr] \, ,
\end{equation}
with $\Delta\langle \sigma_{j} \rangle$ given in (\ref{29012021_0814}).

The contribution of $\breve{\mathscr{D}}_{ab}^{(m)}$ to the $n$-point correlation function is thus accounted for by
\begin{equation}
\label{ }
D_{ab}^{(m)} \Lbag \mathfrak{D}^{(m)} \Rbag_{\eta_{1},\tau_{1}; \dots ;\eta_{n},\tau_{n}} \, .
\end{equation}
The integral with respect to rapidities can be straightforwardly computed and it yields
\begin{equation}
\begin{aligned}
\label{}
\Lbag \mathfrak{D}^{(m)} \Rbag_{\eta_{1},\tau_{1}; \dots ;\eta_{n},\tau_{n}} & = \mathcal{G}_{n-m}(\bm{x}_{1},\dots,\bm{x}_{n-m}) \, ,
\end{aligned}
\end{equation}
hence, the matrix element $\breve{\mathscr{D}}_{ab}^{(m)}$ originates the contribution
\begin{equation}
\label{ }
\left( G_{n}(\bm{x}_{1},\dots,\bm{x}_{n}) \right)^{[ \breve{\mathscr{D}}_{ab}^{(m)} ]} = D_{ab}^{(m)} \mathcal{G}_{n-m}(\bm{x}_{1}, \dots, \bm{x}_{n-m})
\end{equation}
to the $n$-point correlation function.

It is interesting to observe how the pictorial representation of matrix elements provides insights on the structure of $G_{n}$. A matrix element represented by a diagram in which the spin fields $\sigma_{1}\dots,\sigma_{n-m}$ are connected yields a cluster function $\mathcal{G}_{n-m}(\bm{x}_{1}, \dots, \bm{x}_{n-m})$ which depends on the spatial coordinates carried by those spin fields which constitute the connected part of the diagram. The spatial coordinates relative to disconnected spin fields do not report in the resulting cluster function.

It is now evident how to construct \emph{all} the disconnected diagrams belonging to the family $\mathcal{D}_{ab}^{(m)}$. Firstly, we observe that the number of such diagram is $\# \mathcal{D}_{ab}^{(m)} = 2^{m} \binom{n}{m}$ because each spin field can be disconnected either passing left or right aside it, hence the factor $2^{m}$ follows. Thus, up to left/right combinatorics, there is one ($\binom{n}{0}$) fully connected diagram, there are $\binom{n}{1}=n$ diagrams with one disconnected spin field, and so on. Note that, the total number of diagrams is $\sum_{m=0}^{n}\binom{n}{m}=2^{n}$. It is thus clear how Hasse diagrams can be used in order to classify the disconnected diagrams too.

%==================================================================================
\subsection{Full result and specific cases}
\label{sec_2_4}
We are now in the position to construct the full correlation function $G_{n}$. An arbitrary disconnected diagram can be identified by specifying the labels of those spin fields which are disconnected. Let $\mathcal{V}_{m}$ be the \emph{ordered} set of vertices which composes the connected part of the diagram with $m$ disconnected spin fields and let us denote its vertices with the labels $j_{1},\dots,j_{n-m}$. Analogously, let $\overline{\mathcal{V}}_{m}$ be the set of vertices which compose the disconnected part. Clearly, for any $m$, $\mathcal{V}_{m} \cup \overline{\mathcal{V}}_{m} = \{1,\dots,n\}$. The contribution stemming from the diagrams with $m$ disconnected spin fields reads
\begin{equation}
\label{ }
\left( G_{n}(\bm{x}_{1},\dots,\bm{x}_{n}) \right)^{[ \mathcal{D}_{ab}^{(m)} ]} = \sum_{ \mathcal{V}_{m} } \prod_{ i \in \overline{\mathcal{V}}_{m} } \widetilde{\langle \sigma_{i} \rangle} \prod_{ j \in \mathcal{V}_{m} } (- \widehat{\langle\sigma_{j}\rangle} ) \biggl[ \sum_{ j_{1} < \dots < j_{n} \in \mathcal{V}_{m}} \mathcal{G}_{n-m}( \bm{x}_{j_{1}},\dots, \bm{x}_{j_{n-m}} ) \biggr] \, ;
\end{equation}
we recall that $\widehat{\langle\sigma_{j}\rangle}$ is given by (\ref{29012021_0848}). Thanks to (\ref{18012021_1317}), the $n$-point spin correlator $G_{n}$ is given by
\begin{equation}
\label{ }
G_{n}(\bm{x}_{1},\dots,\bm{x}_{n}) = \sum_{m=0}^{n} \sum_{ \mathcal{V}_{m} } \prod_{ i \in \overline{\mathcal{V}}_{m} } \widetilde{\langle \sigma_{i} \rangle} \prod_{ j \in \mathcal{V}_{m} } (- \widehat{\langle\sigma_{j}\rangle} ) \biggl[ \sum_{ j_{1} < \dots < j_{n} \in \mathcal{V}_{m}} \mathcal{G}_{n-m}( \bm{x}_{j_{1}},\dots, \bm{x}_{j_{n-m}} ) \biggr] \, .
\end{equation}

It is instructive to consider some examples. The case $n=1$ gives the magnetization profile
\begin{equation}
\begin{aligned}
\label{29012021_0952}
\langle \sigma_{1}(x,y) \rangle_{ab} & = \widetilde{\langle \sigma_{1} \rangle}  - \widehat{\langle\sigma_{j}\rangle} \mathcal{G}_{1}(\bm{x}_{1}) + \Os(R^{-1/2}) \\
& = \widetilde{\langle \sigma_{1} \rangle}  - \widehat{\langle\sigma_{j}\rangle} \textrm{erf}(\chi_{1}) + \Os(R^{-1/2}) \, .
\end{aligned}
\end{equation}
The above agrees with the result of \cite{DV}. In the last line follows by using the expression of the one-body cluster function $\mathcal{G}_{1}$ given below (\ref{17012021_1200}). We also observe that (\ref{29012021_0952}) retrieves the known magnetization profile for the Ising model \cite{AR_1974_phasesep} as a particular case. Ising symmetry requires $\langle \sigma \rangle_{+} = - \langle \sigma \rangle_{-}=-M$ with $M$ the spontaneous magnetization. Then, $\widetilde{\langle \sigma \rangle} = 0$ and $\widehat{\langle\sigma\rangle} = -M$ yield the profile $\langle \sigma(x,y) \rangle_{-+} = M \textrm{erf}(\chi) + \Os(R^{-1})$; the correction at order $R^{-1/2}$ vanishes (see (\ref{11042021_1928})).

Let us consider the case $n=2$ corresponding to the pair correlation function of the order parameter. The connected part yields $\widehat{\langle\sigma_{1}\rangle} \widehat{\langle\sigma_{2}\rangle} \mathcal{G}_{2}(\bm{x}_{1},\bm{x}_{2})$. The disconnected parts with $m=1$ give $- \widetilde{\langle \sigma_{1} \rangle} \widehat{\langle\sigma_{2}\rangle} \mathcal{G}_{1}(\bm{x}_{2}) - \widetilde{\langle \sigma_{2} \rangle} \widehat{\langle\sigma_{1}\rangle} \mathcal{G}_{1}(\bm{x}_{1})$ and the fully disconnected part contributes with $\widetilde{\langle \sigma_{1} \rangle} \widetilde{\langle \sigma_{2} \rangle}$. Collecting the various pieces, we find 
\begin{equation}
\label{29012021_1001}
\langle \sigma_{1}(\bm{x}_{1}) \sigma_{2}(\bm{x}_{2}) \rangle_{ab} = \widehat{\langle\sigma_{1}\rangle} \widehat{\langle\sigma_{2}\rangle} \mathcal{G}_{2}(\bm{x}_{1},\bm{x}_{2}) - \widehat{\langle\sigma_{1}\rangle} \widetilde{\langle\sigma_{2}\rangle} \mathcal{G}_{1}(\bm{x}_{1}) - \widehat{\langle\sigma_{2}\rangle} \widetilde{\langle\sigma_{1}\rangle} \mathcal{G}_{1}(\bm{x}_{2}) + \widetilde{\langle\sigma_{1}\rangle} \widetilde{\langle\sigma_{2}\rangle}  + \Os(R^{-1/2}) \, ,
\end{equation}
which perfectly matches with the findings of \cite{DS_twopoint}. As a further example, the three-point correlation function is given by
\begin{equation}
\begin{aligned}
\label{29012021_1002}
\langle \sigma_{1}(\bm{x}_{1}) \sigma_{2}(\bm{x}_{2}) \sigma_{3}(\bm{x}_{3}) \rangle_{ab} & = - \widehat{\langle\sigma_{1}\rangle} \widehat{\langle\sigma_{2}\rangle} \widehat{\langle\sigma_{3}\rangle} \mathcal{G}_{3}(\bm{x}_{1},\bm{x}_{2},\bm{x}_{3}) + \widehat{\langle\sigma_{1}\rangle} \widehat{\langle\sigma_{2}\rangle} \widetilde{\langle\sigma_{3}\rangle} \mathcal{G}_{2}(\bm{x}_{1},\bm{x}_{2}) \\
& + \widehat{\langle\sigma_{1}\rangle} \widehat{\langle\sigma_{3}\rangle} \widetilde{\langle\sigma_{2}\rangle} \mathcal{G}_{2}(\bm{x}_{1},\bm{x}_{3}) + \widehat{\langle\sigma_{2}\rangle} \widehat{\langle\sigma_{3}\rangle} \widetilde{\langle\sigma_{1}\rangle} \mathcal{G}_{2}(\bm{x}_{2},\bm{x}_{3}) \\
& - \widehat{\langle\sigma_{1}\rangle} \widetilde{\langle\sigma_{2}\rangle} \widetilde{\langle\sigma_{3}\rangle} \mathcal{G}_{1}(\bm{x}_{1}) - \widehat{\langle\sigma_{2}\rangle} \widetilde{\langle\sigma_{1}\rangle} \widetilde{\langle\sigma_{3}\rangle} \mathcal{G}_{1}(\bm{x}_{2}) \\
& - \widehat{\langle\sigma_{3}\rangle} \widetilde{\langle\sigma_{1}\rangle} \widetilde{\langle\sigma_{2}\rangle} \mathcal{G}_{1}(\bm{x}_{3}) + \widetilde{\langle\sigma_{1}\rangle} \widetilde{\langle\sigma_{2}\rangle} \widetilde{\langle\sigma_{3}\rangle} + \Os(R^{-1/2}) \, .
\end{aligned}
\end{equation}

The explicit expressions (\ref{29012021_0952})-(\ref{29012021_1002}) allow for a direct check of the clustering properties for $n=1$, $n=2$, and $n=3$. The corresponding statement for arbitrary $n$ is the content of the following theorem 
\begin{theorem}
\label{clusteringGn}
Let $1 \leqslant j \leqslant n$. The $n$-point correlation function of the spin field satisfies the clustering properties
\begin{equation}
\label{ }
\lim_{ x_{j} \rightarrow - \infty} \langle \sigma_{1}(\bm{x}_{1}) \cdots \sigma_{n}(\bm{x}_{n}) \rangle_{ab} = \langle \sigma_{j} \rangle_{a} \langle \sigma_{1}(\bm{x}_{1}) \cdots \widehat{\sigma_{j}(\bm{x}_{j})} \cdots \sigma_{n}(\bm{x}_{n}) \rangle_{ab} \, ,
\end{equation}
and
\begin{equation}
\label{ }
\lim_{ x_{j} \rightarrow + \infty} \langle \sigma_{1}(\bm{x}_{1}) \cdots \sigma_{n}(\bm{x}_{n}) \rangle_{ab} = \langle \sigma_{j} \rangle_{b} \langle \sigma_{1}(\bm{x}_{1}) \cdots \widehat{\sigma_{j}(\bm{x}_{j})} \cdots \sigma_{n}(\bm{x}_{n}) \rangle_{ab} \, ,
\end{equation}
where $\widehat{\sigma_{j}(\bm{x}_{j})}$ stands for the omission of $\sigma_{j}(\bm{x}_{j})$ in the correlation function.
\end{theorem}
The proof follows by using the results of the (clustering) Theorem \ref{thm_clustering}.

%==================================================================================
\subsection{The limit $R\rightarrow \infty$}
\label{sec_2_5}
In this case, $\tau_{j}=2y_{y}/R \rightarrow 0$, meaning that all correlation coefficients tend to unity, i.e.,
\begin{equation}
\label{10042021_0927}
\lim_{R \rightarrow \infty} \rho_{ij} = 1 \, .
\end{equation}
Correspondingly, the correlation matrix reduces to a $n \times n$ matrix whose entries consists of all $1$s; we denote such a matrix with $\textsf{J}_{n}$. Analogously, the limit $R \rightarrow \infty$ projects the variables $\chi_{j}$, which encode the dependence through the coordinates $x_{j}$ and $y_{j}$, to the origin, i.e, 
\begin{equation}
\label{10042021_0928}
\lim_{R \rightarrow \infty} \chi_{j} = 0 \, .
\end{equation}
In both the limits (\ref{10042021_0927}) and (\ref{10042021_0928}) the coordinates $x_{j},y_{j}$ with $j=1,\dots,n$ are fixed. The limit (\ref{10042021_0928}) implies that the cumulative distribution functions which appear in the block functions are evaluated at the origin. Said differently, the cumulative distribution functions become the so-called \emph{orthant probabilities} \cite{Tong, AG}.

The calculation of orthant probabilities is a notoriously difficult problem. The first few orthant probabilities are:
\begin{equation}
\begin{aligned}
\label{10042021_0935}
\int_{-\infty}^{0}\textrm{d}u_{1}\, \Pi_{1}(u_{1}) & = \frac{1}{2} \, , \\
\int_{-\infty}^{0}\textrm{d}u_{1} \int_{-\infty}^{0}\textrm{d}u_{2} \, \Pi_{2}(u_{1},u_{2} \vert \textsf{R}_{12} ) & = \frac{1}{4} + \frac{1}{2\pi} \sin^{-1}(\rho_{12}) \, , \\
\int_{-\infty}^{0}\textrm{d}u_{1} \int_{-\infty}^{0}\textrm{d}u_{2} \int_{-\infty}^{0}\textrm{d}u_{3} \, \Pi_{3}(u_{1},u_{2},u_{3} \vert \textsf{R}_{123}) & = \frac{1}{8} + \frac{1}{4\pi} \left( \sin^{-1}(\rho_{12}) + \sin^{-1}(\rho_{13}) + \sin^{-1}(\rho_{23}) \right) \, .
\end{aligned}
\end{equation}
The above result may indicate a general pattern for the orthant of the $n$-variate normal distribution. This is actually not the case, as it is revealed by the orthant of the quadrivariate normal distribution \cite{Cheng}.

The limit (\ref{10042021_0927}) comes in our rescue. Although the orthant of the $n$-variate normal distribution is a complicated function of the correlation coefficients $\rho_{ij}$, the orthant drastically simplifies when $\rho_{ij}=1$, and the corresponding result is simply $\nicefrac{1}{2}$. With this in mind, the cumulative function given in (\ref{06012021_1747}) reduces to $\nicefrac{1}{2}$ and the block function (\ref{06012021_1748}) reduces to $\nicefrac{1}{2}$ times the number of elements in the level $p \geqslant1$ of the Hasse diagram $\mathcal{H}_{n}$, therefore
\begin{equation}
\label{10042021_1015}
\mathcal{B}_{p,n}(0,\dots, 0\vert \textsf{J}_{n} ) = 
\begin{cases}
\frac{1}{2} \binom{n}{p} \, ,      & p \geqslant 1\\
1 \, ,     & p=0 \, .
\end{cases}
\end{equation}
Thanks to (\ref{18012021_1027}) and (\ref{10042021_1015}) the cluster function thus reduces to
\begin{equation}
\begin{aligned}
\label{10042021_1021}
\lim_{R \rightarrow \infty} \mathcal{G}_{n}(\bm{x}_{1},\dots,\bm{x}_{n}) & = \mathscr{G}_{n}(0,\dots,0 \vert \textsf{J}_{n}) \\
& = \sum_{p=0}^{n} (-1)^{p} 2^{n-p} \mathcal{B}_{n-p,n}(0,\dots, 0\vert \textsf{J}_{n} ) \\
& = (-1)^{n} + 2^{n-1} \sum_{p=0}^{n-1} \binom{n}{n-p} \left( - \frac{1}{2} \right)^{p} \\
& = \frac{ 1 + (-1)^{n} }{ 2 } \, .
\end{aligned}
\end{equation}
We are now in the position to compute the full correlation function in the limit $R \rightarrow \infty$. The case which we are going to examine is the one where spin fields entering the correlation functions are all identical, i.e., $\sigma_{i} \equiv \sigma$. Thanks to the selection rule (\ref{10042021_1021}), the matrix elements with $m$ disconnected legs contribute with the term
\begin{equation}
\label{ }
\binom{n}{m}\left(-\widehat{\langle\sigma\rangle} \right)^{n-m} \widetilde{\langle\sigma\rangle}^{m} \biggl[ \lim_{R \rightarrow \infty} \mathcal{G}_{n-m}(\bm{x}_{1},\dots,\bm{x}_{n}) \biggr] \, ,
\end{equation}
the full correlation function is obtained by summing the above terms with respect to $m$ from $m=0$ (fully connected term) to $m=n$ (fully disconnected term). The result reads
\begin{equation}
\begin{aligned}
\label{10042021_0935}
\lim_{R \rightarrow \infty} G_{n}(\bm{x}_{1},\dots,\bm{x}_{n}) & = \sum_{m=0}^{n} \binom{n}{m}  \widehat{\langle\sigma\rangle}^{n-m} \widetilde{\langle\sigma\rangle}^{m} \frac{1+(-1)^{n-m}}{2} \\
& = \frac{ (\widetilde{\langle\sigma\rangle} + \widehat{\langle\sigma\rangle})^{n} + (\widetilde{\langle\sigma\rangle} - \widehat{\langle\sigma\rangle})^{n} }{ 2 } \\
& = \frac{ \langle\sigma\rangle_{a}^{n} + \langle\sigma\rangle_{b}^{n} }{ 2 } \, .
\end{aligned}
\end{equation}
The interface separating phases $a$ and $b$ is characterized by midpoint fluctuations of order $R^{1/2}$ along the $x$-axis. For $R \rightarrow \infty$ the unbounded interfacial fluctuations yield the averaging over the phases $a$ and $b$ given by (\ref{10042021_0935}). For the Ising model, the averaging property (\ref{10042021_0935}) is known from rigorous result \cite{Abraham_review}. The above derivation shows that the averaging (\ref{10042021_0935}) is actually a more general feature.

%==================================================================================
\section{Large-$R$ expansion}
\label{sec_3}
In this section, we examine the correlation function $G_{n}$ including the leading-order corrections in finite size. By extending the approach of \cite{DV, DS_twopoint} the correction at order $R^{-1/2}$ is interpreted in terms of a probabilistic picture.

%==================================================================================
\subsection{Correction at order $R^{-1/2}$: connected part}
To be definite, we begin by examining the connected part of $G_{n}$. The treatment of finite-size corrections for disconnected parts will be facilitated by the treatment of the connected part.

The large-$R$ expansion of $G_{n}$ can be studied in a systematic fashion by expanding the numerator of (\ref{02}), as well as the partition function $\mathcal{Z}_{ab}(R)$, in powers of the small parameter $R^{-1/2}$. The large-$R$ expansion of the numerator in (\ref{02}) proceeds by Taylor expanding the integrand in (\ref{10}) at small rapidities. Retaining the connected part, the integrand in (\ref{10}) reads
\begin{equation}
\label{ }
f_{ab}^{*}(\theta_{1}) f_{ab}(\theta_{n+1}) \prod_{j=1}^{n} F_{aba}^{\sigma_{j}}(\theta_{j}-\theta_{j+1}+\im \pi) \equiv 
\mathcal{C}(\theta_{1},\dots,\theta_{n+1}) \, .
\end{equation}
The function $\mathcal{C}(\theta_{1},\dots,\theta_{n+1})$ is expanded at small rapidities and the corresponding result is organized as follows
\begin{equation}
\label{ }
\mathcal{C}(\theta_{1},\dots,\theta_{n+1}) = \sum_{\Delta=-n}^{\infty} \mathcal{C}_{\Delta}(\theta_{1},\dots,\theta_{n+1}) \, ,
\end{equation}
where $\mathcal{C}_{\Delta}$ is a homogeneous function of order $\Delta$, i.e., $\mathcal{C}_{\Delta}(\alpha \theta_{1},\dots, \alpha \theta_{n+1}) = \alpha^{\Delta}\mathcal{C}_{\Delta}(\theta_{1},\dots,\theta_{n+1})$, $\alpha>0$. It is simple to check that terms in the aforementioned series with homogeneity exponent $\Delta$ contributes to the correlation function $G_{n}(\bm{x}_{1}, \dots, \bm{x}_{n})$ at order $R^{-(\Delta+n)/2}$. The leading order term in the large-$R$ expansion is thus generated by the function $\mathcal{C}_{-n}$ and its corresponding expression is provided in (\ref{29012021_1058}). The function $\mathcal{C}_{\Delta}$ with $\Delta=-n+1$ gives the first subleading correction which occurs at order $R^{-1/2}$.

Regarding the large-$R$ expansion of the partition function, we denote the leading order expression (\ref{05}) with $\mathcal{Z}_{ab}^{(0)}(R)$. Therefore
\begin{equation}
\label{ }
\mathcal{Z}_{ab}(R) = \mathcal{Z}_{ab}^{(0)}(R) \bigl[ 1 + \Os(R^{-1}) \bigr] \, .
\end{equation}
The correction at order $\Os(R^{-1/2})$ is absent since the low-energy expansion of $f_{ab}(\theta)$ does not exhibit odd powers of $\theta$. As a result, the first subleading correction for the $n$-point correlation function is, in general, proportional to $\Os(R^{-1/2})$ and is entirely originated by the matrix element $\mathcal{C}_{-n+1}$. The latter is obtained by multiplying $n-1$ kinematical poles with one of the $c_{ab}^{ (\sigma_{k}) }$ factors appearing in the low-energy expansion (\ref{27052020_01}) and then summing over permutations of labels. We have
\begin{equation}
\label{ }
\mathcal{C}_{-n+1} = 2^{n-1} \sum_{k=1}^{n} I_{k,n} \mathfrak{D}_{k,n}(\theta_{1},\dots,\theta_{n+1}) \, ,
\end{equation}
with $I_{k,n}$ an overall factor which depends on the vacuum expectation values,
\begin{equation}
\label{ }
I_{k,n} = c_{ab}^{ (\sigma_{k}) } \prod_{j=1, j \neq k}^{n} ( - \widehat{\sigma}_{j} ) \, , %\OK
\end{equation}
and $\mathfrak{D}_{k,n}$ the following function of the rapidities
\begin{equation}
\label{ }
\mathfrak{D}_{k,n}(\theta_{1},\dots,\theta_{n+1}) = \prod_{j=1, j \neq k}^{n} \frac{ - \im  }{ \theta_{j}-\theta_{j+1} } \, . %\OK
\end{equation}
Note how $\mathfrak{D}_{k,n}$ can be obtained from the connected matrix element
\begin{equation}
\label{ }
\mathfrak{M}_{n}(\theta_{1},\dots,\theta_{n+1}) = \prod_{j=1}^{n} \frac{ - \im  }{ \theta_{j}-\theta_{j+1} }
\end{equation}
simply by removing one annihilation pole. The removal of the pole $1/(\theta_{j}-\theta_{j+1})$ in the matrix element $\mathfrak{M}_{n}$ can be achieved thanks to a differentiation with respect to the horizontal coordinate $x_{j}$ conjugated to the rapidity difference $\theta_{j}-\theta_{j+1}$; hence,
\begin{equation}
\begin{aligned}
\label{07012021_0909}
\Lbag \mathfrak{D}_{k,n} \Rbag & = \partial_{\eta_{k}} \Lbag \mathfrak{M}_{n} \Rbag \, , \\
& = 2^{-n} \partial_{\eta_{k}} \mathcal{G}_{n}(\bm{x}_{1}, \dots, \bm{x}_{n}) \, . %\OK
\end{aligned}
\end{equation}
The last equality brings in touch the first subleading correction to connected matrix elements with the cluster functions introduced in Sec.~\ref{sec_2}.

We are now in the position to write the correction at order $R^{-1/2}$. The large-$R$ expansion of the $n$-point connected correlation function can be written as follows
\begin{equation}
\label{}
G_{n}^{\rm CP} = \sum_{\ell=0}^{\infty} \bigl[ G_{n}^{\rm CP} \bigr]_{\ell} \, ,
\end{equation}
with $\bigl[ G_{n}^{\rm CP} \bigr]_{\ell} = \Os(R^{-\ell/2})$. The term with $\ell=0$ is the one computed in Sec.~\ref{sec_2}. According to the above discussion, the term with $\ell=1$ reads
\begin{equation}
\begin{aligned}
\label{29012021_1117}
\bigl[ G_{n}^{\rm CP} \bigr]_{1} & = \frac{1}{\sqrt{2mR}} \sum_{k=1}^{n} I_{k,n} \partial_{\eta_{k}} \mathcal{G}_{n}(\bm{x}_{1}, \dots, \bm{x}_{n}) \, .
\end{aligned}
\end{equation}
Note that (\ref{29012021_1117}) is proportional to $c_{ab}^{(\sigma_{j})}$, therefore it vanishes for the Ising model \cite{DV}.

%==================================================================================
\subsection{Correction at order $R^{-1/2}$: disconnected parts and full result}
Once we have established how the large-$R$ expansion is implemented for the connected part, the analysis of the  disconnected parts follows from the diagrammatic construction of matrix elements. Let us consider $n=1$ as the first example. In this case the disconnected term coincides with the fully disconnected one, the latter simplifies with the partition function up to the factor $\widetilde{\langle\sigma_{1}\rangle}$. The correction is thus entirely due to the connected part. As a result, the expansion of the magnetization profile reads
\begin{equation}
\begin{aligned}
G_{1}(\bm{x}) & = \widetilde{\langle\sigma_{1}\rangle}  + \bigl[ G_{1}^{\rm CP}(\bm{x}) \bigr]_{0} + \bigl[ G_{1}^{\rm CP}(\bm{x}) \bigr]_{1} + \Os(R^{-1}) \, ,
\end{aligned}
\end{equation}
with the leading-order ($\ell=0$) connected part given by
\begin{equation}
\bigl[ G_{1}^{\rm CP}(\bm{x}) \bigr]_{0} = - \widehat{ \langle \sigma_{1} \rangle } \mathcal{G}_{1}(\chi) \, .
\end{equation}
According to (\ref{29012021_1117}), the first subleading correction is given by
\begin{equation}
\begin{aligned}
\label{08042021_1125}
\bigl[ G_{1}^{\rm CP}(\bm{x}) \bigr]_{1} & = \frac{ c_{ab}^{(\sigma_{1})} }{ m } P_{1}(x,y) \, .
\end{aligned}
\end{equation}
The above perfectly matches the results established in \cite{DV,DS_bubbles}.

Let us consider the case $n=2$. The diagrams with the field $\sigma_{1}$ disconnected contribute to $G_{2}$ with $\widetilde{\langle\sigma_{1}\rangle} \Bigl[ G_{1}^{\rm CP}(\bm{x}_{2}) \Bigr]_{1}$ and analogously when the labels $1$ and $2$ are interchanged. The large-$R$ expansion of the pair correlation function reads
\begin{equation}
\begin{aligned}
G_{2}(\bm{x}_{1}, \bm{x}_{2}) & = \bigl[ G_{2}(\bm{x}_{1}, \bm{x}_{2}) \bigr]_{0} + \bigl[ G_{2}^{\rm CP}(\bm{x}_{1}, \bm{x}_{2}) \bigr]_{1} \\
& + \widetilde{\langle\sigma_{1}\rangle} \bigl[ G_{1}^{\rm CP}(\bm{x}_{2}) \bigr]_{1} + \widetilde{\langle\sigma_{2}\rangle} \bigl[ G_{1}^{\rm CP}(\bm{x}_{1}) \bigr]_{1} + \Os(R^{-1}) \, ,
\end{aligned}
\end{equation}
where $\bigl[ G_{2}(\bm{x}_{1}, \bm{x}_{2}) \bigr]_{0}$ is the expression given in (\ref{29012021_1001}). The terms due to disconnected matrix elements are those multiplied by factors $\widetilde{\langle\sigma_{j}\rangle}$. Subleading corrections for $n=2$ are computed from (\ref{29012021_1117}) and expressed as follows
\begin{equation}
\label{ }
\bigl[ G_{2}^{\rm CP}(\bm{x}_{1},\bm{x}_{2}) \bigr]_{1} = \frac{1}{\sqrt{2mR}} \bigl[ c_{ab}^{ (\sigma_{1}) } \widehat{\sigma}_{2} \partial_{\eta_{1}} + c_{ab}^{ (\sigma_{2}) } \widehat{\sigma}_{1} \partial_{\eta_{2}}\bigr] \mathcal{G}_{2}(\bm{x}_{1}, \bm{x}_{2}) \, .
\end{equation}
This result can be written in a more explicit way by carrying first derivatives with respect to $x_{1}$ and $x_{2}$ of the cluster function $\mathcal{G}_{2}$. It is simple to show that
\begin{equation}
\begin{aligned}
\label{}
\partial_{x_{1}} \mathcal{G}_{2}(\bm{x}_{1}, \bm{x}_{2}) & = 2 P_{1}(x_{1},y_{1}) \textrm{erf}\left( \frac{ \chi_{2} - \rho_{12} \chi_{1} }{ \sqrt{1-\rho_{12}^{2}} } \right) \, , \\
\partial_{x_{2}} \mathcal{G}_{2}(\bm{x}_{1}, \bm{x}_{2}) & = 2 P_{1}(x_{2},y_{2}) \textrm{erf}\left( \frac{ \chi_{1} - \rho_{12} \chi_{2} }{ \sqrt{1-\rho_{12}^{2}} } \right) \, .
\end{aligned}
\end{equation}
Grouping together corrections at order $R^{-1/2}$ stemming from both connected and disconnected parts, we find
\begin{equation}
\begin{aligned}
\label{08042021_0954}
[ G_{2}(\bm{x}_{1},\bm{x}_{2}) ]_{1} & = \frac{ c_{ab}^{(\sigma_{1})} }{ m } P_{1}(x_{1},y_{1}) \biggl[ \widetilde{\langle\sigma_{2}\rangle} - \widehat{\langle\sigma_{2}\rangle} \textrm{erf}\left( \frac{ \chi_{2} - \rho_{12} \chi_{1} }{ \sqrt{1-\rho_{12}^{2}} } \right) \biggr] + \\
& + \frac{ c_{ab}^{(\sigma_{2})} }{ m } P_{1}(x_{2},y_{2}) \biggl[ \widetilde{\langle\sigma_{1}\rangle}  - \widehat{\langle\sigma_{1}\rangle} \textrm{erf}\left( \frac{ \chi_{1} - \rho_{12} \chi_{2} }{ \sqrt{1-\rho_{12}^{2}} } \right) \biggr] \, .
\end{aligned}
\end{equation}
The above expressions coincide with the results given in \cite{DS_twopoint}. We further stress how the clustering of the two-point correlation function is satisfied at order $R^{-1/2}$. This can be easily inspected by considering the following limits
\begin{equation}
\begin{aligned}
\label{08042021_1121}
\lim_{x_{1} \rightarrow \mp \infty} [ G_{2}(\bm{x}_{1},\bm{x}_{2}) ]_{1} & = \langle \sigma_{1} \rangle_{a [b]} \frac{ c_{ab}^{(\sigma_{2})} }{ m } P_{1}(x_{2},y_{2}) \, , \\
\lim_{x_{2} \rightarrow \mp \infty} [ G_{2}(\bm{x}_{1},\bm{x}_{2}) ]_{1} & = \langle \sigma_{2} \rangle_{a [b]} \frac{ c_{ab}^{(\sigma_{1})} }{ m } P_{1}(x_{1},y_{1}) \, ,
\end{aligned}
\end{equation}
which are in agreement with (\ref{08042021_1125}).

By carrying on the above procedure, the three-point correlation function expands as follows
\begin{equation}
\begin{aligned}
G_{3}(\bm{x}_{1}, \bm{x}_{2}, \bm{x}_{3}) & = \bigl[ G_{3}(\bm{x}_{1}, \bm{x}_{2}, \bm{x}_{3}) \bigr]_{0} + \bigl[ G_{3}^{\rm CP}(\bm{x}_{1}, \bm{x}_{2}, \bm{x}_{3}) \bigr]_{1} \\
&+ \widetilde{\langle\sigma_{1}\rangle} \bigl[ G_{2}^{\rm CP}(\bm{x}_{2}, \bm{x}_{3}) \bigr]_{1} + \widetilde{\langle\sigma_{2}\rangle} \bigl[ G_{2}^{\rm CP}(\bm{x}_{1}, \bm{x}_{3}) \bigr]_{1} + \widetilde{\langle\sigma_{3}\rangle} \bigl[ G_{2}^{\rm CP}(\bm{x}_{1}, \bm{x}_{2}) \bigr]_{1} \\
& + \widetilde{\langle\sigma_{1}\rangle} \widetilde{\langle\sigma_{2}\rangle} \bigl[ G_{1}^{\rm CP}(\bm{x}_{3}) \bigr]_{1} + \widetilde{\langle\sigma_{1}\rangle} \widetilde{\langle\sigma_{3}\rangle} \bigl[ G_{1}^{\rm CP}(\bm{x}_{2}) \bigr]_{1} + \widetilde{\langle\sigma_{2}\rangle} \widetilde{\langle\sigma_{3}\rangle} \bigl[ G_{1}^{\rm CP}(\bm{x}_{1}) \bigr]_{1} \\
& + \Os(R^{-1}) \, ,
\end{aligned}
\end{equation}
with the leading-order term $\bigl[ G_{3}(\bm{x}_{1}, \bm{x}_{2}, \bm{x}_{3}) \bigr]_{0}$ given by (\ref{29012021_1002}) and the terms at order $R^{-1/2}$ given by (\ref{29012021_1117}). The result for arbitrary $n$ follows along the same lines.

%==================================================================================
\subsection{Probabilistic interpretation}
\label{sec33}
It is possible to reconstruct the $n$-point correlation function within a probabilistic interpretation in which the interface is regarded as a fluctuating line with fixed extremities. Let
\begin{equation}
\label{ }
P_{n}(x_{1},y_{1}; \dots ; x_{n},y_{n})\textrm{d}x_{1}\cdots \textrm{d}x_{n}
\end{equation}
be the probability that the interface crosses the intervals $(x_{j} , x_{j} + \textrm{d}x_{j})$ at time $y_{j}$. Then, let
\begin{equation}
\label{29012021_1613}
\sigma_{j}(x_{j} \vert u_{j}) = \widetilde{\langle \sigma_{j} \rangle} - \widehat{\langle \sigma_{j} \rangle} \textrm{sign}(x_{j}-u_{j}) + A_{ab}^{(\sigma_{j})}\delta(x_{j} - u_{j}) + \dots \, ,
\end{equation}
be the magnetization profile at in the point $x_{j}$ and $u_{j}$ the abscissa in which the interface crosses the horizontal line $y=y_{j}$. The first two terms in the right hand side of (\ref{29012021_1613}) account for coexisting phases sharply separated by a structureless interface. Endowing the interface with interface structure amounts to the subsequent terms beyond the sharp picture, as indicated in (\ref{29012021_1613}). 

The sum over interfacial configurations which define the  $n$-point correlation function is formulated as follows
\begin{equation}
\label{06012021_2016}
G_{n}(\bm{x}_{1},\dots,\bm{x}_{n}) = \int_{\mathbb{R}^{n}}\textrm{d}u_{1} \dots \textrm{d}u_{n} \, P_{n}(u_{1},y_{1}; \dots ; u_{n},y_{n}) \prod_{j=1}^{n} \sigma_{j}(x_{j} \vert u_{j}) \, .
\end{equation}
The fact that $P_{n}$ occurring in (\ref{06012021_2016}) is the expression found in field theory can be established by matching (\ref{06012021_2016}) with the field-theoretical calculation for arbitrary $n$. This is what we will do in the following.

We begin by focusing on the leading order in the large-$R$ expansion which is captured by the first two terms in (\ref{29012021_1613}). The development of the product appearing in (\ref{06012021_2016}) yields $2^{n}$ terms whose integral with respect to $u_{1},\dots,u_{n}$ reproduces the cluster functions introduced in Sec.~\ref{sec_2}. Proving the above statement is a simple matter. We denote the $n$-fold integral with respect to horizontal coordinates $\{u_{j}\}_{j=1,\dots,n}$ with the compact notation
\begin{equation}
\label{ }
\llbracket f(u_{1},\dots,u_{n}) \rrbracket \equiv \int_{\mathbb{R}^{n}}\textrm{d}u_{1} \dots \textrm{d}u_{n} \, f(u_{1},\dots,u_{n}) P_{n}(u_{1},y_{1}; \dots ; u_{n},y_{n}) \, .
\end{equation}
Then, we employ the following abbreviation for the sign function $\mathcal{s}_{j} \equiv \textrm{sign}(x_{j}-u_{j}) = 2\vartheta(x_{j}-u_{j})-1$ and $\vartheta_{j} \equiv \vartheta(x_{j}-u_{j})$, with $\vartheta(\cdots)$ Heaviside theta function. Block functions are expressed as follows
\begin{equation}
\begin{aligned}
\label{}
\llbracket \vartheta_{1} \rrbracket & =
\begin{tikzpicture}[baseline={([yshift=-.6ex]current bounding box.center)}, vertex/.style={anchor=center, circle, rounded corners=3, fill=yellow!30, minimum size=4mm, inner sep=0pt}]
\draw[rounded corners=8pt,  fill=black!10]
(0,0) rectangle ++(0.8, 0.8);
\node[vertex,draw, line width=0.5pt, fill=colordiagram1!100] (G1) at (0.4, 0.4)  { {\footnotesize{1}} };
\end{tikzpicture}
\\
\llbracket \vartheta_{1} \vartheta_{2} \rrbracket & =
\begin{tikzpicture}[baseline={([yshift=-.6ex]current bounding box.center)}, vertex/.style={anchor=center, circle, rounded corners=3, fill=yellow!30, minimum size=4mm, inner sep=0pt}]
\draw[rounded corners=8pt,  fill=black!10]
(0,0) rectangle ++(1.4, 0.8);
\node[vertex,draw, line width=0.5pt, fill=colordiagram1!100] (G1) at (0.4, 0.4)  { {\footnotesize{1}} };
\node[vertex,draw, line width=0.5pt, fill=colordiagram1!100] (G2) at (1.0,0.4)   { {\footnotesize{2}} };
\end{tikzpicture}
\\
\llbracket \vartheta_{1} \vartheta_{2} \vartheta_{3} \rrbracket & =
\begin{tikzpicture}[baseline={([yshift=-.6ex]current bounding box.center)}, vertex/.style={anchor=center, circle, rounded corners=3, fill=yellow!30, minimum size=4mm, inner sep=0pt}]
\draw[rounded corners=8pt,  fill=black!10]
(0,0) rectangle ++(2.0, 0.8);
\node[vertex,draw, line width=0.5pt, fill=colordiagram1!100] (G1) at (0.4, 0.4)  { {\footnotesize{1}} };
\node[vertex,draw, line width=0.5pt, fill=colordiagram1!100] (G2) at (1.0,0.4)   { {\footnotesize{2}} };
\node[vertex,draw, line width=0.5pt, fill=colordiagram1!100] (G2) at (1.6,0.4)   { {\footnotesize{3}} };
\end{tikzpicture}
\,\, .
\end{aligned}
\end{equation}
Consequently, cluster functions admit the following representation
\begin{equation}
\begin{aligned}
\label{}
\llbracket \mathcal{s}_{1} \rrbracket & = \mathcal{G}_{1}(\bm{x}_{1}) \\
\llbracket \mathcal{s}_{1} \mathcal{s}_{2} \rrbracket & = \mathcal{G}_{2}(\bm{x}_{1},\bm{x}_{2}) \\
\llbracket \mathcal{s}_{1} \mathcal{s}_{2} \mathcal{s}_{3} \rrbracket & = \mathcal{G}_{3}(\bm{x}_{1},\bm{x}_{2},\bm{x}_{3}) \, ;
\end{aligned}
\end{equation}
thus, for an arbitrary $1 \leqslant m \leqslant n$, we have
\begin{equation}
\label{ }
\Bigl\llbracket \prod_{j=1}^{m} \mathcal{s}_{j} \Bigr\rrbracket = \mathcal{G}_{m}(\bm{x}_{1},\dots,\bm{x}_{m}) \, .
\end{equation}
For $m=0$ the normalization condition gives $\mathcal{G}_{0}=\llbracket 1 \rrbracket=1$, as we also stipulated Sec.~\ref{sec_2}. The matching between the field theoretic calculation and the probabilistic interpretation is thus completely characterized at the leading order in the large-$R$ expansion.

We can now establish the matching at order $R^{-1/2}$. In order to do this, we focus on the connected part of the correlation function, $[ G_{n}^{\rm CP}(\bm{x}_{1},\dots,\bm{x}_{n}) ]_{1}$. The latter can be extracted from the probabilistic picture (\ref{06012021_2016}) by removing the offset values $\widetilde{\langle \sigma_{j} \rangle}$ in (\ref{29012021_1613}), namely
\begin{equation}
\begin{aligned}
\label{}
[ G_{n}(\bm{x}_{1},\dots,\bm{x}_{n}) ]_{1} & = \sum_{k=1}^{n} A_{ab}^{(\sigma_{k})}  \int_{\mathbb{R}^{n}}\textrm{d}u_{1} \dots \textrm{d}u_{n} \, P_{n}(x_{1},y_{1}; \dots ; x_{n},y_{n}) \delta(x_{k} - u_{k}) \\
& \times \prod_{j=1, j \neq k}^{n} \left( \widetilde{\langle \sigma_{j} \rangle} - \frac{1}{2} \Delta \langle \sigma_{j} \rangle \textrm{sign}(x_{j}-u_{j}) \right) \, .
\end{aligned}
\end{equation}
By applying the multiple derivative $\partial_{x_{1}}\cdots\partial_{x_{n}}$, one finds
\begin{equation}
\begin{aligned}
\label{29012021_1628}
\partial_{x_{1}}\cdots\partial_{x_{n}} [ G_{n}(\bm{x}_{1},\dots,\bm{x}_{n}) ]_{1} & = \sum_{k=1}^{n} A_{ab}^{(\sigma_{k})} \prod_{j=1, j \neq k}^{n} \left( - \Delta \langle \sigma_{j} \rangle \right) \partial_{x_{k}} P_{n}(x_{1},y_{1}; \dots ; x_{n},y_{n}) \, .
\end{aligned} 
\end{equation}
On the other hand, field theory yields
\begin{equation}
\begin{aligned}
\label{29012021_1629}
\partial_{x_{1}}\cdots\partial_{x_{n}} [ G_{n}(\bm{x}_{1},\dots,\bm{x}_{n}) ]_{1} & = \frac{2^{n}}{\sqrt{2mR}} \sum_{k=1}^{n} I_{k,n}  \partial_{\eta_{k}} P_{n}(x_{1},y_{1}; \dots ; x_{n},y_{n}) \, , \\
& = \frac{1}{m} \sum_{k=1}^{n} c_{ab}^{ (\sigma_{k}) } \prod_{j=1, j \neq k}^{n} ( - \Delta\langle\sigma\rangle_{j} )  \partial_{x_{k}} P_{n}(x_{1},y_{1}; \dots ; x_{n},y_{n}) \, .
\end{aligned}
\end{equation}
By matching (\ref{29012021_1628}) and (\ref{29012021_1629}), we readily extract the structure amplitudes
\begin{equation}
\label{29012021_1630}
A_{ab}^{(\sigma_{k})} = \frac{c_{ab}^{(\sigma_{k})} }{ m } \, .
\end{equation}
As consistency requires, (\ref{29012021_1630}) agrees with calculations based on $n=1$ and $n=2$, respectively in \cite{DV} and \cite{DS_twopoint}.

It is actually possible to cast the above results within the diagrammatic framework which we employed in the previous section. We note the following property
\begin{equation}
\label{07012021_0929}
\Lbag \mathfrak{M}_{n} \Rbag = 2^{-n} \mathcal{G}_{n}(\bm{x}_{1}, \dots, \bm{x}_{n}) \, ,
\end{equation}
which expresses the cluster function in terms of the matrix element generated by the product of $n$ kinematical poles. Then, the cluster function $\mathcal{G}_{n}$ admits the following diagrammatic representation in terms of block diagrams
\begin{equation}
\label{07012021_0930}
\mathcal{G}_{n}(\bm{x}_{1}, \dots, \bm{x}_{n}) \equiv \sum_{k=0}^{n} 2^{k} (-1)^{n-k} \left( \sum_{ 1 \leqslant i_{1} < \cdots < i_{k} \leqslant n}
\begin{tikzpicture}[baseline={([yshift=-.5ex]current bounding box.center)}, vertex/.style={anchor=center, circle, rounded corners=3, fill=yellow!30, minimum size=4mm, inner sep=0pt}]
\draw[rounded corners=8pt,  fill=black!10]
(0,0) rectangle ++(2+1.4, 0.8);
\node[vertex,draw, line width=0.5pt, fill=colordiagram1!100] (G1) at (0.4, 0.4)  { {\footnotesize{$i_{1}$}} };
\node[vertex,draw, line width=0.5pt, fill=colordiagram1!100] (G1) at (1.0, 0.4)  { {\footnotesize{$i_{2}$}} };
\node[] (G1) at (2.00, 0.4)  { {\footnotesize{$\cdots$}} };
\node[vertex,draw, line width=0.5pt, fill=colordiagram1!100] (G2) at (2+1.0,0.4)   { {\footnotesize{$i_{k}$}} };
\end{tikzpicture}
\right) \, .
\end{equation}
In turn, the relationships (\ref{07012021_0929}) and (\ref{07012021_0930}) allow us to connect the calculation of the $(n+1)$-fold integral with respect to rapidities $\Lbag \mathfrak{M}_{n} \Rbag$ to a diagrammatic expansion. Such a relationship turns out to be extendable to the instance in which one of the simple poles of $\Lbag \mathfrak{M}_{n} \Rbag$ is replaced by $1$, which is precisely the construction which leads to the matrix element $\mathfrak{D}_{k,n}$. Note that each circle appearing in the block diagrams is in one-to-one correspondence with a Heaviside theta function, as we have shown. 

The diagrammatic representation of block function is extended by introducing the modified diagrams in which one of the Heaviside theta is replaced by a Dirac delta and the latter is represented with an orange diamond; thus
\begin{equation}
\begin{aligned}
\label{29012021_1303}
\llbracket \delta_{1} \rrbracket & =
\begin{tikzpicture}[baseline={([yshift=-.6ex]current bounding box.center)}, vertex/.style={anchor=center, diamond, rounded corners=3, fill=yellow!30, minimum size=6mm, inner sep=0pt}]
\draw[rounded corners=8pt,  fill=black!10]
(0,0) rectangle ++(0.8, 0.8);
\node[vertex,draw, line width=0.5pt, fill=colordiagram2!100] (G1) at (0.4, 0.4)  { {\footnotesize{1}} };
\end{tikzpicture}
=
\partial_{x_{1}}
\begin{tikzpicture}[baseline={([yshift=-.6ex]current bounding box.center)}, vertex/.style={anchor=center, circle, rounded corners=3, fill=yellow!30, minimum size=4mm, inner sep=0pt}]
\draw[rounded corners=8pt,  fill=black!10]
(0,0) rectangle ++(0.8, 0.8);
\node[vertex,draw, line width=0.5pt, fill=colordiagram1!100] (G1) at (0.4, 0.4)  { {\footnotesize{1}} };
\end{tikzpicture}
\\
\llbracket \delta_{1} \vartheta_{2} \rrbracket & =
\begin{tikzpicture}[baseline={([yshift=-.6ex]current bounding box.center)}, vertex/.style={anchor=center, circle, rounded corners=3, fill=yellow!30, minimum size=4mm, inner sep=0pt}, vertex2/.style={anchor=center, diamond, rounded corners=3, fill=yellow!30, minimum size=6mm, inner sep=0pt}]
\draw[rounded corners=8pt,  fill=black!10]
(0,0) rectangle ++(1.4, 0.8);
\node[vertex2,draw, line width=0.5pt, fill=colordiagram2!100] (G1) at (0.4, 0.4)  { {\footnotesize{1}} };
\node[vertex,draw, line width=0.5pt, fill=colordiagram1!100] (G2) at (1.0,0.4)   { {\footnotesize{2}} };
\end{tikzpicture}
=
\partial_{x_{1}}
\begin{tikzpicture}[baseline={([yshift=-.6ex]current bounding box.center)}, vertex/.style={anchor=center, circle, rounded corners=3, fill=yellow!30, minimum size=4mm, inner sep=0pt}, vertex2/.style={anchor=center, diamond, rounded corners=3, fill=yellow!30, minimum size=6mm, inner sep=0pt}]
\draw[rounded corners=8pt,  fill=black!10]
(0,0) rectangle ++(1.4, 0.8);
\node[vertex,draw, line width=0.5pt, fill=colordiagram1!100] (G1) at (0.4, 0.4)  { {\footnotesize{1}} };
\node[vertex,draw, line width=0.5pt, fill=colordiagram1!100] (G2) at (1.0,0.4)   { {\footnotesize{2}} };
\end{tikzpicture}
\\
\llbracket \delta_{1} \vartheta_{2} \vartheta_{3} \rrbracket & =
\begin{tikzpicture}[baseline={([yshift=-.6ex]current bounding box.center)}, vertex/.style={anchor=center, circle, rounded corners=3, fill=yellow!30, minimum size=4mm, inner sep=0pt}, vertex2/.style={anchor=center, diamond, rounded corners=3, fill=yellow!30, minimum size=6mm, inner sep=0pt}]
\draw[rounded corners=8pt,  fill=black!10]
(0,0) rectangle ++(2.0, 0.8);
\node[vertex2,draw, line width=0.5pt, fill=colordiagram2!100] (G1) at (0.4, 0.4)  { {\footnotesize{1}} };
\node[vertex,draw, line width=0.5pt, fill=colordiagram1!100] (G2) at (1.0,0.4)   { {\footnotesize{2}} };
\node[vertex,draw, line width=0.5pt, fill=colordiagram1!100] (G2) at (1.6,0.4)   { {\footnotesize{3}} };
\end{tikzpicture}
=
\partial_{x_{1}}
\begin{tikzpicture}[baseline={([yshift=-.6ex]current bounding box.center)}, vertex/.style={anchor=center, circle, rounded corners=3, fill=yellow!30, minimum size=4mm, inner sep=0pt}, vertex2/.style={anchor=center, diamond, rounded corners=3, fill=yellow!30, minimum size=6mm, inner sep=0pt}]
\draw[rounded corners=8pt,  fill=black!10]
(0,0) rectangle ++(2.0, 0.8);
\node[vertex,draw, line width=0.5pt, fill=colordiagram1!100] (G1) at (0.4, 0.4)  { {\footnotesize{1}} };
\node[vertex,draw, line width=0.5pt, fill=colordiagram1!100] (G2) at (1.0,0.4)   { {\footnotesize{2}} };
\node[vertex,draw, line width=0.5pt, fill=colordiagram1!100] (G2) at (1.6,0.4)   { {\footnotesize{3}} };
\end{tikzpicture}
\,\, ,
\end{aligned}
\end{equation}
where $\delta_{j} \equiv \delta(x_{j}-u_{j})$. The second equalities in (\ref{29012021_1303}) follow by virtue of $\partial_{x_{j}} \vartheta_{j} = \delta_{j}$. As an example, the connected pair correlation function at order $R^{-1/2}$ reads
\begin{equation}
\begin{aligned}
\label{08042021_0949}
[G_{2}^{\rm CP}(\bm{x}_{1},\bm{x}_{2})]_{1} & = - \widehat{\langle\sigma_{1}\rangle} A_{ab}^{(\sigma_{2})} \biggl[ 
2 \,\,
\begin{tikzpicture}[baseline={([yshift=-.6ex]current bounding box.center)}, vertex/.style={anchor=center, circle, rounded corners=3, fill=yellow!30, minimum size=4mm, inner sep=0pt}, vertex2/.style={anchor=center, diamond, rounded corners=3, fill=yellow!30, minimum size=6mm, inner sep=0pt}]
\draw[rounded corners=8pt,  fill=black!10]
(0,0) rectangle ++(1.4, 0.8);
\node[vertex,draw, line width=0.5pt, fill=colordiagram1!100] (G1) at (0.4, 0.4)  { {\footnotesize{1}} };
\node[vertex2,draw, line width=0.5pt, fill=colordiagram2!100] (G2) at (1.0,0.4)   { {\footnotesize{2}} };
\end{tikzpicture}
\,\,
-
\,\,
\begin{tikzpicture}[baseline={([yshift=-.6ex]current bounding box.center)}, vertex/.style={anchor=center, circle, rounded corners=3, fill=yellow!30, minimum size=4mm, inner sep=0pt}, vertex2/.style={anchor=center, diamond, rounded corners=3, fill=yellow!30, minimum size=6mm, inner sep=0pt}]
\draw[rounded corners=8pt,  fill=black!10]
(0,0) rectangle ++(0.8, 0.8);
\node[vertex2,draw, line width=0.5pt, fill=colordiagram2!100] (G1) at (0.4, 0.4)  { {\footnotesize{2}} };
\end{tikzpicture}
\,\,
\biggr]
\,\,
- \widehat{\langle\sigma_{2}\rangle} A_{ab}^{(\sigma_{1})} \biggl[ 
2 \,\,
\begin{tikzpicture}[baseline={([yshift=-.6ex]current bounding box.center)}, vertex/.style={anchor=center, circle, rounded corners=3, fill=yellow!30, minimum size=4mm, inner sep=0pt}, vertex2/.style={anchor=center, diamond, rounded corners=3, fill=yellow!30, minimum size=6mm, inner sep=0pt}]
\draw[rounded corners=8pt,  fill=black!10]
(0,0) rectangle ++(1.4, 0.8);
\node[vertex2,draw, line width=0.5pt, fill=colordiagram2!100] (G1) at (0.4, 0.4)  { {\footnotesize{1}} };
\node[vertex,draw, line width=0.5pt, fill=colordiagram1!100] (G2) at (1.0,0.4)   { {\footnotesize{2}} };
\end{tikzpicture}
\,\,
-
\,\,
\begin{tikzpicture}[baseline={([yshift=-.6ex]current bounding box.center)}, vertex/.style={anchor=center, circle, rounded corners=3, fill=yellow!30, minimum size=4mm, inner sep=0pt}, vertex2/.style={anchor=center, diamond, rounded corners=3, fill=yellow!30, minimum size=6mm, inner sep=0pt}]
\draw[rounded corners=8pt,  fill=black!10]
(0,0) rectangle ++(0.8, 0.8);
\node[vertex2,draw, line width=0.5pt, fill=colordiagram2!100] (G1) at (0.4, 0.4)  { {\footnotesize{1}} };
\end{tikzpicture}
\,\,
\biggr] \, .
\end{aligned}
\end{equation}
The diagrams appearing in the above can be easily computed. Focusing on the first two diagrams, the results are:
\begin{equation}
\begin{aligned}
\label{}
\begin{tikzpicture}[baseline={([yshift=-.6ex]current bounding box.center)}, vertex/.style={anchor=center, circle, rounded corners=3, fill=yellow!30, minimum size=4mm, inner sep=0pt}, vertex2/.style={anchor=center, diamond, rounded corners=3, fill=yellow!30, minimum size=6mm, inner sep=0pt}]
\draw[rounded corners=8pt,  fill=black!10]
(0,0) rectangle ++(0.8, 0.8);
\node[vertex2,draw, line width=0.5pt, fill=colordiagram2!100] (G1) at (0.4, 0.4)  { {\footnotesize{1}} };
\end{tikzpicture}
& = P_{1}(x_{1},y_{1}) \\
\begin{tikzpicture}[baseline={([yshift=-.6ex]current bounding box.center)}, vertex/.style={anchor=center, circle, rounded corners=3, fill=yellow!30, minimum size=4mm, inner sep=0pt}, vertex2/.style={anchor=center, diamond, rounded corners=3, fill=yellow!30, minimum size=6mm, inner sep=0pt}]
\draw[rounded corners=8pt,  fill=black!10]
(0,0) rectangle ++(1.4, 0.8);
\node[vertex2,draw, line width=0.5pt, fill=colordiagram2!100] (G1) at (0.4, 0.4)  { {\footnotesize{1}} };
\node[vertex,draw, line width=0.5pt, fill=colordiagram1!100] (G2) at (1.0,0.4)   { {\footnotesize{2}} };
\end{tikzpicture}
& = \frac{1}{2} P_{1}(x_{1},y_{1}) + \frac{1}{2} P_{1}(x_{1},y_{1}) \textrm{erf}\left(\frac{\chi_{2}-\rho_{12}\chi_{1}}{\sqrt{1-\rho_{12}^{2}}}\right) \, ;
\end{aligned}
\end{equation}
and analogous results are obtained for the other diagrams. It is thus evident that (\ref{08042021_0949}) can be written in the following explicit form
\begin{equation}
\begin{aligned}
\label{}
[G_{2}^{\rm CP}(\bm{x}_{1},\bm{x}_{2})]_{1} & = - \widehat{\langle\sigma_{1}\rangle} A_{ab}^{(\sigma_{2})} P_{1}(x_{2},y_{2}) \textrm{erf}\left(\frac{\chi_{1}-\rho_{12}\chi_{2}}{\sqrt{1-\rho_{12}^{2}}}\right) \\
& - \widehat{\langle\sigma_{2}\rangle} A_{ab}^{(\sigma_{1})} P_{1}(x_{1},y_{1}) \textrm{erf}\left(\frac{\chi_{2}-\rho_{12}\chi_{1}}{\sqrt{1-\rho_{12}^{2}}}\right) \, ,
\end{aligned}
\end{equation}
this result perfectly matches with the connected part of the expression (\ref{08042021_0954}) obtained from the field theoretical calculation. Such a connected part can be selected simply by removing the terms proportional to the offsets $\widetilde{\langle\sigma_{1}\rangle}$ and $\widetilde{\langle\sigma_{2}\rangle}$.

%==================================================================================
\subsection{Triplet correlations}
\label{sec_triplet}
As an application of the formal results derived in the previous sections, here we consider the explicit form of the three-point correlation function
\begin{equation}
\label{29012021_1758}
\mathcal{g}_{3}(x,y) \equiv \langle \sigma(0,y) \sigma(x,0) \sigma(0,-y) \rangle_{ab} \, .
\end{equation}
The integrals which define the corresponding cluster functions can be evaluated in closed form. We leave the technical calculations in Appendix \ref{Appendix_B} and here we recall the main notations. Thus, we introduce the rescaled coordinates $\eta=x/\lambda$, $\tau=2y/R$, the correlation coefficient
\begin{equation}
\label{ }
\rho_{12}(\tau) = \sqrt{\frac{1-\tau}{1+\tau}} \, ,
\end{equation}
and the parameter $r(\tau)=\rho_{12}(\tau)/\sqrt{1-\rho_{12}^{2}(\tau)}$. The correlation function (\ref{29012021_1758}) reads
\begin{equation}
\begin{aligned}
\label{29012021_1804}
\mathcal{g}_{3}(x,y) & = - \left( \frac{\langle \sigma \rangle_{a} - \langle \sigma \rangle_{b}}{2} \right)^{3} \mathcal{Y}(\eta, \tau) + \frac{(\langle \sigma \rangle_{a} - \langle \sigma \rangle_{b})^{2}(\langle \sigma \rangle_{a} + \langle \sigma \rangle_{b})}{8} \mathcal{K}(\eta, \tau) \\
& - \frac{(\langle \sigma \rangle_{a} - \langle \sigma \rangle_{b})(\langle \sigma \rangle_{a} + \langle \sigma \rangle_{b})^{2}}{8} \textrm{erf}(\eta) + \left( \frac{\langle \sigma \rangle_{a} + \langle \sigma \rangle_{b}}{2} \right)^{3} + \Os(R^{-1/2}) \, ,
\end{aligned}
\end{equation}
where $\mathcal{Y}(\eta, \tau)$ and $\mathcal{K}(\eta, \tau)$ are the functions given by
\begin{equation}
\label{ }
\mathcal{Y}(\eta, \tau) = \frac{2}{\sqrt{\pi}} \int_{0}^{\eta} \textrm{d}u \, \textrm{e}^{-u^{2}} \textrm{erf}^{2}(r u) \, ,
\end{equation}
and
\begin{equation}
\label{ }
\mathcal{K}(\eta, \tau) = \frac{2}{\pi} \sin^{-1}(\rho_{12}^{2}) + 8 T(\sqrt{2}\eta, r(\tau)) \, ,
\end{equation}
where $T$ is Owen's function \cite{Owen1980}
\begin{equation}
\label{ }
T(\sqrt{2}\eta, r) = \frac{1}{2\pi} \int_{0}^{r} \textrm{d}u \, \frac{\textrm{e}^{-(1+u^{2})\eta^{2}}}{1+u^{2}} \, .
\end{equation}

Let us comment on some general properties. In the limit $x \rightarrow \pm \infty$ the triplet correlation function reduces to the pair correlation function with spin fields placed along the interface and symmetrically displaced with respect to the horizontal axis. In that limit, one finds
\begin{equation}
\begin{aligned}
\label{}
\lim_{x \rightarrow \mp \infty} \langle \sigma(0,y) \sigma(x,0) \sigma(0,-y) \rangle_{ab} & = \langle \sigma \rangle_{a[b]} \langle \sigma(0,y) \sigma(0,-y) \rangle_{ab} \, ;
\end{aligned}
\end{equation}
the quantity in the right hand side is the two-point correlation function with spin fields along the interface
\begin{equation}
\begin{aligned}
\label{}
\langle \sigma(0,y) \sigma(0,-y) \rangle_{ab} & = \left( \frac{ \langle \sigma \rangle_{a} + \langle \sigma \rangle_{b} }{ 2 } \right)^{2} + \left( \frac{ \langle \sigma \rangle_{a} - \langle \sigma \rangle_{b} }{ 2 } \right)^{2} \sin^{-1}(\rho_{12}^{2}) \, .
\end{aligned}
\end{equation}

Interestingly enough, for $\tau=\nicefrac{1}{3}$ -- corresponding to $y=R/6$, and $r=1$ -- the special functions in (\ref{29012021_1804}) reduce to powers of the error function, in particular
\begin{equation}
\label{ }
\mathcal{Y}(\eta, \nicefrac{1}{3}) = \frac{1}{3} \textrm{erf}^{3}(\eta) \, , \qquad T(\sqrt{2}\eta,1) = \frac{1}{8} \Bigl[ 1 - \textrm{erf}^{2}(\eta) \Bigr] \, .
\end{equation}
Lastly, the correlation function with the three spins placed along the line which joins the pinning points reads
\begin{equation}
\begin{aligned}
\label{09042021_0931}
\mathcal{g}_{3}(0,y) & = \frac{(\langle \sigma \rangle_{a} - \langle \sigma \rangle_{b})^{2}(\langle \sigma \rangle_{a} + \langle \sigma \rangle_{b})}{4\pi} \biggl[ 2 \sin^{-1}(\rho_{12}) + \sin^{-1}(\rho_{12}^{2}) \biggr] + \left( \frac{\langle \sigma \rangle_{a} + \langle \sigma \rangle_{b}}{2} \right)^{3} \\
& + [\mathcal{g}_{3}(0,y)]_{1} + \Os(R^{-1}) \, ,
\end{aligned}
\end{equation}
where $[\mathcal{g}_{3}(0,y)]_{1}=\Os(R^{-1/2})$ is the subdominant correction due to interface structure effects. The result for $[\mathcal{g}_{3}(0,y)]_{1}$ can be obtained by taking $y_{1}=y$, $y_{2}=0$ and $y_{3}=-y$ in thee expression for the correction at order $R^{-1/2}$ of the correlation function $\langle \sigma_{c}(0,y_{1}) \sigma_{c}(0,y_{2}) \sigma_{c}(0,y_{3}) \rangle_{ab}$, which is calculated in Appendix \ref{appendix_B2}. By taking the above limiting case in (\ref{08042021_2254}), we obtain the subdominant correction
\begin{equation}
\begin{aligned}
\label{09042021_1009}
[\mathcal{g}_{3}(0,y)]_{1} & = A_{ab}^{(\sigma)} \sqrt{\frac{2m}{\pi R}} \biggl[ \left( \frac{ \langle \sigma \rangle_{a} + \langle \sigma \rangle_{b} }{2} \right)^{2} \left( 1 + \frac{2}{\kappa} \right) + \frac{ \left( \langle \sigma \rangle_{a} - \langle \sigma \rangle_{b} \right)^{2} }{ 2\pi \kappa } \tan^{-1}\rho_{12} \biggr] \, ,
\end{aligned}
\end{equation}
with $\kappa=\sqrt{1-\tau^{2}}$.

The occurrence of long-range interfacial correlations can be verified by expanding the correlation coefficient $\rho_{12}$ and $\kappa$ for small $\tau$. Focusing on the leading-order tern, we find the asymptotic behavior
\begin{equation}
\begin{aligned}
\label{29012021_1836}
\mathcal{g}_{3}(0,y) & \simeq \frac{(\langle \sigma \rangle_{a} - \langle \sigma \rangle_{b})^{2}(\langle \sigma \rangle_{a} + \langle \sigma \rangle_{b})}{4\pi} \biggl[ \frac{3\pi}{2} - (4+2\sqrt{2}) \sqrt{\frac{y}{R}} \biggr] + \left( \frac{\langle \sigma \rangle_{a} + \langle \sigma \rangle_{b}}{2} \right)^{3} \, ,
\end{aligned}
\end{equation}
with $\xi_{\rm b} \ll y \ll R$. An analogous expansion can be performed for the interface structure correction given by (\ref{09042021_1009}).

The term proportional to $\sqrt{y}$ in (\ref{29012021_1836}) is the signature of long range interfacial correlations. This power-law behavior in the direction parallel to the interface has to be compared with the exponential decay of correlations which characterizes the transverse direction. This feature can be neatly appreciated simply by evaluating the derivative of $\mathcal{g}_{3}(x,y)$ with respect to $x$; a simple calculation gives
\begin{equation}
\begin{aligned}
\label{08042021_1153}
\partial_{x}\mathcal{g}_{3}(x,y) & = - \frac{ \langle \sigma \rangle_{a} - \langle \sigma \rangle_{b} }{4\sqrt{\pi}\lambda} \textrm{e}^{-\eta^{2}} \biggl[ \left( \langle \sigma \rangle_{a} - \langle \sigma \rangle_{b} \right)^{2} \textrm{erf}^{2}(r(\tau)\eta) \\
& + 2 \left( \langle \sigma \rangle_{a}^{2} - \langle \sigma \rangle_{b}^{2} \right) \textrm{erf}(r(\tau)\eta) + \left( \langle \sigma \rangle_{a} + \langle \sigma \rangle_{b} \right)^{2} \biggr] + \Os(R^{-1}) \, .
\end{aligned}
\end{equation}
The exponential factor $\textrm{e}^{-\eta^{2}}$ produces the confinement within the interfacial region of the long-range fluctuations of the order parameter.

Let us us consider now the Ising model. Denoting the spontaneous magnetization with $M=\langle \sigma \rangle_{+}$, the leading-order form of the triplet correlation becomes
\begin{equation}
\label{29012021_1837}
\mathcal{g}_{3}(x,y) = M^{3} \mathcal{Y}(\eta,r) \, .
\end{equation}
The above vanishes for $x=0$. However, going away from $x=0$ the decay of correlations along the direction parallel to the interface exhibits a long-range character analogous to (\ref{29012021_1836}) as well as an additional dependence on $x$ with the anisotropic features (i.e., dependence on $x$ and $y$) discussed above. A detailed asymptotic analysis of (\ref{29012021_1837}) and the comparison with results obtained with Monte Carlo simulations is carried out in a forthcoming publication \cite{ST_threepoint}.

Finally, we discuss the case of the $q$-state Potts model \cite{Wu}. For ferromagnetic interactions and with $q \leqslant4$ the model exhibits a continuous phase transition \cite{Baxter}. In the low-temperature phase there are $q$ degenerate ground states and, in the scaling limit, phase separation between them is described by field theory \cite{DV}. Thanks to permutational symmetry, the vacuum expectation values of the order parameter field satisfy
\begin{equation}
\label{ }
\langle \sigma_{c} \rangle_{a} = \frac{q \delta_{ac}-1}{q-1} M \, ,
\end{equation}
with $M=\langle\sigma_{a}\rangle_{a}$ the spontaneous magnetization. The formalism presented in this paper allows for a characterization of triplet correlations of the generic form $\langle \sigma_{c}(0,y) \sigma_{d}() \sigma_{e}(\bm{x}_{3}) \rangle_{ab}$ with $c,d,e \in \{1,\dots,q\}$. Focusing on the simplest case in which the three spins entering the correlation function have the same component, i.e., $c=d=e$, the three-point correlation function (\ref{29012021_1804}) takes the form
\begin{equation}
\begin{aligned}
\label{29012021_1239}
\langle \sigma_{c}(0,y) \sigma_{c}(x,0) \sigma_{c}(0,-y) \rangle_{ab} & = \frac{q^{3}M^{3}}{8(q-1)^{3}}\left( \delta_{ac} - \delta_{bc} \right)^{3} \mathcal{Y}(\eta, \tau) \\
& - \frac{q^{2}M^{3}}{8(q-1)^{3}} \left( \delta_{ac} - \delta_{bc} \right)^{2} \bigl[ q\left( \delta_{ac} + \delta_{bc} \right) -2 \bigr] \mathcal{K}(\eta, \tau) \\
& +\frac{qM^{3}}{8(q-1)^{3}} \left( \delta_{ac} - \delta_{bc} \right) \bigl[ q\left( \delta_{ac} + \delta_{bc} \right) -2 \bigr]^{2} \textrm{erf}(\eta) \\
& -\frac{M^{3}}{8(q-1)^{3}} \bigl[ q\left( \delta_{ac} + \delta_{bc} \right) -2 \bigr]^{3} + \Os(R^{-1/2}) \, .
\end{aligned}
\end{equation}
We observe that when $c$ equals one of the two boundary colors, e.g. $c=b$ (with $c\neq a$), the correlation function (\ref{29012021_1239}) reduces to a particularly simple expression
\begin{equation}
\begin{aligned}
\label{29012021_1254}
\langle \sigma_{c}(0,y) \sigma_{c}(x,0) \sigma_{c}(0,-y) \rangle_{ab} & = \frac{M^{3}}{8(q-1)^{3}} \biggl[ q^{3} \mathcal{Y}(\eta, \tau) - q^{2}(q-2) \mathcal{K}(\eta, \tau) + q (q-2)^{2} \textrm{erf}(\eta) \\
& - (q-2)^{3} \biggr] + \Os(R^{-1/2}) \, .
\end{aligned}
\end{equation}
It has to be observed how the result corresponding to the Ising model given in (\ref{29012021_1837}) is retrieved in the limit $q\rightarrow2$.

For the $q$-state Potts model with $q=3$ and $q=4$ it is possible to consider the correlations between non-boundary colors. By taking $c \neq a,b$ the first three terms in the right hand side of (\ref{29012021_1239}) vanish and one finds a term proportional to $1/(q-1)^{3}$ up to corrections proportional to $R^{-1/2}$. This feature is actually expected because the non-boundary color contributes in a nontrivial way to the magnetization profile at order $R^{-1/2}$, the same happens for correlation functions. As we are going to show in an explicit fashion, the term proportional to $R^{-1/2}$ depends on $y$. We can compute the correction at order $R^{-1/2}$ in (\ref{29012021_1239}) by adopting the probabilistic interpretation illustrated in Sec.~\ref{sec33}. For the sake of simplicity we show the specific form of these corrections for the special case $x=0$. The subleading correction is given by (\ref{09042021_1009}) with the structure amplitude for the $q$-state Potts field theory given by
\begin{equation}
\label{09042021_1016}
A_{ab}^{(\sigma_{c})} = \frac{2-q (\delta_{ca}+\delta_{cb})}{2} \frac{B_{q}M}{m} \, ,
\end{equation}
where $B_{3}=1/(2\sqrt{3})$ and $B_{4}=2/(3\sqrt{3})$ \cite{DV}. By inserting (\ref{09042021_1016}) into (\ref{09042021_1009}), we find the subleading correction
\begin{equation}
\begin{aligned}
\label{09042021_1021}
[\mathcal{g}_{3}(0,y)]_{1} & = - \frac{M^{3}}{\sqrt{2\pi mR}} \frac{[q(\delta_{ca}+\delta_{cb})-2]B_{q}}{(q-1)^{2}} \biggl\{ \bigl[ q(\delta_{ca}+\delta_{cb})-2 \bigr]^{2} \left( \frac{1}{4}+\frac{1}{2\kappa} \right) \\
& + q^{2} \left( \delta_{ca} - \delta_{cb} \right)^{2} \frac{\tan^{-1}\rho_{12}}{\pi\kappa} \biggr\} \, .
\end{aligned}
\end{equation}
By performing a small-$y$ expansion it is possible to show that (\ref{09042021_1021}) exhibits power-law correlations which are analogous to those obtained at the leading order in (\ref{29012021_1836}). For $q=3$ and $q=4$ the correlations of the non-boundary color are characterized by a non-vanishing amplitude. For $q=2$ the color $c$ must coincide either with $a$ or $b$ and the amplitude vanishes. In this case, we expect the first correction to occur at order $R^{-1}$. All these features are actually shared by the interface structure correction of the magnetization profile and can be interpreted as the formation of isolated droplets of phase $c$ adsorbed along the $ab$ interface \cite{DV}.

%==================================================================================
\section{Corrections at order $R^{-1}$}
\label{sec_4}
We have seen that finite-size corrections proportional to $R^{-1/2}$ computed within field theory match with a calculation based on the probabilistic interpretation. We show in this section that corrections at order $R^{-1}$ for the magnetization profile can be interpreted within the probabilistic picture by allowing certain structure amplitudes to be $y$-dependent. However, by using such an information gained for $n=1$, the case $n=2$ does not necessarily lead to a matching between the two formulations. We will cover these aspects by focusing on the explicit example of the Ising model.

%==================================================================================
\subsection{Magnetization profile}
The techniques developed in Sec.~\ref{sec_2} and Sec.~\ref{sec_3} can be straightforwardly applied to the case in question. We start by considering the low-rapidity expansion
\begin{equation}
\label{ }
f_{-+}^{*}(\theta_{1}) F^{\sigma}(\theta_{12}+\im\pi) f_{-+}(\theta_{2}) = \sum_{n=0}^{\infty} Q_{2n-1}(\theta_{1},\theta_{2}) \, ,
\end{equation}
where $Q_{2n-1}(\theta_{1},\theta_{2})$ are homogeneous functions of degree $2n-1$ in the rapidity variables, i.e., $Q_{2n-1}(\alpha\theta_{1},\alpha\theta_{2})=\alpha^{2n-1} Q_{2n-1}(\theta_{1},\theta_{2})$ for $\alpha>0$. The occurrence of odd powers follows because $f_{-+}(\theta) = f_{-+}(-\theta)$ and $F^{\sigma}(\theta+\im \pi) = - F^{\sigma}(-\theta+\im \pi)$; in particular, $F^{\sigma}(\theta+\im \pi) = -\im M \coth(\theta/2)$ \cite{YZ}, with $M=\langle \sigma \rangle_{+}>0$ the spontaneous magnetization. The boundary amplitude $f_{-+}(\theta)$ is known exactly for the Ising model but for the purpose of this paper it is sufficient to take\footnote{Without loss of generality, we can take $f_{-+}(0)=1$ because $f_{-+}(\theta)$ appears both at the numerator and denominator and the low-rapidity asymptotic is needed.} $f_{-+}(\theta) = 1 + f_{2} \theta^{2} + \Os(\theta^{4})$. The large-$R$ expansion reads
\begin{equation}
\begin{aligned}
\label{}
\langle \sigma(x,y) \rangle_{-+} & = - 2\im M \frac{ \Lbag Q_{-1} \Rbag + \frac{2}{mR} \Lbag Q_{1} \Rbag + \Os(R^{-2}) }{ 1 + \frac{2f_{2}}{mR} + \Os(R^{-2}) } \\
& = - 2 \im M \biggl[ \Lbag Q_{-1} \Rbag + \frac{2}{mR} \left( \Lbag Q_{1} \Rbag - f_{2} \Lbag Q_{-1} \Rbag \right)  + \Os(R^{-2}) \biggr] \, ,
\end{aligned}
\end{equation}
with
\begin{equation}
\begin{aligned}
\label{}
Q_{-1} & = \frac{1}{\theta_{12}} \, , \qquad Q_{1} & = f_{2} \frac{ \theta_{1}^{2} + \theta_{2}^{2} }{ \theta_{12} } + \frac{1}{12} \theta_{12} \, .
\end{aligned}
\end{equation}
We have
\begin{eqnarray}
\label{A008}
\Lbag 1 \Rbag & = & \frac{ 1 }{ \sqrt{\pi}\kappa } \textrm{e}^{-\chi^{2}} \\
\Lbag \theta_{1}^{2}+\theta_{2}^{2} \Rbag & = & \frac{ 2 }{ \sqrt{\pi} \kappa^{3}} \Bigl[ 1-\left( 1+\tau^{2} \right) \chi^{2} \Bigr] \textrm{e}^{-\chi^{2}} \\
\Lbag \theta_{12} \Rbag & = & \frac{ 2\textrm{i}\chi }{ \sqrt{\pi}\kappa^{2}} \textrm{e}^{-\chi^{2}} \, .
\end{eqnarray}
Then,
\begin{eqnarray}
\label{A007}
\scaleleftright[1.2ex]{\Lbag}{    \frac{ 1}{\theta_{12}}    }{\Rbag} & = & \textrm{i}  \int\textrm{d}\eta \, \Lbag 1 \Rbag =  \frac{ \textrm{i} }{ 2 } \textrm{erf}(\chi) \, ,\\
\label{A007b}
\scaleleftright[1.2ex]{\Lbag}{    \frac{ \theta_{1}^{2}+\theta_{2}^{2} }{\theta_{12}}    }{\Rbag} & = & \textrm{i}  \int\textrm{d}\eta \, \Lbag \theta_{1}^{2}+\theta_{2}^{2} \Rbag = \frac{ \textrm{i} }{ 2 } \textrm{erf}(\chi) + \frac{ \textrm{i} }{ \sqrt{\pi}\kappa^{2} } \left( 1+\tau^{2} \right) \chi \textrm{e}^{-\chi^{2}} \, .
\end{eqnarray}
The magnetization profile reads
\begin{equation}
\begin{aligned}
\label{11042021_1928}
\langle \sigma(x,y) \rangle_{-+} & = M \textrm{erf}(\chi) + \frac{M}{mR} \biggl[ \frac{2}{3} + 4f_{2}(1+\tau^{2}) \biggr] \chi \textrm{e}^{-\chi^{2}} + \Os(R^{-2}) \, .
\end{aligned}
\end{equation}
It has to be noticed how the expansion of the numerator originates an extended profile proportional to $\textrm{erf}(\chi)$ through the function $\Lbag (\theta_{1}^{2}+\theta_{2}^{2})/\theta_{12} \Rbag$ (see $Q_{1}$) and that such contribution is canceled by the subtraction of $f_{2} \Lbag Q_{-1} \Rbag$. As a result, the correction at order $R^{-1}$ is a localized profile proportional to $\chi \textrm{e}^{-\chi^{2}} \propto \partial_{x} P_{1}(x,y)$. This remark actually indicates that the above result can be obtained within the probabilistic description by averaging the sharp magnetization profile
\begin{equation}
\label{ }
\sigma(x \vert u) = -M + 2M \theta(x-u) + A_{1}(y) \partial_{x} \delta(x-u) + \dots \, ,
\end{equation}
and the matching yields the structure amplitude
\begin{equation}
\label{ }
A_{1}(y) = - \frac{M}{m^{2}} \biggl[ \frac{1}{6} + f_{2}(1+\tau^{2}) \biggr] \, ,
\end{equation}
which depends on $y=R\tau/2$.

%==================================================================================
\subsection{Two-point correlation function}
The calculation proceeds as follows
\begin{equation}
\label{ }
f_{-+}^{*}(\theta_{1}) F_{-+-}^{\sigma}(\theta_{12}+\im \pi) F_{-+-}^{\sigma}(\theta_{23}+\im \pi) f_{-+}(\theta_{3}) = \sum_{n=0}^{\infty} I_{2n-2}(\theta_{1},\theta_{2},\theta_{3}) \, ,
\end{equation}
with $I_{2n-2}$ a homogeneous function of order $2n-2$. The expansion yields
\begin{equation}
\begin{aligned}
\label{}
I_{-2}(\theta_{1},\theta_{2},\theta_{3}) & = \frac{1}{\theta_{12}\theta_{23}} \, , \\
I_{0}(\theta_{1},\theta_{2},\theta_{3}) & = \frac{1}{\theta_{12}\theta_{23}} \biggl[ f_{2} \left( \theta_{1}^{2} + \theta_{3}^{2} \right) + \frac{1}{12} \left( \theta_{12}^{2} + \theta_{23}^{2} \right) \biggr] \, .
\end{aligned}
\end{equation}
The spin-spin correlation function expands as follows
\begin{equation}
\label{ }
G_{2} = - 4M^{2} \frac{ \Lbag I_{-2} \Rbag + \frac{2}{mR} \Lbag I_{0} \Rbag + \Os\left( R^{-2} \right) }{ 1 + \frac{2f_{2}}{mR} + \Os\left( R^{-2} \right) } \, ,
\end{equation}
thus
\begin{equation}
\label{ }
G_{2} = - 4M^{2} \biggl[ \Lbag I_{-2} \Rbag + \frac{2}{mR} \left( \Lbag I_{0} \Rbag - f_{2} \Lbag I_{-2} \Rbag \right) \biggr] + \Os(R^{-2}) \, .
\end{equation}

In order to simplify the analysis, we restrict ourselves to the parallel correlation function, $G_{2}\vert_{\parallel} \equiv G_{2}(x,y;x,-y)$. As a further simplification for our considerations, we take the double derivative $\partial_{x_{1}}\partial_{x_{2}}$ and evaluate it for spin field in the parallel arrangement defined above. Therefore, we examine
\begin{equation}
\label{ }
\partial_{x_{1}x_{2}}^{2} G_{2}(x_{1},y_{1}; x_{2},y_{2}) \Big\vert_{\parallel} \equiv \partial_{x_{1}x_{2}}^{2} G_{2} \Big\vert_{\parallel} \, ,
\end{equation}
the subscript $\parallel$ means that $x_{1}=x_{2} \equiv x, y_{1}=-y_{2}\equiv y$ are set afterwards the application of $\partial_{\eta_{1}} \partial_{\eta_{2}}$. Now, we look more closely to $\partial_{\eta_{1}\eta_{2}}^{2} \Lbag I_{n} \Rbag$, for $n=-2$ and $n=0$. We have
\begin{equation}
\begin{aligned}
\label{}
\partial_{\eta_{1} \eta_{2}}^{2} \Lbag I_{-2} \Rbag & = - \Lbag 1 \Rbag \, ,
\end{aligned}
\end{equation}
with $\Lbag 1 \Rbag = \lambda^{2} P_{2}(x_{1},y_{1};x_{2},y_{2})$ and 
\begin{equation}
\begin{aligned}
\label{}
\partial_{\eta_{1} \eta_{2}}^{2} \Lbag I_{0} \Rbag & = - f_{2} \Lbag \theta_{1}^{2} + \theta_{3}^{2} \Rbag - \frac{1}{12} \Lbag \theta_{12}^{2} + \theta_{23}^{2} \Rbag \, .
\end{aligned}
\end{equation}
The second term can be written as follows
\begin{equation}
\begin{aligned}
\Lbag \theta_{12}^{2} + \theta_{23}^{2} \Rbag & = - \left( \partial_{\eta_{1}}^{2} + \partial_{\eta_{2}}^{2} \right)\Lbag 1 \Rbag \, .
\end{aligned}
\end{equation}
The above admits a simple expression for spin fields in parallel arrangement, i.e., with $x_{1}=x_{2} \equiv x, y_{1}=-y_{2}\equiv y$, in particular
\begin{equation}
\label{ }
- \left( \partial_{\eta_{1}}^{2} + \partial_{\eta_{2}}^{2} \right)\Lbag 1 \Rbag_{\parallel} = B(\eta,\tau) \Lbag 1 \Rbag_{\parallel} \, ,
\end{equation}
with
\begin{equation}
\label{ }
B(\eta,\tau) = \frac{ 1-\tau^{2}-2\tau\eta^{2} }{ \tau(1-\tau)^{2} } \, .
\end{equation}
The other term is arranged in a similar way. By writing
\begin{equation}
\label{ }
\Lbag \theta_{1}^{2} + \theta_{3}^{2} \Rbag_{\parallel} = E(\eta,\tau) \Lbag 1 \Rbag_{\parallel} \, .
\end{equation}
with
\begin{equation}
\label{ }
E(\eta,\tau) = \frac{ 2-2\tau-2\eta^{2} }{ (1-\tau)^{2} } \, .
\end{equation}
Summing up all the pieces, we obtain
\begin{equation}
\begin{aligned}
\label{12012021_1658}
\partial_{x_{1}x_{2}}^{2} G_{2} \Big\vert_{\parallel} & = 4 M^{2} P_{2}(x,y;x,-y) \\
& + \frac{M^{2}}{mR} \biggl\{ \frac{2}{3} B(\eta,\tau) - 8f_{2} \bigl[ 1-E(\eta,\tau) \bigr] \biggr\} P_{2}(x,y;x,-y) + \Os(\lambda^{-2} (mR)^{-2}) \, .
\end{aligned}
\end{equation}
The factor $\lambda^{-2}$ in the $\Os$-symbol is due to the differentiation with respect to $x_{1}=\lambda\eta_{1}$ and $x_{2}=\lambda\eta_{2}$. The calculation within the probabilistic interpretation follows straightforwardly and reads
\begin{equation}
\begin{aligned}
\label{}
\partial_{x_{1}x_{2}}^{2} G_{2}^{(\rm prob.)}(x_{1},y_{1}; x_{2},y_{2}) & = 4 M^{2} P_{2}(x_{1},y_{1}; x_{2},y_{2}) \\
& + 2M \bigl[ A_{1}(\tau_{1}) \partial_{x_{1}}^{2} + A_{1}(\tau_{2}) \partial_{x_{2}}^{2}  \bigr] P_{2}(x_{1},y_{1}; x_{2},y_{2}) + \Os(\lambda^{-2} (mR)^{-2}) \, ,
\end{aligned}
\end{equation}
the superscript ``prob.'' stresses that such an expression has been derived within the probabilistic interpretation. The second term can be written in a form which is similar to (\ref{12012021_1658})
\begin{equation}
\begin{aligned}
\label{12012021_1703}
\partial_{x_{1}x_{2}}^{2} G_{2}^{(\rm prob.)} \Big\vert_{\parallel} & = 4 M^{2} P_{2}(x,y;x,-y) \\
& + \frac{M^{2}}{mR} \biggl\{ \frac{2}{3} B(\eta,\tau) + 4f_{2} (1+\tau^{2}) B(\eta,\tau) \bigr] \biggr\} P_{2}(x,y;x,-y) + \Os(\lambda^{-2} (mR)^{-2}) \, .
\end{aligned}
\end{equation}
We see that the field-theoretic calculation (\ref{12012021_1658}) and the one carried out within the probabilistic interpretation, (\ref{12012021_1703}), agree at the leading order but disagree at the order $R^{-1}$. The disagreement is actually caused by the term proportional to $f_{2}$ and is ultimately originated by the features of the boundary field theory. On the other hand, the term proportional to the factor $\nicefrac{2}{3}$, which emerges from the low-energy properties of the bulk form factor, does not originate the disagreement.

The above analysis suffices in order to provide an example which exhibits an explicit breakdown of the matching between probabilistic interpretation and field theory. As we have already proved, the probabilistic approach has to be limited to corrections at order $R^{-1/2}$ in which the low-energy behavior of the boundary features does not report.

%==================================================================================
\subsection{Stress tensor trace and mixed correlation functions}
The calculation method illustrated in the previous sections can be applied to mixed correlation functions involving both the trace of the stress tensor and spin fields. In this conclusive section, we give an account on this aspect. Denoting the stress tensor trace with $\Theta(\bm{x})$, the counterpart of the matrix element decomposition (\ref{14012021_1619}) reads
\begin{equation}
\label{18012021_1441}
\mathcal{M}_{ab}^{\Theta}(\theta_{j} \vert \theta_{j+1}) \,\,
=
\vcenter{\hbox{
\begin{tikzpicture}[baseline={([yshift=-.6ex]current bounding box.center)},vertex/.style={anchor=base, circle, minimum size=50mm, inner sep=0pt}]
\draw[-,black]   (0, -1.8) -- (0, -0.6);
\draw[-,black]   (0, 0.6) -- (0, 1.8);
\path (0,0.0) node[circle, draw, fill=myyellow!60] (s1) {$\Theta$};
\path (0,0.0) node () {} (-0.75,0) node (s4) { ${\color{blue}{a}}$ };
\path (0,0.0) node () {} (0.75,0) node (s4) { ${\color{red}{b}}$ };
\path (0,0.0) node () {} (0.26,1.6) node (s4) { ${\color{black}{\theta_{j}}}$ };
\path (0,0.0) node () {} (0.47,-1.6) node (s4) { ${\color{black}{\theta_{j+1}}}$ };
\end{tikzpicture}
}}
=
\vcenter{\hbox{
\begin{tikzpicture}[baseline={([yshift=-.6ex]current bounding box.center)},vertex/.style={anchor=base, circle, minimum size=50mm, inner sep=0pt}]
\draw[-,black]   (0, -1.8) -- (0, 1.8);
\path (0,0.0) node[circle, draw, fill=myyellow!60] (s1) {$\Theta$};
\path (0,0.0) node () {} (-0.75,0) node (s4) { ${\color{blue}{a}}$ };
\path (0,0.0) node () {} (0.75,0) node (s4) { ${\color{red}{b}}$ };
\path (0,0.0) node () {} (0.26,1.6) node (s4) { ${\color{black}{\theta_{j}}}$ };
\path (0,0.0) node () {} (0.47,-1.6) node (s4) { ${\color{black}{\theta_{j+1}}}$ };
\end{tikzpicture}
}}
\quad
+
\vcenter{\hbox{
\begin{tikzpicture}[baseline={([yshift=-.6ex]current bounding box.center)},vertex/.style={anchor=base, circle, minimum size=50mm, inner sep=0pt}]
\def\u{0.95}
\def\v{0.65}
\path (0,0.0) node[circle, draw, fill=myyellow!60] (s1) {$\Theta$};
\path (0,0.0) node () {} (-0.75,0) node (s4) { ${\color{blue}{a}}$ };
\path (0,0.0) node () {} (1.05,0) node (s4) { ${\color{red}{b}}$ };
\path (0,0.0) node () {} (0.26,1.6) node (s4) { ${\color{black}{\theta_{j}}}$ };
\path (0,0.0) node () {} (0.47,-1.6) node (s4) { ${\color{black}{\theta_{j+1}}}$ };
\draw[]
(0, -1.8) ..controls +(0, \u) and ( $(0.8, 0) - (0, +\v)$ )..
(0.8, 0) ..controls +(0, \v) and ( $(0, 1.8) - (0, +\u)$ )..
(0, 1.8);
\end{tikzpicture}
}} \, ,
\end{equation}
or equivalently,
\begin{equation}
\label{18012021_1448}
\mathcal{M}_{ab}^{\Theta}(\theta_{j} \vert \theta_{j+1}) = F_{aba}^{\Theta}(\theta_{j}-\theta_{j+1}+\im\pi) + 2\pi \langle \Theta \rangle_{a} \delta(\theta_{j}-\theta_{j+1}) \, .
\end{equation}
Contrary to the spin field, for the stress tensor $\langle \Theta \rangle_{a} = \langle \Theta \rangle_{b} \equiv \langle \Theta \rangle$; thus, its two-particle form factor can be expanded as follows
\begin{equation}
\label{18012021_1449}
F_{aba}^{\Theta}(\theta_{j}-\theta_{j+1}+\im\pi) = F_{aba}^{\Theta}(\im\pi) + \Os((\theta_{j}-\theta_{j+1})^{2}) \, .
\end{equation}
The normalization of $\Theta$ actually implies $F_{aba}^{\Theta}(\im\pi) = 2\pi m^{2}$ \cite{MS}.

Let us consider the $n$-point correlation function of the stress tensor trace, $\langle \Theta(\bm{x}_{1}) \cdots \Theta(\bm{x}_{n}) \rangle_{ab}$. The connected part follows straightforwardly
\begin{equation}
\label{18012021_1500}
\langle \Theta(\bm{x}_{1}) \cdots \Theta(\bm{x}_{n}) \rangle_{ab}^{\rm CP} = \frac{\left( F_{aba}^{\Theta}(\im\pi) \right)^{n}}{m^{n}} P_{n}(x_{1},y_{1}; \dots x_{n},y_{n}) \, ,
\end{equation}
up to subdominant large-$R$ corrections. Therefore, the joint $n$-intervals passage probability is proportional to the connected part of the $n$-point correlation function of the stress tensor.

It has to be observed that (\ref{18012021_1500}) scales as $R^{-n/2}$. As a first consequence of (\ref{18012021_1449}), the leading term in the large-$R$ expansion is the one which counts the maximum number of disconnected pieces. The fully disconnected term cancels exactly the partition function at the denominator and yields a spatially-independent offset given by $\langle \Theta \rangle^{n}$. The next-to-leading term comes at order $R^{-1/2}$ and it is due to the contraction of $n-1$ disconnected pieces with one connected matrix element; such terms are captured by the matrix element
\begin{equation}
\label{ }
\mathfrak{M}_{n}^{\Theta}(\theta_{1},\dots,\theta_{n+1}) = \sum_{i=1}^{n} F_{aba}^{\Theta}(\im \pi) \prod_{j \neq i}^{n} 2\pi \langle \Theta \rangle \delta(\theta_{j}-\theta_{j+1}) \, .
\end{equation}
The corresponding result for the $n$-point correlations of $\Theta$ reads
\begin{equation}
\label{30012021_1618}
\langle \Theta(\bm{x}_{1}) \cdots \Theta(\bm{x}_{n}) \rangle_{ab} = \langle \Theta \rangle^{n} + \langle \Theta \rangle^{n-1} \frac{F_{aba}^{\Theta}(\im \pi)}{m} \sum_{i=1}^{n} P_{1}(x_{i},y_{i}) + \Os(R^{-1}) \, .
\end{equation}

We can now consider a mixed correlation function which involves $n-m$ spin fields. The leading-order term follows by contracting the product of $(n-m)$ kinematical poles with $m$ Dirac deltas stemming from the stress tensor matrix element (\ref{18012021_1448}). The corresponding result reads
\begin{equation}
\begin{aligned}
\label{30012021_1614}
\langle \sigma_{1}(\bm{x}_{1}) \cdots \sigma_{n-m}(\bm{x}_{n-m}) \Theta(\bm{x}_{n-m+1}) \cdots \Theta(\bm{x}_{n}) \rangle_{ab}^{\rm CP} & = \langle \Theta \rangle^{m} \left( \prod_{j=1}^{n-m} (-\widehat{\langle\sigma_{j}\rangle}) \right) \mathcal{G}_{n-m}(\bm{x}_{1},\dots,\bm{x}_{n-m}) \\
& + \Os(R^{-1/2}) \, .
\end{aligned}
\end{equation}
As a consistency check we consider two limiting cases. For $m=0$ the above reduces to the connected $n$-point spin correlation function and the corresponding result (\ref{07062020_02}) is found as a limiting case. For $m=n$ we retrieve the connected $n$-point correlator of $\Theta$ given by the first term in the right hand side of (\ref{30012021_1618}).

%==================================================================================
\section{Conclusions}
\label{sec_5}
In this paper, we considered the scaling limit of a generic two-dimensional ferromagnetic system at phase coexistence near a second order phase transition point. We showed how field theory provides exact results for $n$-point spin and stress tensor trace correlation functions in presence of a fluctuating interface. More specifically, the system we considered is defined on an infinite strip of width $R$ much larger than the bulk correlation length. Boundary conditions are used in order to enforce phase separation through an interface which spans between the two edges and whose endpoints are pinned.

By extending the field-theoretical technique developed for one- and two-point correlation functions, respectively in \cite{DV} and \cite{DS_twopoint}, we have been able to find the exact analytic form of order parameter and stress tensor correlation functions. Analogously to the case of the two-point correlation function, we have showed that, as long as $R$ is finite, the $n$-point correlation function is characterized by long-range correlations in the direction parallel to the interface. The spatial extent of the interface midpoint fluctuations grows as $R^{1/2}$ and, for $R=\infty$, these unbounded fluctuations lead to an exponential decay of bulk correlations averaged over the two coexisting phases separated by the interface.

More technically, these results follow by exploiting general low-energy properties of two-dimensional field theory whose excitations -- in two dimensions -- are topological (kink) particles. We found that the leading asymptotic form of correlation functions involving spin fields is completely codified by the kinematical pole singularity exhibited by matrix element of the order parameter field. 

Among our findings, the dominant asymptotic form of $n$-point correlation functions is expressed in terms of $n$-body cluster functions which are constructed out of cumulative distribution functions of the $n$-variate gaussian distribution. The first subleading finite-size correction, which is proportional to $R^{-1/2}$ and arises from effects due to interface structure, depends on the bulk universality class only. Specificities related to the boundaries, which are incorporated in the low-energy behavior of matrix elements of boundary changing operators, do not report at order $R^{-1/2}$, but appear at order $R^{-1}$. Both the leading term and the first subleading corrections can be interpreted within a probabilistic picture in which the interface is regarded as the worldline of a particle which propagates randomly between the pinning points by undergoing a Brownian bridge. By using the Ising model as a specific example, we also show that the subleading correction at order $R^{-1}$ does not  necessarily emerge from the probabilistic description. We identify the origin of the mismatch as a specificity arising from matrix elements of the boundary condition changing operators.

Throughout this manuscript we have introduced a diagrammatic notation (block diagrams) for $n$-body cluster functions which facilitates the handling of expressions at both the leading and first subleading orders. Such  a notation proved to be useful also in establishing a graphical connection between disconnected matrix elements and their contribution to the correlation function.

We conclude by discussing some interesting perspectives. The reconstruction of $n$-point correlation function through the probabilistic interpretation, which we have shown to be correct at both the leading order and including corrections at order $R^{-1/2}$, can be used in order to find exact results in closed form once the passage probability is known. This is indeed the case for $n=3$ \cite{ST_threepoint} and $n=4$ \cite{ST_fourpoint} in which numerical simulations confirm the analytic results. The extension of the techniques developed in this paper has been merged with the techniques of \cite{DS_wetting} in a companion paper for the study of correlations in the half-plane. There, explicit results for the spin-spin correlation function on the half-plane with boundary conditions enforcing a droplet have been found and successfully tested by means of high-precision Monte Carlo simulations \cite{ST_droplet}. 

%==================================================================================
\section*{Acknowledgements}
I am grateful to Gesualdo Delfino for his valuable comments. I also thank Douglas B. Abraham for many interesting discussions and for collaborations on closely related topics. It is also a pleasure to acknowledge the Galileo Galilei Institute for Theoretical Physics (Arcetri, Florence) for hospitality received in the germinal stages of this work during the event \emph{``SFT 2019: Lectures on Statistical Field Theories''}.

\appendix
%==================================================================================
\section{Brownian bridges}
\label{Appendix_Bridges}
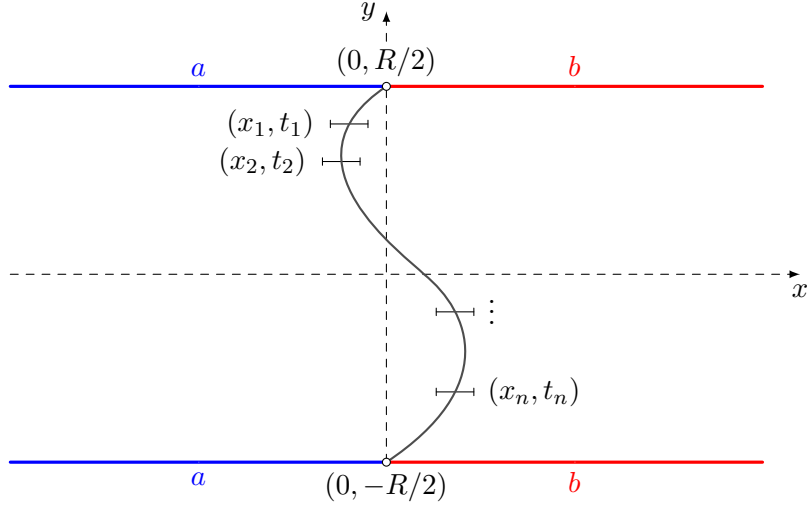
\begin{figure}[htbp]
\centering
\begin{tikzpicture}[thick, line cap=round, >=latex, scale=0.5]
\tikzset{fontscale/.style = {font=\relsize{#1}}}
%\draw[thin, gray!20, help lines] (-10 ,-5) grid (10 ,5);
\draw[thin, dashed, ->] (-10, 0) -- (11, 0) node[below] {$x$};
%\draw[thin, dashed, ->] (2, 2) -- (3, 2) node[below] {$x$};
\def\w{0.1}

\draw[thin, -] (-0.99-0.5, 4) -- (-0.99+0.5 , 4)  node[left] {};
\draw[thin, -] ($(-0.99-0.5, 4) + (0,-\w)$) -- ($(-0.99-0.5, 4) + (0,\w)$);
\draw[thin, -] ($(-0.99+0.5, 4) + (0,-\w)$) -- ($(-0.99+0.5, 4) + (0,\w)$);
\node at (-0.99-2.1, 4) [fontscale=0.04] {${(x_{1},t_{1})}$};

\draw[thin, -] (-1.2-0.5, 3) -- (-1.2+0.5 , 3)  node[left] {};
\draw[thin, -] ($(-1.2-0.5, 3) + (0,-\w)$) -- ($(-1.2-0.5, 3) + (0,\w)$);
\draw[thin, -] ($(-1.2+0.5, 3) + (0,-\w)$) -- ($(-1.2+0.5, 3) + (0,\w)$);
\node at (-1.2-2.1, 3) [fontscale=0.04] {${(x_{2},t_{2})}$};

\draw[thin, -] (1.825-0.5, -1) -- (1.825+0.5 , -1)  node[left] {};
\draw[thin, -] ($(1.825-0.5, -1) + (0,-\w)$) -- ($(1.825-0.5, -1) + (0,\w)$);
\draw[thin, -] ($(1.825+0.5, -1) + (0,-\w)$) -- ($(1.825+0.5, -1) + (0,\w)$);
\node at (1.825+0.98, -0.79) [fontscale=0.04] {$\vdots$};

\draw[thin, -] (1.825-0.5, -3.13) -- (1.825+0.5 , -3.13)  node[left] {};
\draw[thin, -] ($(1.825-0.5, -3.13) + (0,-\w)$) -- ($(1.825-0.5, -3.13) + (0,\w)$);
\draw[thin, -] ($(1.825+0.5, -3.13) + (0,-\w)$) -- ($(1.825+0.5, -3.13) + (0,\w)$);
\node at (1.825+2.1, -3.13) [fontscale=0.04] {${(x_{n},t_{n})}$};
  
\draw[thin, dashed, -] (0, -5) -- (0, 5) node[left] {};
\draw[thin, dashed, ->] (0, 6.2) -- (0, 7) node[left] {$y$};
\draw[very thick, red, -] (0, 5) -- (10, 5) node[] {};
\draw[very thick, red, -] (0, -5) -- (10, -5) node[] {};
\draw[very thick, blue, -] (-10, 5) -- (0, 5) node[] {};
\draw[very thick, blue, -] (-10, -5) -- (0, -5) node[] {};
\draw[black!70]
(0, -5) ..controls +(3.0, 2.0) and ( $(1, 0) + (1.25, -1.0195)$ )..
(1, 0) ..controls +(-0.5, 0.5) and ( $(0, 5) - (3.0, 2.0)$ )..
(0, 5);
\draw[thin, fill=white] (0, -5) circle (3pt) node[below] {$(0,-R/2)$};;
\draw[thin, fill=white] (0, 5) circle (3pt) node[above] {$(0,R/2)$};;
\draw[thin, fill=white] (-5, 5) circle (0pt) node[above] {${\color{blue}{a}}$};;
\draw[thin, fill=white] (-5, -5) circle (0pt) node[below] {${\color{blue}{a}}$};;
\draw[thin, fill=white] (5, 5) circle (0pt) node[above] {${\color{red}{b}}$};;
\draw[thin, fill=white] (5, -5) circle (0pt) node[below] {${\color{red}{b}}$};;
\end{tikzpicture}
\caption{The multi-interval construction of the passage probability $P_{n}$ with the time $t_{j}$ identified according to (\ref{19012021_1823}).}
\label{fig_passage}
\end{figure}

In this appendix we recall how to compute the probability density function of a Brownian bridge and show how it relates to the $(n+1)$-fold integral (\ref{27052020_05}). The Brownian bridge is defined as a Brownian motion which is constrained to come back to the initial position after a fixed amount of time $T$. To be definite, let us consider the origin $x=0$ as the initial position of a Brownian motion which moves in one spatial dimension. The diffusion equation is solved by the transition probability \cite{Gardiner}
\begin{equation}
\label{29012021_1924}
W(x_{1},t_{1} \vert x_{0},t_{0}) = \frac{1}{\sqrt{4\pi D (t_{1}-t_{0})}} \exp\biggl[ - \frac{(x_{1}-x_{0})^{2}}{4D (t_{1}-t_{0})} \biggr]
\end{equation}
where $D$ is the diffusion coefficient, $(x_{0},t_{0})$ defines the initial state and $(x_{1},t_{1})$ the final one. Since (\ref{29012021_1924}) is a probability density, $\int_{\mathbb{R}}\textrm{d}x_{1} \, W(x_{1},t_{1} \vert x_{0},t_{0})=1$. Let $I_{j}=(x_{j},x_{j}+{\rm d }x_{j})$ be a space interval at time $t_{j}$ as shown in Fig.~\ref{fig_passage}. 

The above problem can be brought in touch with the passage probability which appears in (\ref{27052020_05}) by shifting and rescaling the time by means of
\begin{equation}
\label{19012021_1823}
\frac{ t_{j} }{ T } = \frac{y_{j}}{R}+\frac{1}{2} = \frac{1+\tau_{j}}{2}
\end{equation}
and by setting $DT=\lambda^{2}$. The net probability for the Brownian walker to cross \emph{all} intervals $\{ I_{j} \}_{j=1,\dots,n}$ and come back to $x=0$ at time $t=T$ reads $P_{n}(x_{1},y_{1}; \dots ;x_{n},y_{n}) \textrm{d}x_{1} \cdots \textrm{d}x_{n}$, with
\begin{equation}
\label{30012021_0848}
P_{n}(x_{1},y_{1}; \dots ;x_{n},y_{n}) = \frac{ W(0,T \vert x_{1}, t_{1}) \left( \prod_{j=1}^{n-1} W(x_{j},t_{j} \vert x_{j+1}, t_{j+1}) \right) W(x_{n},t_{n} \vert 0, 0) }{ W(0, T \vert 0,0) } \, .
\end{equation}
. We note that for $n=1$ the passage probability is
\begin{equation}
\label{ }
P_{1}(x,y) = \frac{1}{\sqrt{\pi}\kappa\lambda}\textrm{e}^{-\chi^{2}}
\end{equation}
and for $n=2$, we have
\begin{equation}
\label{ }
P_{2}(x_{1},y_{1} ; x_{2},y_{2}) = \frac{1}{\pi\lambda^{2}\sqrt{2(1-\tau_{1})(\tau_{1}-\tau_{2})(1+\tau_{2})}} \exp\biggl[ - \frac{1}{2} \left( \frac{ \eta_{1}^{2} }{ 1-\tau_{1} } + \frac{ (\eta_{1}-\eta_{2})^{2} }{ \tau_{1}-\tau_{2} } + \frac{ \eta_{2}^{2} }{ 1+\tau_{2} } \right) \biggr] \, .
\end{equation}

We can now relate the field-theoretical calculation with the passage probability of the Brownian bridge computed from the transition probability. The $(n+1)$-fold integral in (\ref{27052020_05}) has the structure
\begin{equation}
\label{30012021_0912}
\Lbag 1 \Rbag_{\eta_{1},\tau_{1}; \dots ;\eta_{n},\tau_{n}} = N_{n} \textrm{e}^{-B_{n}} \, ,
\end{equation}
with
\begin{equation}
\label{ }
B_{n} = \frac{1}{2} \sum_{j=0}^{n} \frac{ (\eta_{j}-\eta_{j+1})^{2} }{ \tau_{j}-\tau_{j+1} } \, ,
\end{equation}
the normalization factor
\begin{equation}
\label{ }
N_{n} = 2\sqrt{\pi} (2\pi)^{-(n+1)/2} \prod_{j=0}^{n} (\tau_{j}-\tau_{j+1})^{-1/2} \, ,
\end{equation}
and $(\eta_{0},\tau_{0})=(0,1)$, $(\eta_{n+1},\tau_{n+1})=(0,-1)$. The quantity $B_{n}$ is a quadratic form of the coordinates $\eta_{j}$. By defining the column vector $\bm{\eta} = (\eta_{1},\dots\eta_{n})^{T}$, one can write
\begin{equation}
\label{ }
B_{n} = \bm{\eta}^{T} \cdot \textsf{B} \cdot \bm{\eta}
\end{equation}
where $\textsf{B}$ is the symmetric matrix whose entries in the upper triangle and main diagonal are given by
\begin{equation}
\label{ }
(\textsf{B} )_{ij}
= 
\begin{cases}
2^{-1} (\tau_{i-1}-\tau_{i})^{-1} + 2^{-1}(\tau_{i}-\tau_{i+1})^{-1} \, ,      & j = i \\
-2^{-1}(\tau_{i}-\tau_{j})^{-1} \, ,      & j = i+1 \\
0 \, ,      & j \geqslant i+1 \, .
%(\textsf{B}_{n})_{ji} \, , & j < i \, .
\end{cases}
\end{equation}
The following properties are easily established
\begin{equation}
\label{ }
\left( \textsf{B}^{-1} \right)_{ij}
=
\begin{cases}
(1-\tau_{i})(1+\tau_{j}) \, ,      & j \geqslant i \\
(1-\tau_{j})(1+\tau_{i}) \, , & j \leqslant i \, ,
\end{cases}
\end{equation}
and
\begin{equation}
\label{30012021_0936}
\det \textsf{B} = 2^{-(n-1)} \prod_{j=0}^{n} \left( \tau_{j} - \tau_{j+1} \right)^{-1} \, .
\end{equation}
It is convenient to express the passage probability in terms of the rescaled variables $\chi_{j}$ defined in the main body of the paper. To this end, we introduce the column vector $\bm{\chi} = (\chi_{1},\dots,\chi_{n})^{T}$ with $\chi_{j}=\eta_{j}/\kappa_{j}$ and $\kappa_{j}=\sqrt{1-\tau_{j}^{2}}$. The change of basis from $\eta_{j}$ variables to $\chi_{j}$ variables is implemented by the (diagonal) matrix $\textsf{U}=\textrm{diag}(\kappa_{1},\dots,\kappa_{n})$; hence $\bm{\eta} = \textsf{U} \bm{\chi}$. The quadratic form $B_{n}$ becomes
\begin{equation}
\begin{aligned}
\label{}
B_{n} & = \bm{\chi}^{T} \cdot \textsf{U} \textsf{B} \textsf{U} \cdot \bm{\chi} \, .
\end{aligned}
\end{equation}

The $n$-variate normal distribution with zero mean has the density
\begin{equation}
\label{30012021_0913}
\Pi_{n}( u_{1}, \dots, u_{n} \vert \textsf{R} ) = \frac{1}{(2\pi)^{n/2}\sqrt{\det\textsf{R}}} \textrm{e}^{- \frac{1}{2} \bm{u}^{T} \cdot \textsf{R}^{-1} \cdot \bm{u}} \, ,
\end{equation}
with $\bm{u}=(u_{1},\dots,u_{n})^{T}$ and $\textsf{R}$ the correlation coefficient.

In order to cast (\ref{30012021_0912}) in to the form (\ref{30012021_0913}), we identify the correlation coefficient $\textsf{R}$ as follows
\begin{equation}
\label{ }
\textsf{R} = (\textsf{U} \textsf{B} \textsf{U})^{-1} \, .
\end{equation}
The correlation matrix is evidently symmetric and its matrix elements for $i \leqslant j$ are $(\textsf{R})_{ij} = \rho_{ij}$ with $\rho_{ii}=1$ and
\begin{equation}
\label{19012021_1924}
\rho_{ij} = \frac{(1-\tau_{i})(1+\tau_{j})}{\kappa_{i}\kappa_{j}} \, ;
\end{equation}
recalling that $\kappa_{j}=\sqrt{1-\tau_{j}^{2}}$, (\ref{19012021_1924}) agrees with (\ref{06012021_1031}). For any $i < k < j$ the Markov property \cite{Doob,McFadden}
\begin{equation}
\label{19012021_1925}
\rho_{ij} = \rho_{ik} \rho_{kj} \, ,
\end{equation}
follows as a direct consequence of (\ref{19012021_1924}). While the most generic correlation matrix is characterized by $n(n-1)/2$ entries, thanks to the Markov property the number of independent correlation coefficients is lowered to $n-1$. Thanks to (\ref{30012021_0936}), we have
\begin{equation}
\label{ }
\det \textsf{R} = 2^{(n-1)} \left( \prod_{j=1}^{n} \kappa_{j}^{-2} \right) \left( \prod_{j=0}^{n} \left( \tau_{j} - \tau_{j+1} \right) \right) \, ,
\end{equation}
from which we can read the normalization factor $N_{n}$
\begin{equation}
\label{ }
N_{n} = \frac{1}{(2\pi)^{n/2}\sqrt{\det\textsf{R}}} \prod_{j=1}^{n}(\sqrt{2}\kappa_{j}) \, .
\end{equation}
The result (\ref{30012021_0912}) is thus 
\begin{equation}
\begin{aligned}
\label{30012021_0947}
\Lbag 1 \Rbag_{\eta_{1},\tau_{1}; \dots ;\eta_{n},\tau_{n}} & = \left( \prod_{j=1}^{n} \sqrt{2}\kappa_{j} \right) \frac{1}{(2\pi)^{n/2}\sqrt{\det\textsf{R}}}  \textrm{e}^{- \bm{\chi}^{T} \cdot \textsf{R} \cdot \bm{\chi} } \\
& = \lambda^{n} \left( \prod_{j=1}^{n} \frac{\sqrt{2}\kappa_{j}}{\lambda} \right) \Pi_{n}(\sqrt{2}\bm{\chi} \vert \textsf{R}) \\
& = \lambda^{n} P_{n}(x_{1},y_{1}; \dots, x_{n},y_{n}) \, ,
\end{aligned}
\end{equation}
and the passage probability reads
\begin{equation}
\label{30012021_0950}
P_{n}(x_{1},y_{1}; \dots, x_{n},y_{n}) =  \left( \prod_{j=1}^{n} \frac{\sqrt{2}\kappa_{j}}{\lambda} \right) \Pi_{n} (\sqrt{2}\bm{\chi} \vert \textsf{R}) \, ;
\end{equation}
the above coincides with (\ref{27052020_05}) provided in the main body of the paper.

As a further consistency check, we compute the normalization. Let us consider the left hand side of (\ref{30012021_0947}). We can perform the $n$-fold integral with respect to the coordinates $x_{j}$ directly in the function $\Lbag 1 \Rbag_{\eta_{1},\tau_{1}; \dots ;\eta_{n},\tau_{n}}$. Since
\begin{equation}
\int_{\mathbb{R}}\textrm{d}x_{j} \, \textrm{e}^{ \im \eta_{j} (\theta_{j}-\theta_{j+1})} = 2\pi \lambda \delta(\theta_{j}-\theta_{j+1}) \, ,
\end{equation}
we find
\begin{equation}
\begin{aligned}
\int_{\mathbb{R}^{n}} \textrm{d}x_{1} \cdots \textrm{d}x_{n} \, \Lbag 1 \Rbag_{\eta_{1},\tau_{1}; \dots ;\eta_{n},\tau_{n}} & = (2\pi\lambda)^{n} \scaleleftright[1.2ex]{\Lbag}{    \prod_{j=1}^{n} \delta(\theta_{j}-\theta_{j+1})    }{\Rbag} \\
& = \lambda^{n} \frac{1}{\sqrt{\pi}} \int_{\mathbb{R}}\textrm{d}\theta \, \textrm{e}^{-\theta^{2}} \\
& = \lambda^{n} \, .
\end{aligned}
\end{equation}
Let us consider the right hand side of (\ref{30012021_0947}). By applying the rescaling of integration integration variables $\sqrt{2}\chi_{j}=u_{j}$ in (\ref{30012021_0950}), we find the normalization
\begin{equation}
\begin{aligned}
\int_{\mathbb{R}^{n}} \textrm{d}x_{1} \cdots \textrm{d}x_{n} \, P_{n}(x_{1},y_{1}; \dots, x_{n},y_{n}) & = \int_{\mathbb{R}^{n}} \textrm{d}u_{1} \cdots \textrm{d}u_{n} \, \Pi_{n} (\bm{u} \vert \textsf{R}) \\
& = 1 \, ,
\end{aligned}
\end{equation}
which completes the check of (\ref{30012021_0947}).

%==================================================================================
\section{Triplet correlations}
\label{Appendix_B}

%=====================================================================================
\subsection{Leading order}
The results given in Sec.~\ref{sec_triplet} are derived in this appendix. The one-body cluster functions for the correlation function (\ref{29012021_1758}) are: $\mathscr{G}_{1}(\bm{x}_{1}) = \mathscr{G}_{1}(\bm{x}_{3}) =0$, and $\mathscr{G}_{1}(\bm{x}_{2}) = \textrm{erf}(\eta)$, with $\eta=x/\lambda$. In order to find the two-body cluster functions, we need to recall the following identity
\begin{equation}
\begin{aligned}
\Phi_{2}(\sqrt{2}\eta,0 \vert \rho_{12}) & = \int_{-\infty}^{\sqrt{2}\eta}\textrm{d}u_{1} \int_{-\infty}^{0}\textrm{d}u_{2} \, \Pi_{2}(u_{1},u_{2} \vert \rho_{12}) \\
& = \frac{1}{2\sqrt{\pi}} \int_{-\infty}^{\eta}\textrm{d}t \, \textrm{e}^{-t^{2}} \textrm{erfc}(rt) \, , \qquad r \equiv \frac{\rho_{12}}{\sqrt{1-\rho_{12}^{2}}} \\
& = \frac{1}{4} + \frac{1}{4} \textrm{erf}(\eta) + T(\sqrt{2}\eta,r) \, ;
\end{aligned}
\end{equation}
in the last line, we used the following property of Owen's $T$ function \cite{Owen1956,Owen1980}
\begin{equation}
\label{ }
\int_{-\infty}^{\eta} \textrm{d}t \, \textrm{e}^{-t^{2}} \textrm{erf}(rt) = -2\sqrt{\pi} T(\sqrt{2}\eta, r) \, .
\end{equation}
From the above, we find the following two-body cluster functions
\begin{equation}
\begin{aligned}
\label{}
\mathscr{G}_{2}(\bm{x}_{1}, \bm{x}_{2}) & = \mathscr{G}_{2}(\bm{x}_{2}, \bm{x}_{3}) = 4 T(\sqrt{2}\eta,r) \, .
\end{aligned}
\end{equation}
The calculation of $\mathscr{G}_{2}(\bm{x}_{1}, \bm{x}_{3})$ proceeds as follows:
\begin{equation}
\begin{aligned}
\label{}
\mathscr{G}_{2}(\bm{x}_{1}, \bm{x}_{3}) & = 4 \Phi_{2}(0, 0 \vert \rho_{13}) - 4 \Phi_{1}(0) + 1 \\
& = 4 \Phi_{2}(0, 0 \vert \rho_{13}) - 1\\
\end{aligned}
\end{equation}
but since
\begin{equation}
\label{ }
\Phi_{2}(0, 0 \vert \rho_{13}) = \frac{1}{4} + \frac{1}{2\pi} \sin^{-1}(\rho_{13})
\end{equation}
and $\rho_{13}=\rho_{12}^{2}$, we have
\begin{equation}
\begin{aligned}
\label{}
\mathscr{G}_{2}(\bm{x}_{1}, \bm{x}_{3}) & = \frac{2}{\pi} \sin^{-1}(\rho_{12}^{2}) \, .
\end{aligned}
\end{equation}
For the three-body cluster function, we need the following result for the cumulative distribution function of the trivariate normal distribution with correlation coefficients $\rho_{12}=\rho_{23}$, and $\rho_{13}=\rho_{12}^{2}$, 
\begin{equation}
\begin{aligned}
\label{}
\Phi_{3}(0, \sqrt{2} \eta, 0 \vert \rho_{12}, \rho_{12}^{2}, \rho_{12}) & = \int_{-\infty}^{0}\textrm{d}u_{1} \int_{-\infty}^{\sqrt{2}\eta}\textrm{d}u_{2} \int_{-\infty}^{0}\textrm{d}u_{3} \, \Pi_{3}(u_{1},u_{2},u_{3} \vert \rho_{12}, \rho_{12}^{2}, \rho_{12}) \\
& = \frac{1}{4\sqrt{\pi}} \int_{-\infty}^{\eta}\textrm{d}t \, \textrm{e}^{-t^{2}} \textrm{erfc}^{2}(rt) \\
& = \frac{1}{8} + \frac{1}{8} \textrm{erf}(\eta) + T(\sqrt{2}\eta,r) + \frac{1}{8} \mathcal{Y}(+\infty,r) + \frac{1}{8} \mathcal{Y}(\eta,r) \, .
\end{aligned}
\end{equation}
It is then easy to see that
\begin{equation}
\label{ }
\mathcal{Y}(+\infty,r) = \frac{2}{\pi} \sin^{-1}(\rho_{12}^{2}) \, .
\end{equation}
Finally, the three-body cluster function reads $\mathscr{G}_{3}(\bm{x}_{1}, \bm{x}_{2}, \bm{x}_{3}) = \mathcal{Y}(\eta,r)$. Collecting the above findings, the result (\ref{29012021_1804}) given in the main text can be easily assembled.

%=====================================================================================
\subsection{Subleading correction}
\label{appendix_B2}
We provide the analytic form of the three-point correlation function along the interface including corrections at order $\Os(R^{-1/2})$. The calculation of the leading-order term for $x_{1}=x_{2}=x_{3}=0$ can be computed straightforwardly from the probabilistic interpretation. By recalling the expressions for the quadrant and orthant probabilities given in (\ref{10042021_0935}), we find the first term given below
\begin{equation}
\begin{aligned}
\langle \sigma_{c}(0,y_{1}) \sigma_{c}(0,y_{2}) \sigma_{c}(0,y_{3}) \rangle_{ab} & = \left( \frac{ \langle \sigma_{c} \rangle_{a} + \langle \sigma_{c} \rangle_{b} }{2} \right)^{3} + \frac{1}{2\pi} \left( \langle \sigma_{c} \rangle_{a} - \langle \sigma_{c} \rangle_{b}  \right)^{2} \left( \frac{ \langle \sigma_{c} \rangle_{a} + \langle \sigma_{c} \rangle_{b} }{2} \right) \times \\
& \times \biggl[ \sin^{-1}(\rho_{12}) + \sin^{-1}(\rho_{13}) + \sin^{-1}(\rho_{23}) \biggr] \\
& + [\langle \sigma_{c}(0,y_{1}) \sigma_{c}(0,y_{2}) \sigma_{c}(0,y_{3}) \rangle_{ab}]_{1} + \Os(R^{-1}) \, ,
\end{aligned}
\end{equation}
which coincides with the limit $x\rightarrow0$ of (\ref{09042021_0931}), as consistency requires. The correction proportional to $R^{-1/2}$ is given by
\begin{equation}
\begin{aligned}
\label{08042021_2254}
{[ \langle \sigma_{c}(0,y_{1}) \sigma_{c}(0,y_{2}) \sigma_{c}(0,y_{3}) \rangle_{ab} } ]_{1} & = \frac{ A_{ab}^{(\sigma_{c})} }{ \sqrt{\pi}\lambda } \left( \frac{ \langle \sigma_{c} \rangle_{a} + \langle \sigma_{c} \rangle_{b} }{2} \right)^{2} \left( \frac{1}{\kappa_{1}} + \frac{1}{\kappa_{2}} + \frac{1}{\kappa_{3}} \right) \\
& + \frac{ A_{ab}^{(\sigma_{c})} }{ \sqrt{\pi}\lambda } \frac{ \left( \langle \sigma_{c} \rangle_{a} - \langle \sigma_{c} \rangle_{b} \right)^{2} }{ 2\pi } \biggl[ \frac{\tan^{-1}J_{1}}{\kappa_{1}} + \frac{\tan^{-1}J_{3}}{\kappa_{3}} \biggr] \, ,
\end{aligned}
\end{equation}
with $\kappa_{j}=\sqrt{1-\tau_{j}^{2}}$ and
\begin{equation}
J_{1} = \rho_{23} \sqrt{\frac{1-\rho_{12}^{2}}{1-\rho_{23}^{2}}} \, , \qquad J_{3} = \rho_{12} \sqrt{\frac{1-\rho_{23}^{2}}{1-\rho_{12}^{2}}} \, .
\end{equation}
The derivation of (\ref{08042021_2254}) within the probabilistic picture is immediate.

 %==================================================================================
\bibliographystyle{unsrt}
\bibliography{bibliography}{}

%==================================================================================
\end{document}